\documentclass[10pt,english,twocolumn,final,floatfix,showpacs,showkeys,citesort,preprintnumbers,prd,nofootinbib,superscriptaddress,letterpaper]{revtex4-1}
\usepackage[T1]{fontenc}
\usepackage[latin1]{inputenc}
\usepackage{color}
\usepackage{units}
\usepackage{amsmath}
\usepackage{graphicx}
\usepackage{amssymb}

\makeatletter

\newcommand{\noun}[1]{\textsc{#1}}
\providecommand{\tabularnewline}{\\}

 \@ifundefined{textcolor}{}
 {%
   \definecolor{BLACK}{gray}{0}
   \definecolor{WHITE}{gray}{1}
   \definecolor{RED}{rgb}{1,0,0}
   \definecolor{GREEN}{rgb}{0,1,0}
   \definecolor{BLUE}{rgb}{0,0,1}
   \definecolor{CYAN}{cmyk}{1,0,0,0}
   \definecolor{MAGENTA}{cmyk}{0,1,0,0}
   \definecolor{YELLOW}{cmyk}{0,0,1,0}
 }

\usepackage{nicefrac}
\usepackage[justification=centerlast]{caption,subfig}

\@ifundefined{showcaptionsetup}{}{%
 \PassOptionsToPackage{caption=false}{subfig}}
\usepackage{subfig}
\makeatother

\usepackage{babel}

\begin{document}
\newcommand{\gesim}{\,\raisebox{-0.4ex}{$\stackrel{>}{\scriptstyle\sim}$}\,} \newcommand{\lesim}{\,\raisebox{-0.4ex}{$\stackrel{<}{\scriptstyle\sim}$}\,}

\preprint{\vbox{ 
\null
\vspace{0.3in}
\hbox{SMU-HEP-12-06} 
\hbox{KA-TP-08-2012} 
\hbox{LPSC-12-049}
\hbox{MZ-TH/12-11}
}}

\title{\null
\vspace{0.5in}Strange Quark PDFs and Implications for Drell-Yan Boson Production
at the LHC}

\author{A.~Kusina}

\thanks{akusina@smu.edu}

\affiliation{Southern Methodist University, Dallas, TX 75275, USA}

\author{T.~Stavreva}

\thanks{stavreva@lpsc.in2p3.fr}

\affiliation{Laboratoire de Physique Subatomique et de Cosmologie, Universit\'e
Joseph Fourier/CNRS-IN2P3/INPG, \\
 53 Avenue des Martyrs, 38026 Grenoble, France}

\author{S.~Berge}

\thanks{berge@uni-mainz.de}

\affiliation{Institute for Physics (WA THEP), Johannes Gutenberg-Universit\"at,
\\
D-55099 Mainz, Germany}

\author{F.~I.~Olness}

\thanks{olness@smu.edu}

\affiliation{Southern Methodist University, Dallas, TX 75275, USA}

\author{I.~Schienbein}

\thanks{schien@lpsc.in2p3.fr}

\affiliation{Laboratoire de Physique Subatomique et de Cosmologie, Universit\'e
Joseph Fourier/CNRS-IN2P3/INPG, \\
 53 Avenue des Martyrs, 38026 Grenoble, France}

\author{K.~Kova\v{r}\'{\i}k}

\thanks{kovarik@particle.uni-karlsruhe.de}

\affiliation{Institute for Theoretical Physics, Karlsruhe Institute of Technology,
Karlsruhe, D-76128, Germany}

\author{T.~Je\v{z}o}

\thanks{jezo@lpsc.in2p3.fr}

\affiliation{Laboratoire de Physique Subatomique et de Cosmologie, Universit\'e
Joseph Fourier/CNRS-IN2P3/INPG, \\
 53 Avenue des Martyrs, 38026 Grenoble, France}

\author{J.~Y.~Yu}

\thanks{yu@physics.smu.edu}

\affiliation{Southern Methodist University, Dallas, TX 75275, USA}

\affiliation{Laboratoire de Physique Subatomique et de Cosmologie, Universit\'e
Joseph Fourier/CNRS-IN2P3/INPG, \\
 53 Avenue des Martyrs, 38026 Grenoble, France}

\author{K.~Park}

\affiliation{Southern Methodist University, Dallas, TX 75275, USA}


\begin{abstract}
Global analyses of Parton Distribution Functions (PDFs) have provided
incisive constraints on the up and down quark components of the proton,
but constraining the other flavor degrees of freedom is more challenging.
Higher-order theory predictions and new data sets have contributed
to recent improvements. Despite these efforts, the strange quark PDF
has a sizable uncertainty, particularly in the small $x$ region.
We examine the constraints from experiment and theory, and investigate
the impact of this uncertainty on LHC observables.\textbf{ }In particular,
we study $W/Z$ production to see how the $s$-quark uncertainty propagates
to these observables, and examine the extent to which precise measurements
at the LHC can provide additional information on the proton flavor
structure. 
\end{abstract}
\maketitle
\tableofcontents{}

\cleardoublepage{}

\section{Introduction}

\subsection{Motivation}

Parton distribution functions (PDFs) provide the essential link between
the theoretically calculated partonic cross-sections, and the experimentally
measured physical cross-sections involving hadrons and mesons. This
link is crucial if we are to make incisive tests of the standard model,
and search for subtle deviations which might signal new physics.

Recent measurements of charm production in neutrino deeply-inelastic
scattering (DIS), visible as di-muon final states, provide important
new information on the strange quark distribution, $s(x)$, of the
nucleon \cite{Abramowicz:1982zr,Abramowicz:1982re,Abramowicz:1984yk,Berge:1987zw,Aivazis:1993pi,Bazarko:1994tt,Gluck:1996ve,Alton:2000ze,Goncharov:2001qe,Kretzer:2003it,Nadolsky:2002jr,Pumplin:2002vw,Spentzouris:2002va,Tung:2001mv,Tzanov:2005dp,Tzanov:2005kr}.
We show that despite these recent advances in both the precision data
and theoretical predictions, the relative uncertainty on the heavier
flavors remains large. We will focus on the strange quark and show
the impact of these uncertainties on selected LHC processes. 

The production of $W/Z$ bosons is one of the {}``benchmark'' processes
used to calibrate our searches for the Higgs boson and other {}``new
physics'' signals. We will examine how the uncertainty of the strange
quark PDF influences these measurements, and assess how these uncertainties
might be reduced.

\subsection{Outline}

The outline of the presentation is as follows. In Section~2, we examine
the experimental signatures that constrain the strange quark parton
distribution. In Section~3 we consider the impact of $s$-quark PDF
uncertainties on $W/Z$ production at the LHC, and in Section~4 we
summarize our results. Additional details on PDF fits to di-muon data
at next-to-leading order are provided in the Appendix.

\section{Constraining the PDF flavor components}

\subsection{Extracting the Strange Quark PDF}

\begin{figure}
\includegraphics[clip,width=0.45\textwidth]{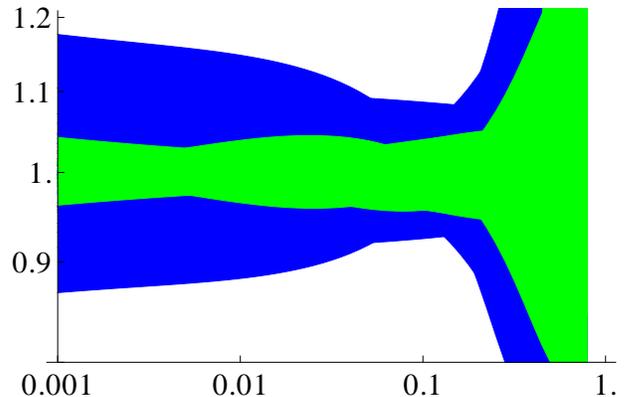}\caption{Relative uncertainty of the strange quark PDF as a function of $x$
for $Q=2$~GeV. The inner band is for the CTEQ6.1 PDF set, and the
outer band is for the CTEQ6.6 PDF set. The band is computed as the
envelope of $s_{i}(x)/s_{0}(x)$ where $s_{0}(x)$ is the central
PDF for each set; for CTEQ6.1, $i=[1,40]$, and for CTEQ6.6, $i=[1,44]$.
\label{fig:sband}}
\end{figure}

In previous global analyses, the predominant information on the strange
quark PDF $s(x)$ came from the difference of (large) inclusive cross
sections for neutral and charged current DIS. For example, at leading-order
(LO) in the parton model one finds that the difference between the
Neutral Current (NC) and Charged Current (CC) DIS $F_{2}$ structure
function is proportional to the strange PDF. Specifically if we neglect
the charm PDF and isospin-violating terms, we have~\cite{Adams:2009kp}
\begin{equation}
\Delta F_{2}=\frac{5}{18}F_{2}^{CC}-F_{2}^{NC}\sim\frac{x}{6}[s(x)+\bar{s}(x)]\quad.\label{eq:deltaf2}\end{equation}
Because the strange distributions are small compared to the large
up and down PDFs, the $s(x)$ extracted from this measurement has
large uncertainties. Lacking better information, it was commonly assumed
the distribution was of the form \begin{equation}
s(x)=\bar{s}(x)\sim\kappa[\bar{u}(x)+\bar{d}(x)]/2\label{eq:kappa}\end{equation}
with $\kappa\sim\nicefrac{1}{2}$. 

This approach was used, for example, in the CTEQ6.1 PDFs~\cite{Stump:2003yu}.
In Figure~\ref{fig:sband} we show the relative uncertainty band
of the strange quark PDF for the 40 CTEQ6.1 PDF error sets relative
to the central value. We observe that over much of the $x$-range
the relative uncertainty on the strange PDF is $\lesim5\%$. The relation
of Eq.~\eqref{eq:kappa} tells us that this uncertainty band in fact
reflects the uncertainty on the up and down sea which is well constrained
by DIS measurements; this does not reflect the true uncertainty of
$s(x)$. 

Beginning with CTEQ6.6 PDFs~\cite{Nadolsky:2008zw} the neutrino--nucleon
dimuon data was included in the global fits to more directly constrain
the strange quark; thus, Eq.~\eqref{eq:kappa} was not used, and
two additional fitting parameters were introduced to allow the strange
quark to vary independently of the up and down sea. We also display
the relative uncertainty band for the CTEQ6.6 PDF set in Fig.~\ref{fig:sband}.
We now observe that the relative error on the strange quark is much
larger than for the CTEQ6.1 set, particularly for $x<0.01$ where
the neutrino--nucleon dimuon data do not provide any constraints.
We expect this is a more accurate representation of the true uncertainty. 

This general behavior is also exhibited in other global PDF sets with
errors~\cite{Martin:2009iq,Ball:2010de,Alekhin:2009ni,JimenezDelgado:2008hf}.
For example, the NNPDF collaboration uses a parameterization-free
method for extracting the PDFs; they observe a large increase in the
$s(x)$ uncertainty in the small $x$ region which is beyond the constraints
of the $\nu$-DIS experiments. (Cf. in particular Fig.~13 of Ref~\cite{Ball:2010de}.)

Thus, there is general agreement that the strange quark PDF is poorly
constrained, particularly in the small $x$ region.

\subsection{Constraints from CCFR and NuTeV}

The primary source of information on the strange quark at present
comes from high-statistics neutrino--nucleon DIS measurements; in
particular, the CCFR and NuTeV dimuon experiments have been used to
determine the strange quark PDF with improved accuracy \cite{Mason:2007zz,Tzanov:2005kr,Zeller:2002du,Goncharov:2001qe,Yang:2000ju,Alton:2000ze,Bazarko:1994tt}.
Neutrino induced dimuon production ($\nu N\rightarrow\mu^{+}\mu^{-}X$)
proceeds primarily through the Cabibbo favored $s\to c$ or $\bar{s}\to\bar{c}$
subprocess. Hence, this provides information on $s$ and $\bar{s}$
directly; this is in contrast to $\Delta F_{2}$ of Eq.~\eqref{eq:deltaf2}.
CCFR has 5030 $\nu$ and 1060 $\bar{\nu}$ di-muon events, and NuTeV
has 5012 $\nu$ and 1458 $\bar{\nu}$ di-muon events, and these cover
the approximate range $x\sim[0.01,0.4]$. Additionally, NuTeV used
a sign-selected beam to separate the $\nu$ and $\bar{\nu}$ events
in order to separately extract $s(x)$ and $\overline{s}(x)$.

\subsubsection{Constraints on $s+\bar{s}$}

\begin{table}
\begin{tabular}{|c|c|c|}
\hline 
\textbf{$\chi^{2}/DoF$} & \textbf{CTEQ6M} & \textbf{Free}\tabularnewline
\hline
\hline 
CCFR $\nu$ dimuon & \emph{1.02} & \emph{0.72}\tabularnewline
\hline 
CCFR $\bar{\nu}$ dimuon & \emph{0.58} & \emph{0.59}\tabularnewline
\hline 
NuTeV $\nu$ dimuon & \emph{1.81} & \emph{1.44}\tabularnewline
\hline 
NuTeV $\bar{\nu}$ dimuon & \emph{1.48} & \emph{1.13}\tabularnewline
\hline
\hline 
BCDMS $F_{2}^{p}$ & 1.11 & 1.11\tabularnewline
\hline 
BCDMS $F_{2}^{d}$ & 1.10 & 1.11\tabularnewline
\hline 
H1 96/97 & 0.94 & 0.94\tabularnewline
\hline 
H1 98/99 & 1.02 & 1.03\tabularnewline
\hline 
ZEUS 96/97 & 1.14 & 1.15\tabularnewline
\hline 
NMC $F_{2}^{p}$ & 1.52 & 1.49\tabularnewline
\hline 
NMC $F_{2}^{d}/F_{2}^{p}$ & 0.91 & 0.91\tabularnewline
\hline 
CCFR $F_{2}$ & 1.70 & 1.88\tabularnewline
\hline 
CCFR $F_{3}$ & 0.42 & 0.42\tabularnewline
\hline 
E605 & 0.82 & 0.83\tabularnewline
\hline 
NA51 & 0.62 & 0.52\tabularnewline
\hline 
CDF $\ell$ Asym & 0.82 & 0.82\tabularnewline
\hline 
E866 & 0.39 & 0.38\tabularnewline
\hline 
D0 Jets & 0.71 & 0.67\tabularnewline
\hline 
CDF Jets & 1.48 & 1.47\tabularnewline
\hline
\hline 
\textbf{TOTAL $\chi^{2}$} & \textbf{2173} & \textbf{2133}\tabularnewline
\hline
\end{tabular}\caption{We display the $\chi^{2}/DoF$ for selected data sets using the CTEQ6M
PDF set~\cite{Pumplin:2002vw}, and a variant of this (labeled {}``Free'')
which allows for a modified strange quark PDF to accommodate the neutrino
dimuon data. \label{tab:fit}}
\end{table}

In Table~\ref{tab:fit} we illustrate how the bulk of the data used
in the global fits are relatively insensitive to the strange quark
distribution. The first column (labeled {}``CTEQ6M'') lists the
$\chi^{2}/DoF$ for a variety of data sets used in the CTEQ6M fit~\cite{Pumplin:2002vw}.
We have also shown the CCFR and NuTeV dimuon data sets in the table,
but these were \emph{not} used in the CTEQ6M fit. The second column
(labeled {}``Free'') lists results of refitting all the data --
including the dimuon data -- with a flexible strange quark PDF instead
of imposing the relation of Eq.~\eqref{eq:kappa}; this allows the
strange quark PDF to accommodate the dimuon data. Comparing the two
columns, we observe that the change of the strange PDF allowed for
a greatly improved fit of the dimuon data, while the other data sets
are virtually insensitive to this change.%
\footnote{The one exception is the CCFR $F_{2}$ which is mildly sensitive to
the strange quark PDF via Eq.~\eqref{eq:deltaf2}.%
} 

This exercise demonstrates that most of the data sets of the global
analysis are insensitive to the details of the strange quark PDF.

\subsubsection{Constraints on $s-\bar{s}$}

\begin{figure}[t]
\includegraphics[width=0.45\textwidth]{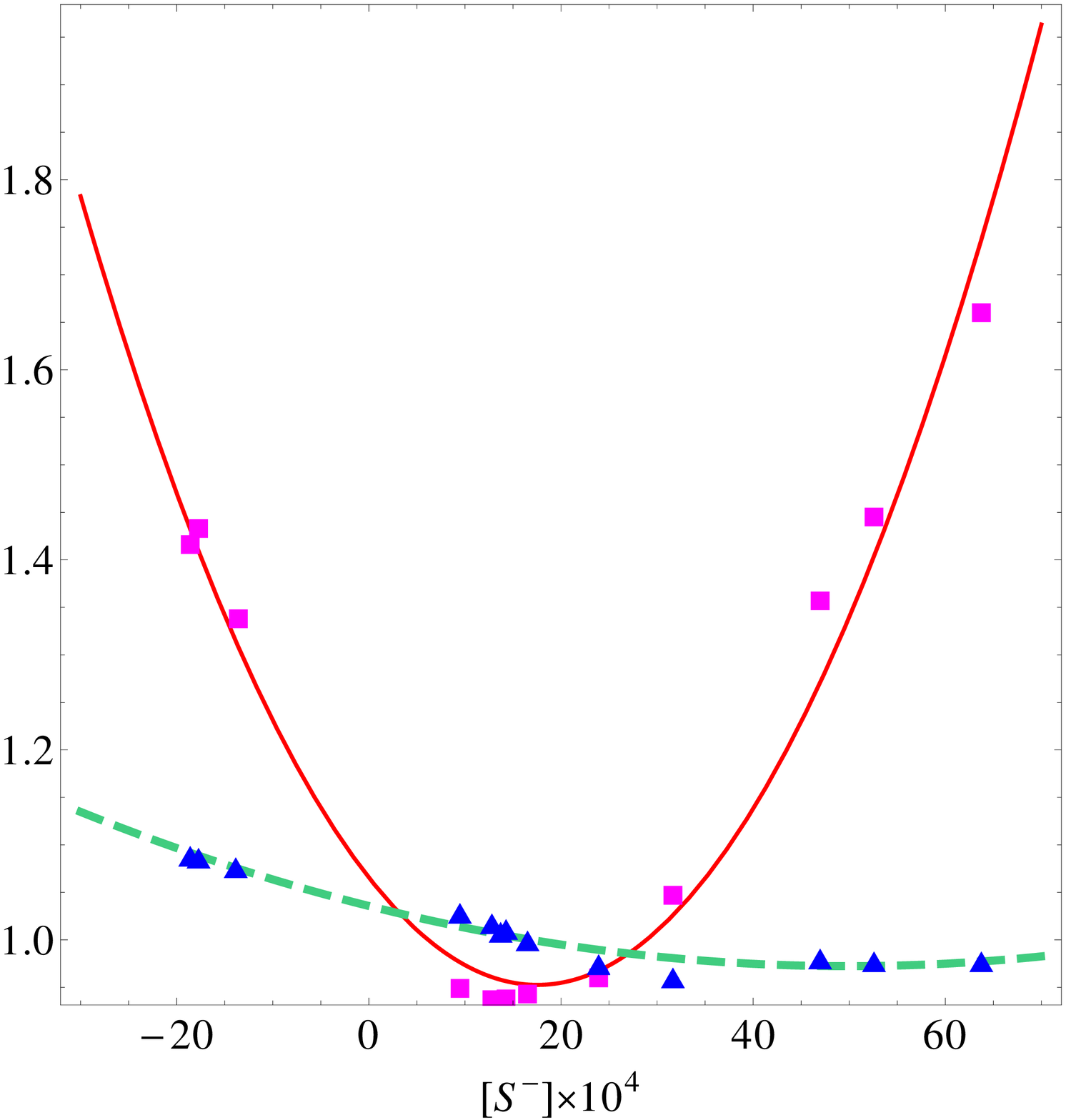}

\caption{We plot $\chi^{2}/\chi_{0}^{2}$ for the dimuon and the {}``Inclusive-I''
data sets evaluated as a function of the strange asymmetry $[S^{-}]\times10^{4}$.
The fits are denoted with \protect\includegraphics[scale=0.25]{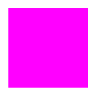}
for the dimuons and \protect\includegraphics[scale=0.2]{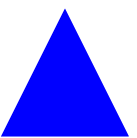}
for ``Inclusive-I''. Quadratic approximations to the fits are displayed
by the solid (red) line for the dimuons and the dashed (green) line
for ``Inclusive-I''.\textcolor{red}{{} \label{fig:chiDelS}}}
\end{figure}

The dimuon data can also provide information on the $s(x)$ and $\bar{s}(x)$
quark PDFs separately. In Fig.~\ref{fig:chiDelS} we display the
relative $\chi^{2}$ for the dimuon and ``Inclusive-I'' data sets
as a function of the strange asymmetry $\left[S^{-}\right]$, where

\begin{equation}
[S^{-}]\equiv\int_{0}^{1}x[s(x)-\bar{s}(x)]\, dx\;.\label{eq:StrMom}\end{equation}
The ``Inclusive-I'' data sets (cf., Ref.~\cite{Olness:2003wz}) contain
the data that is sensitive to $\left[S^{-}\right]$; specifically,
the data sets are a) the neutrino $xF_{3}$ data from CCFR and CDHSW
as this is proportional to the difference of quark and anti-quark
PDFs, and b) the CDF $W$--asymmetry measurement which can receive
contributions from the $sg\to Wc$ subprocess. Figure~\ref{fig:chiDelS}
clearly shows that the dimuon data provides the strongest constraints
on the strange asymmetry $\left[S^{-}\right]$.

\subsection{HERMES }

\begin{figure}
\includegraphics[width=0.95\columnwidth]{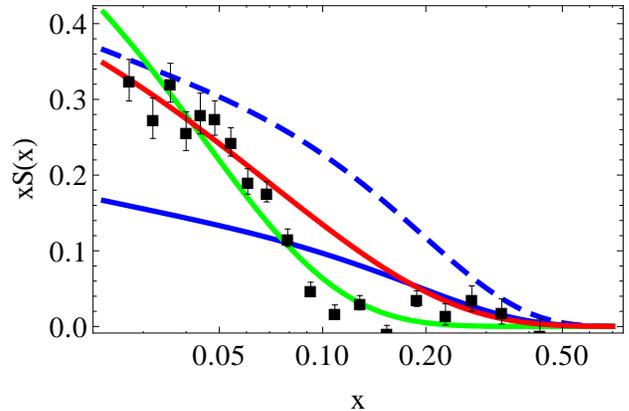}

\caption{The strange parton distribution $xS(x)=x[s(x)+\bar{s}(x)]$ from the
measured Hermes multiplicity for charged kaons evolved to $Q^{2}=2.5\,$GeV$^{2}$.
The solid green curve is a Hermes 3-parameter fit: $S(x)=x^{-0.924}e^{-x/0.0404}(1-x)$,
the dashed blue curve is the sum of light anti-quarks $x(\bar{u}+\bar{d})$
from CTEQ6L, the blue solid curve is $xS(x)$ from CTEQ6L, and the
red solid curve is the $xS(x)$ from CTEQ6.6. Hermes data points and
fit are from Ref.~\textcolor{black}{\cite{Airapetian:2008qf}}.\textcolor{red}{{}
}\label{fig:hermes}}
\end{figure}

The HERMES experiment measured the strange PDF via charged kaon production
in positron--deuteron DIS~\textcolor{black}{\cite{Airapetian:2008qf}},
these results are displayed in Fig.~\ref{fig:hermes}. For comparison,
the strange quark and total sea distributions from CTEQ6L are also
plotted. 

The HERMES data suggests that the $x$-dependence of the strange quark
distribution is quite different from the form assumed for the CTEQ6
set. In particular, they obtain a strange quark distribution that
is suppressed in the region $x\gtrsim0.1$ but then grows quickly
for $x<0.1$ and exceeds the CTEQ6L value in the small $x$ region
by more than a factor of two. 

To gauge the compatibility of this result with the displayed PDFs,
we can replace the initial $s(x)$ distribution with the form preferred
by HERMES, and then evaluate the shift of the $\chi^{2}$ with this
additional constraint. A preliminary investigation with this procedure
indicates that the HERMES $s(x)$ distribution could strongly influence
two data sets of the global fits. The first set is the neutrino-nucleon
dimuon data which controls $s(x)$ in the intermediate $x$ region.
The second set is the HERA measurement of $F_{2}$ in the small $x$
region where the statistical errors are particularly small. 

In Fig.~\ref{fig:hermes} we also show $xS(x)$ from CTEQ6.6; while
the HERMES data are below the CTEQ6.6 result in the $x\sim0.1$ region,
they agree quite well at both the higher and lower $x$ values. 

While these comparisons are sufficient to gauge the general influence
of the Hermes result, a complete analysis that includes the Hermes
data dynamically in the global fit is required to draw quantitative
conclusions.

\subsection{CHORUS }

The CHORUS experiment~\cite{Onengut:2004dy,Onengut:2005kv,KayisTopaksu:2008xp}
measured the neutrino structure functions $F_{2}$, $xF_{3}$, $R$
in collisions of sign selected neutrinos and anti-neutrinos with a
lead target (lead--scintillator CHORUS calorimeter) in the CERN SPS
neutrino beamline. They collected over 3M $\nu_{\mu}$ and 1M $\bar{\nu}_{\mu}$
charged current events in the kinematic range $0.01<x<0.7$, $0.05<y<0.95$,
$10<E_{\nu}<100$. 

This data was analyzed in the context of a global fit in Ref.~\cite{Owens:2007kp}
which was based on the CTEQ6.1 PDFs. This analysis made use of the
correlated systematic errors and found that the CHORUS data is generally
compatible with the other data sets, including the NuTeV data. Thus,
the CHORUS data is consistent with the strange distribution extracted
in CTEQ6.1.

\subsection{NOMAD }

The NOMAD experiment measured neutrino-induced charm dimuon production
to directly probe the $s$-quark PDF~\cite{Astier:2000us,Petti:2006tu,Wu:2007rv}.
Protons from the CERN SPS synchrotron (450~GeV) struck a beryllium
target to produce a neutrino beam with a mean energy of 27~GeV. NOMAD
used an iron-scintillator hadronic calorimeter to collect a very high
statistics (15K) neutrino-induced charm dimuon sample~\cite{Petti:2006tu}. 

Using kinematic cuts of $E_{\mu1},E_{\mu2}>4.5\,$GeV, $15<E_{\nu}<300\,$GeV,
and $Q^{2}>1$~GeV$^{2}$ NOMAD performed a leading-order QCD analysis
of 2714 neutrino- and 115 anti-neutrino--induced opposite sign dimuon
events~\cite{Astier:2000us}. The ratio of the strange to non-strange
sea in the nucleon was measured to be $\kappa=0.48_{-0.07-0.12}^{+0.09+0.17}$;
this is consistent with the values used in the global fits, \emph{c.f.},
Fig.~\ref{fig:kappa}. 

The data analysis is continuing, and it will be very interesting to
include this data set into the global fits as the large dimuon statistics
have the potential to strongly influence the extracted PDFs.

\subsection{MINER$\nu$A }

The cross sections in neutrino DIS experiments from NuTeV, CCFR, CHORUS
and NOMAD have been measured using heavy nuclear targets. In order
to use these measurements in a global analysis of proton PDFs, these
data must be converted to the corresponding proton or isoscalar results
\cite{deFlorian:2003qf,Hirai:2007sx,Eskola:2009uj,Schienbein:2007fs,Schienbein:2009kk,Kovarik:2010uv,Kovarik:2011dr}.
For example, the nuclear correction factors used in the CTEQ6 global
analysis were extracted from $\ell^{\pm}N$ DIS processes on a variety
of nuclei, and then applied to $\nu N$ DIS on heavy nuclear targets.
In a series of recent studies it was found that the $\ell^{\pm}N$
nuclear correction factors could differ substantially from the optimal
$\nu N$ nuclear correction factors~\textcolor{black}{\cite{Schienbein:2009kk,Kovarik:2010uv,Kovarik:2011zz,Kovarik:2011dr,Schienbein:2007fs}.} 

Furthermore, the nuclear corrections depend to a certain degree on
the specific observable as they contain different combinations of
the partons; the nuclear correction factors for dimuon production
will not be exactly the same as the ones for the structure function
$F_{2}$ or $F_{3}$. The impact of varying the nuclear corrections
on the strange quark PDF has to be done in the context of a global
analysis which we leave for a future study.

The MINER$\nu$A experiment has the opportunity to help resolve some
of these important questions as it can measure the neutrino DIS cross
sections on a variety of light and heavy targets. It uses the NuMI
beamline at Fermilab to measure low energy neutrino interactions to
study neutrino oscillations and also the strong dynamics of the neutrino--nucleon
interactions. MINER$\nu$A completed construction in 2010, and they
have begun data collection. MINER$\nu$A can measure neutrino interactions
on a variety of targets including plastic, helium, carbon, water,
iron, and lead. For $4*10^{20}$ Protons on Target (POT) they can
generate over 1M charged current events on plastic. 

These high statistics data on a variety of nuclear targets could allow
us to accurately characterize the nuclear correction factors as a
function of the nuclear $A$ from helium to lead. This data will be
very useful in resolving questions about the nuclear corrections,
and we look forward to the results in the near future.

\subsection{CDF \& DO }

At the Tevatron, the CDF~\cite{Aaltonen:2007dm} and D0~\cite{Abazov:2008qz}
collaborations measured $Wc$ final states in $p\bar{p}$ at $\sqrt{S}=1.96$~TeV
using the semileptonic decay of the charm and the correlation between
the charge of the $W$ and the charm decay. Additionally, a recent
study has investigated the impact of the $W$+dijet cross section
on the strange PDF \cite{Kawamura:2011dt}. These measurement are
especially valuable for two reasons. First, there are no nuclear correction
factors as the initial state is $p$ or $\bar{p}$. Second, this is
in a very different kinematic region as compared to the fixed-target
neutrino experiments. Thus, these have the potential to constrain
the strange quark PDF in a manner complementary to the $\nu N$ DIS
measurements; however, the hadron-hadron initial state is challenging.
Using approximately $1\, fb^{-1}$ of data, both CDF and D0 find their
measurements to be in agreement with theoretical expectations of the
Standard Model. Updated analyses with larger data sets are in progress
and it will be interesting to see the impact of these improved constraints
on the strange quark PDF.

\subsection{Strange Quark Uncertainty}

\begin{figure}
\includegraphics[width=0.45\textwidth]{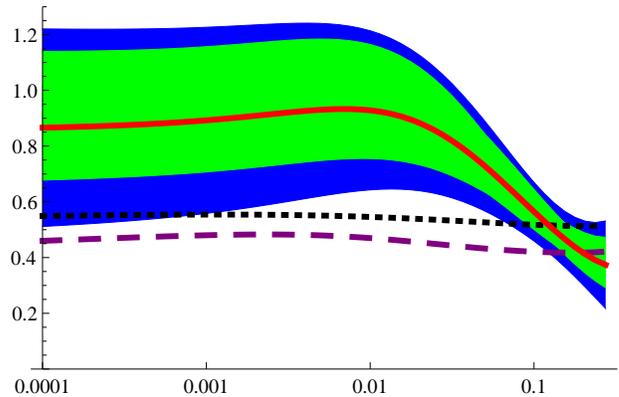}

\caption{$\kappa(x,Q)$ vs. $x$ for $Q=1.5$~GeV for a selection of PDFs,
where $\kappa(x,Q)$ is defined in Eq.~\eqref{eq:kappa2}. The curves
(top to bottom) are CTEQ6.6 (solid, red), CTEQ6.5 (dotted, black)
and CTEQ6.1 (dashed, purple). The wider (blue) band represents the
uncertainty for CTEQ6.6 as computed by Eq.~\eqref{eq:error}, the
inner (green) band represents uncertainty given by the envelope of
$\kappa(x,Q)$ values obtained with the 44 CTEQ6.6 error sets. \label{fig:kappa} }
\end{figure}

\begin{figure}
\includegraphics[width=0.45\textwidth]{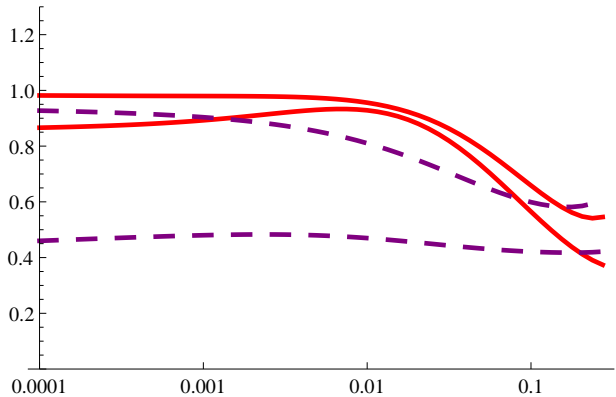}

\caption{$\kappa(x,Q)$ vs. $x$  showing the evolution from low to high scales.
The solid (red) lines are for CTEQ6.6, and the dashed (purple) lines
are for CTEQ6.1. The lower pair of lines (red and purple) are for
$Q=1.5$~GeV and the upper for $Q=80$~GeV. \label{fig:kappaEvolution} }
\end{figure}

The combination of the above results underscores the observation that
our knowledge of the strange quark is limited. To illustrate this
point in another manner, in Fig.~\ref{fig:kappa} we display $\kappa(x,Q)$
for a selection of PDF sets. Here, we define \begin{equation}
\kappa(x,Q)=\frac{s(x,Q)}{[\bar{u}(x,Q)+\bar{d}(x,Q)]/2}\label{eq:kappa2}\end{equation}
which is essentially a differential version of the $\kappa$ parameter
of Eq.~\eqref{eq:kappa}; this allows us to gauge the amount of the
strange PDF inside the proton compared to the average up and down
sea-quark PDFs. If we had exact $SU(3)$ symmetry we would expect
$\bar{u}=\bar{d}=\bar{s}$ and $\kappa(x,Q)\sim1$. As the strange
quark is heavier than the up and down quarks, we expect this component
to be suppressed relative to the up and down quarks, and we would
predict $\bar{u}\simeq\bar{d}>\bar{s}$ which would yield $\kappa(x,Q)<1$.
Thus, $\kappa(x,Q)$ is a measure of the $SU(3)$ breaking across
the $x$ and $Q$ range.

In Fig.~\ref{fig:kappa} we observe that the CTEQ6.1 and CTEQ6.5
PDF sets have $\kappa(x,Q)\sim\nicefrac{1}{2}$; this was by design
as the constraint of Eq.~\eqref{eq:kappa} was used to set the initial
$s(x)$ distribution. The exception is CTEQ6.6 which did not impose
Eq.~\eqref{eq:kappa}; we observe that this set has $\kappa(x,Q)\sim\nicefrac{1}{2}$
for $x\sim0.1$ (where the dimuon DIS data has smaller uncertainties),
but is a factor of two larger than the other PDF sets for small $x$
values. In Fig.~\ref{fig:kappa} we also show the uncertainty on
$s(x)$ computed as \cite{Pumplin:2002vw}

\begin{equation}
\Delta X=\frac{1}{2}\sqrt{\sum_{i=1}^{N_{p}}\left[X(S_{i}^{+})-X(S_{i}^{-})\right]^{2}}\label{eq:error}\end{equation}
which is shown as a (blue) band;%
\footnote{In Eq.~\eqref{eq:error}, $X$ is the observable, $S_{i}^{\pm}$
are the error PDF sets for eigenvalue $i$, and $N_{p}$ is the number
of eigenvalues. For CTEQ6.5 $N_{p}=20$, and for CTEQ6.6 $N_{p}=22$. %
} this results in a band which is larger than simply taking the spread
of the 44 CTEQ6.6 error PDFs (green band).

In order to show the effect of the DGLAP evolution on the strange
distribution, we display $\kappa(x,Q)$ for CTEQ6.1 and CTEQ6.6 at
both a low and high $Q$ scale in Fig.~\ref{fig:kappaEvolution}.
As we explore higher scales, the production of $s(x)$ by gluon splitting
moves $\kappa(x,Q)$ toward the $SU(3)$-symmetric limit. This trend
is especially pronounced at low $x$-values. Thus, as the LHC $W/Z$
production is centered in the range $x\sim0.01$, we will be particularly
interested in the $\kappa(x,Q)$ changes in this region. 

These results reflect the relevant $x$-range of the constraints on
the strange quark PDF, and how they depend on the $Q$-scale. In the
next section we will investigate the implications of this uncertainty
on the Drell-Yan $W/Z$ boson production at the LHC.

\section{Implications for Drell-Yan W/Z Production at the LHC}

\begin{figure}
\includegraphics[width=0.45\textwidth]{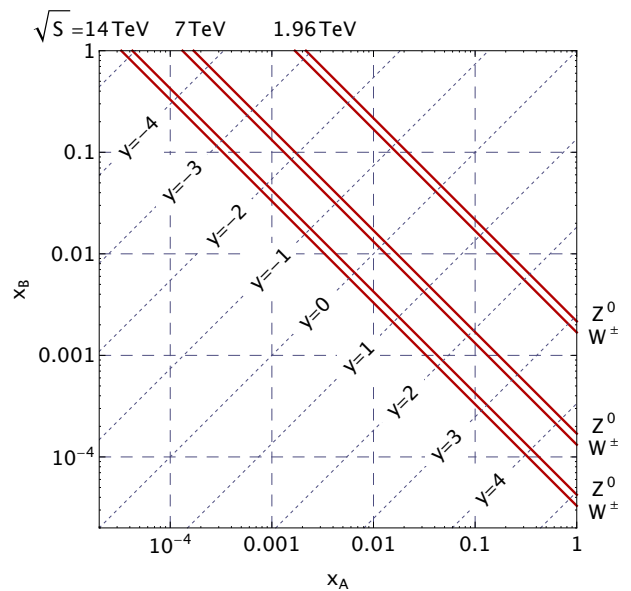}\caption{Parton momentum fractions $x_{A}$ and $x_{B}$ accessible in $W$
and $Z$ boson production in the Tevatron Run-2 ($\sqrt{S}=1.96$
TeV), and at the LHC ($\sqrt{S}=\{7,\,14\}$ TeV). The accessible
ranges of $x_{A}$ and $x_{B}$ are shown by the solid lines. The
contours of the constant rapidity $y$ are shown by the inclined dotted
lines. \label{fig:kinematics}}
\end{figure}

\begin{figure*}[tp]
\subfloat[$d\sigma/dy$ for $W^{-}$ (left), $W^{+}$ (middle), $Z^{0}$ (right)
boson production at the Tevatron.\label{fig:rapTEV}]{\includegraphics[width=0.3\textwidth]{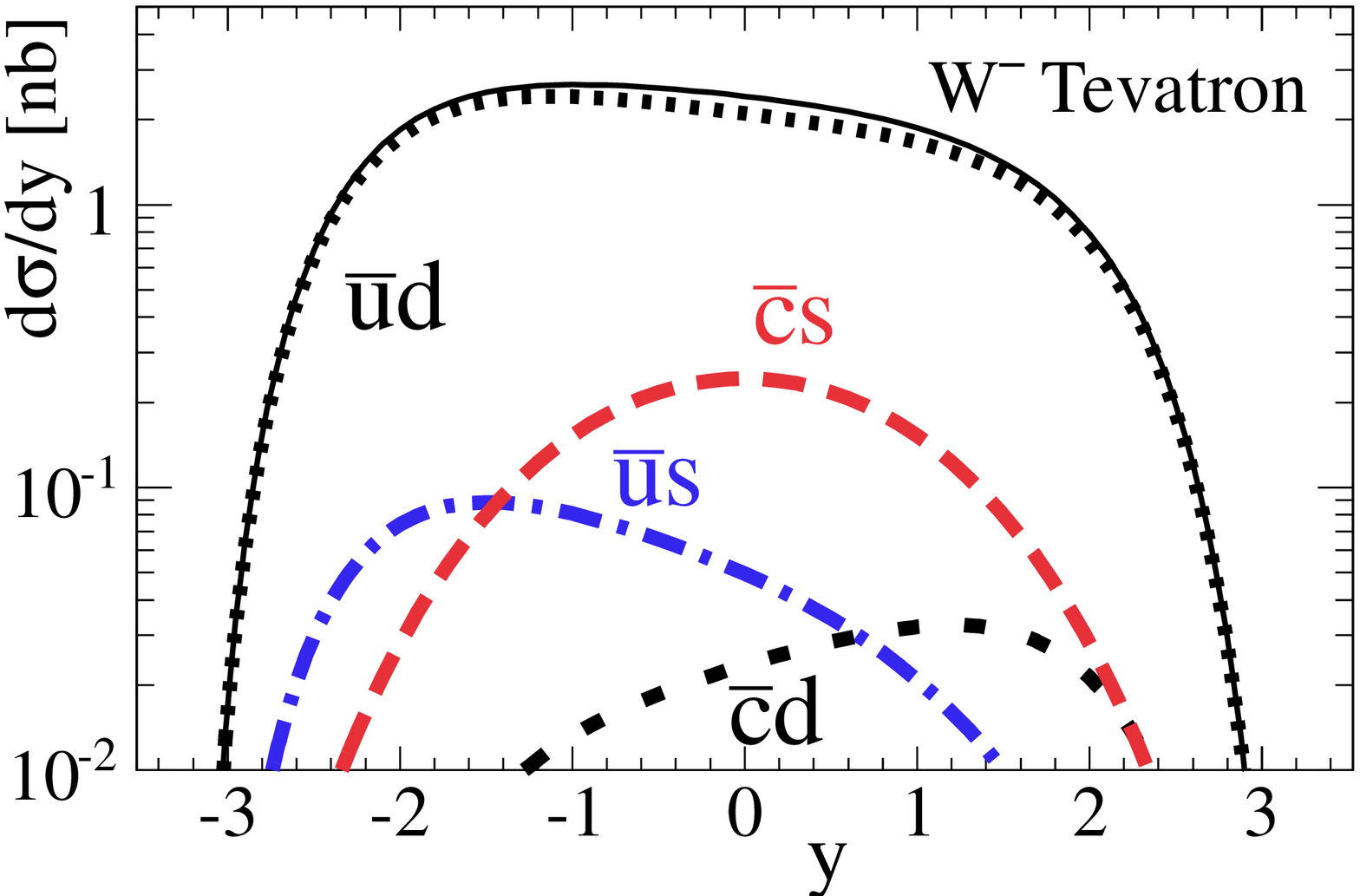}\quad{}\includegraphics[width=0.3\textwidth]{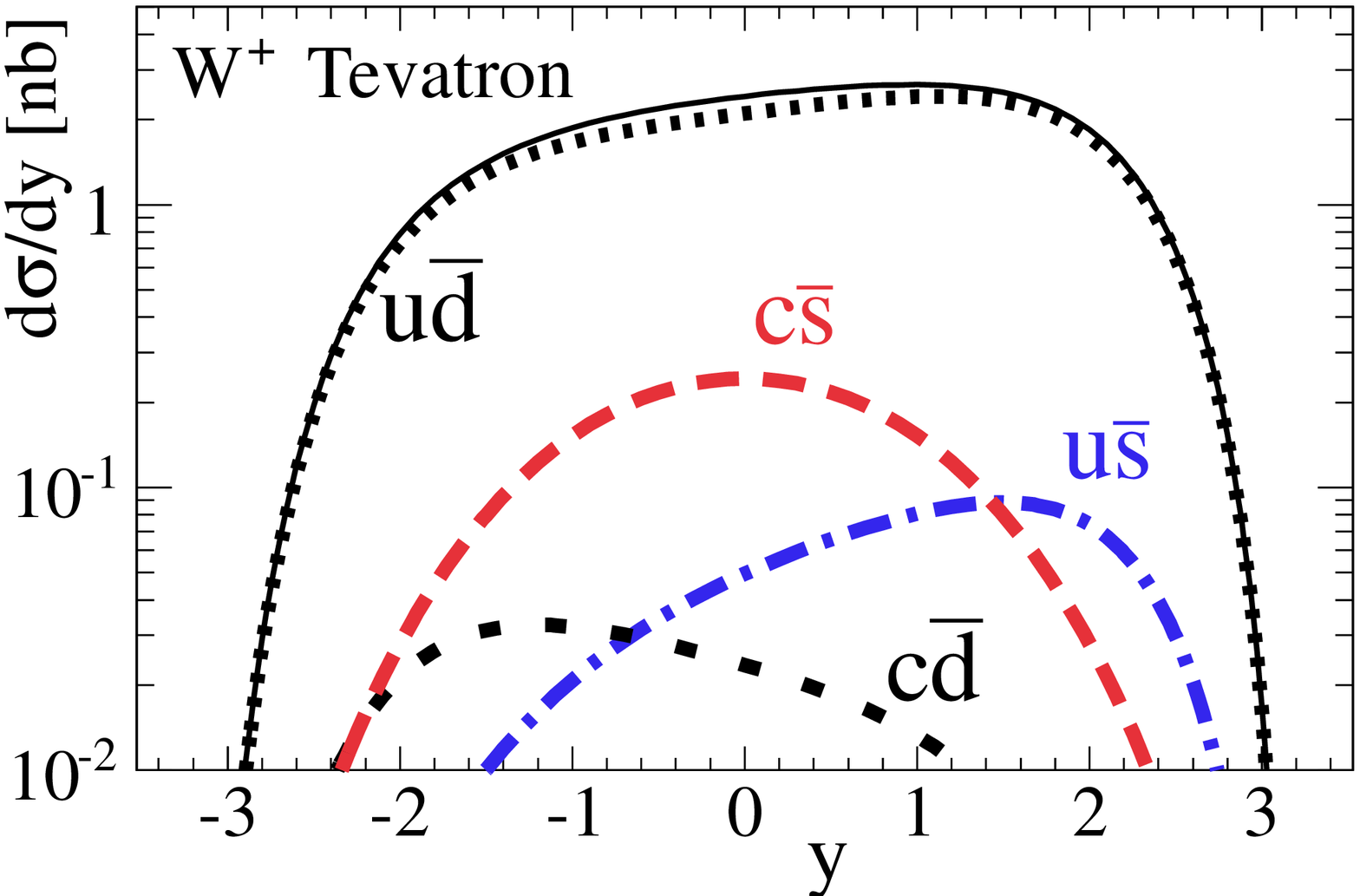}\quad{}\includegraphics[width=0.3\textwidth]{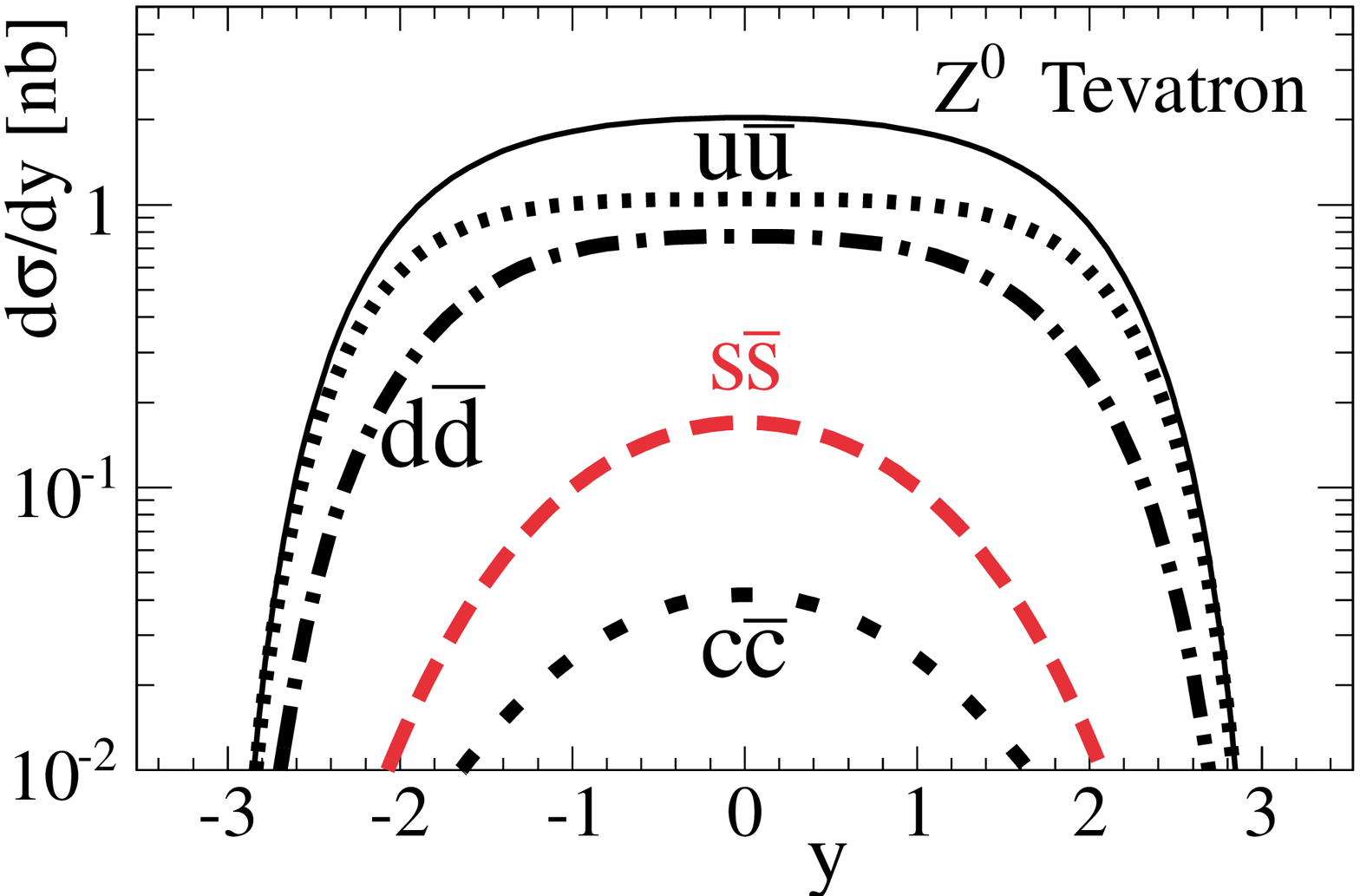}}

\subfloat[$d\sigma/dy$ for $W^{-}$ (left), $W^{+}$ (middle), $Z^{0}$ (right)
boson production at the LHC with $\sqrt{S}=7$ TeV. \label{fig:rapLHC7}]{\includegraphics[width=0.3\textwidth]{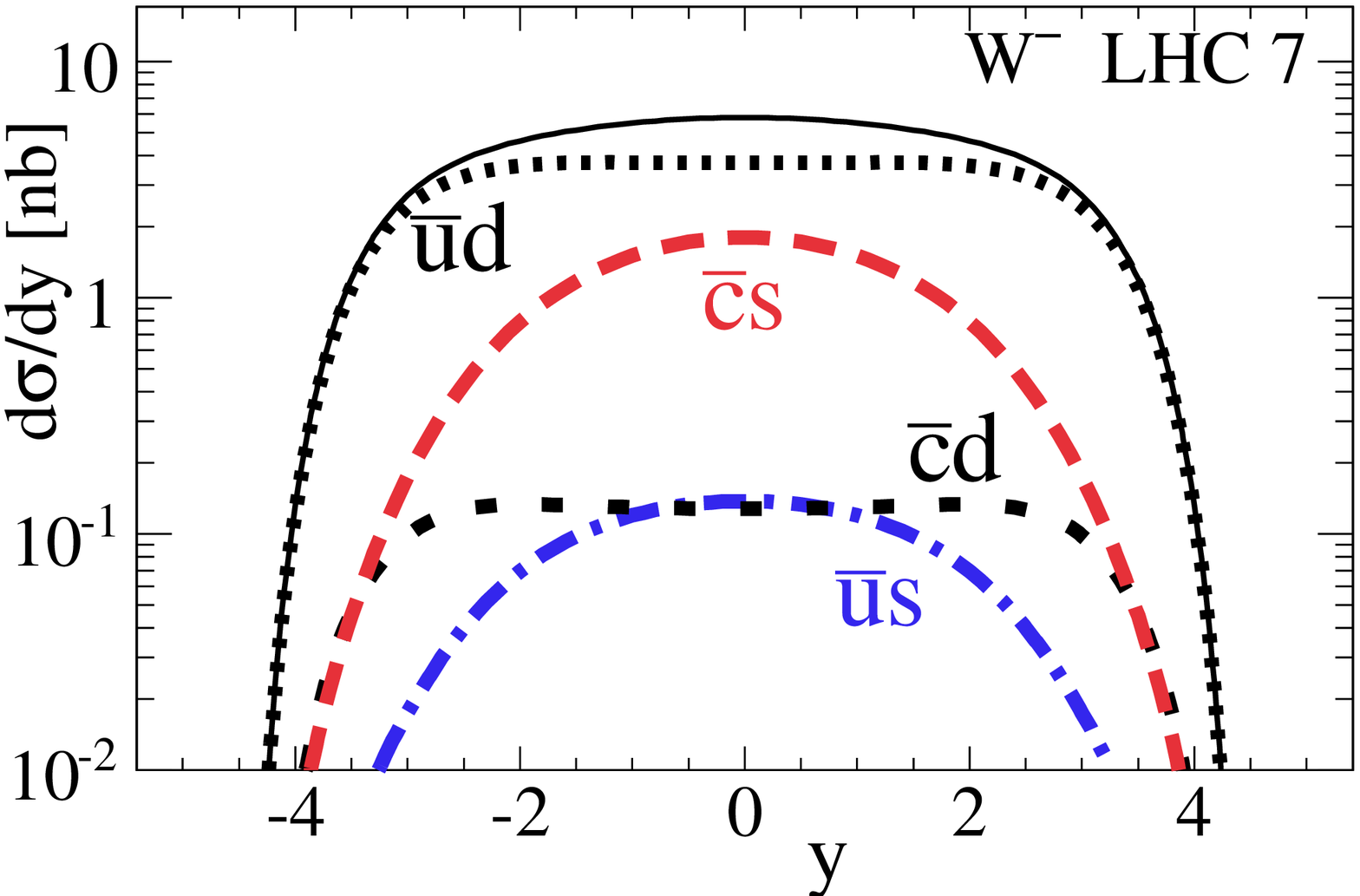}\quad{}\includegraphics[width=0.3\textwidth]{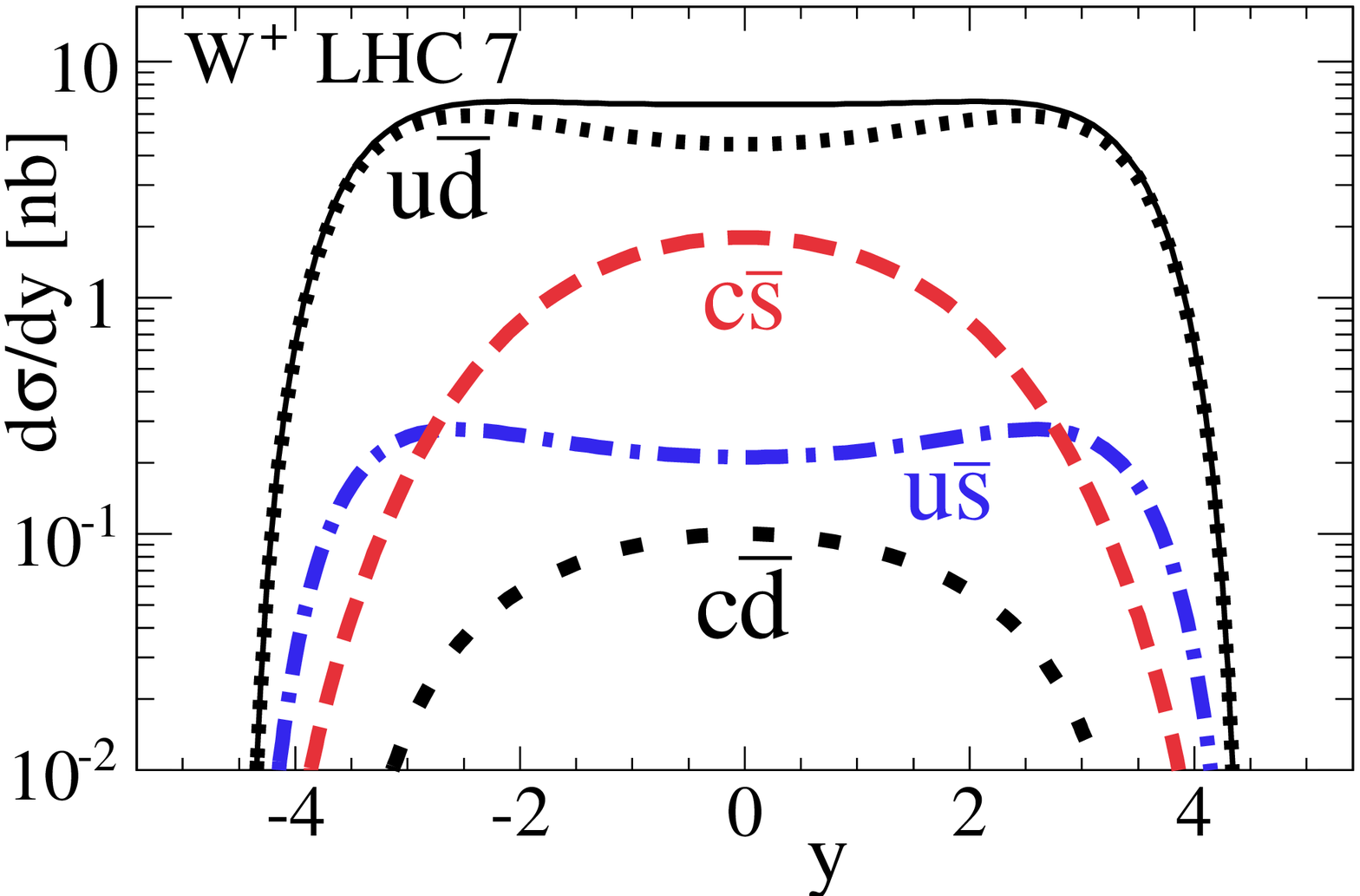}\quad{}\includegraphics[width=0.3\textwidth]{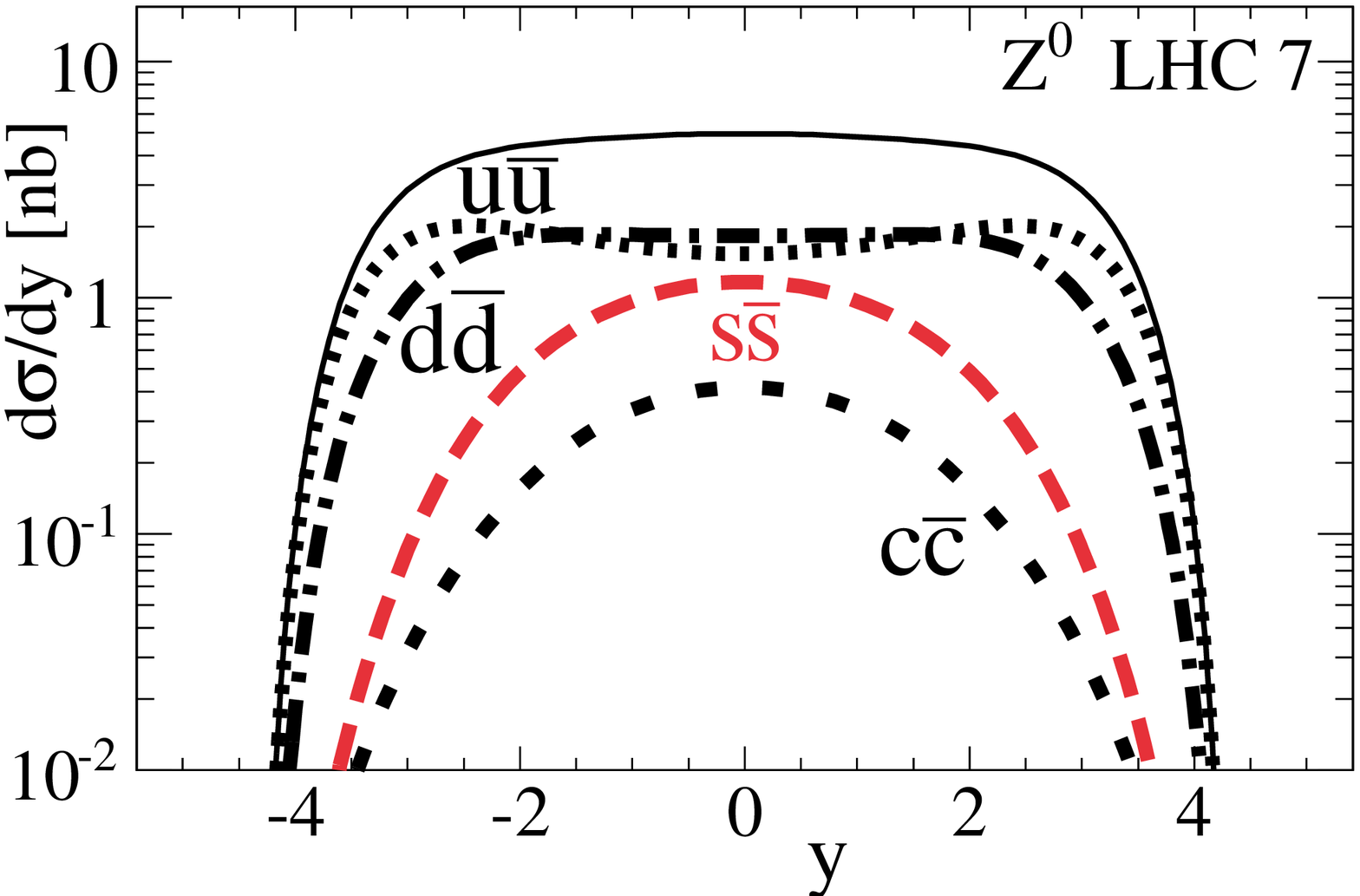}}

\subfloat[$d\sigma/dy$ for $W^{-}$ (left), $W^{+}$ (middle), $Z^{0}$ (right)
boson production at the LHC with $\sqrt{S}=14$ TeV.\label{fig:rapLHC14}]{\includegraphics[width=0.3\textwidth]{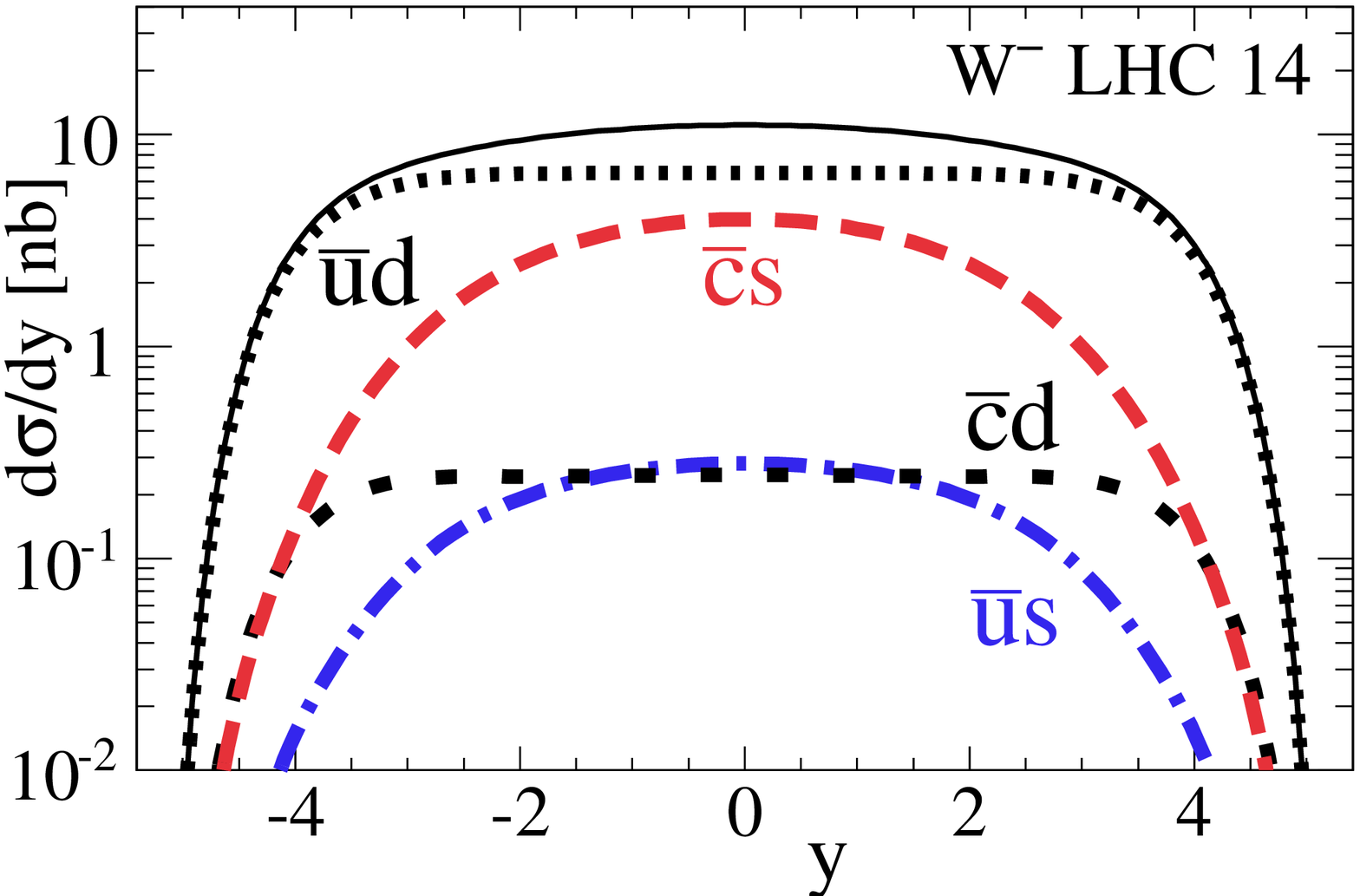}\quad{}\includegraphics[width=0.3\textwidth]{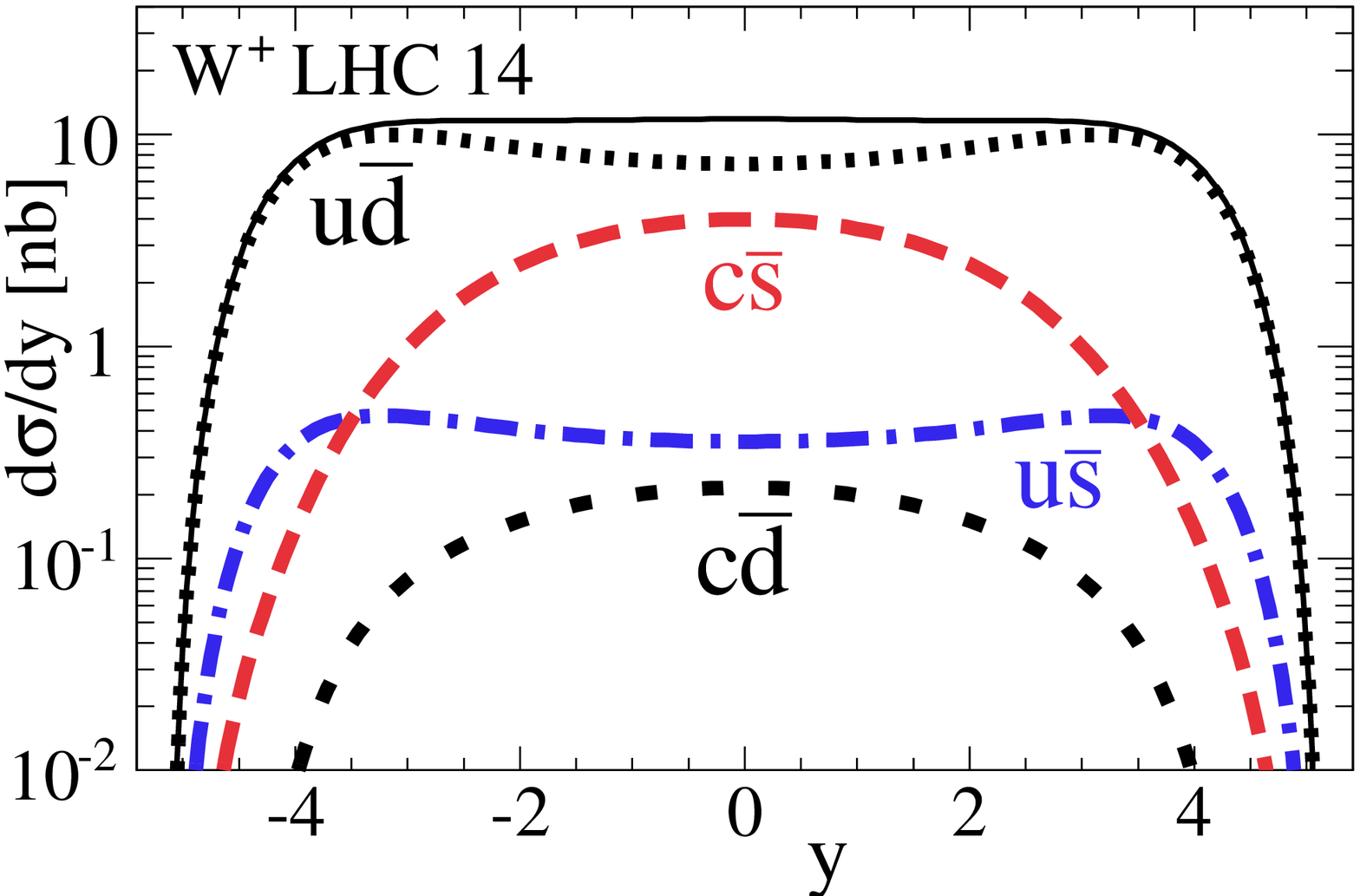}\quad{}\includegraphics[width=0.3\textwidth]{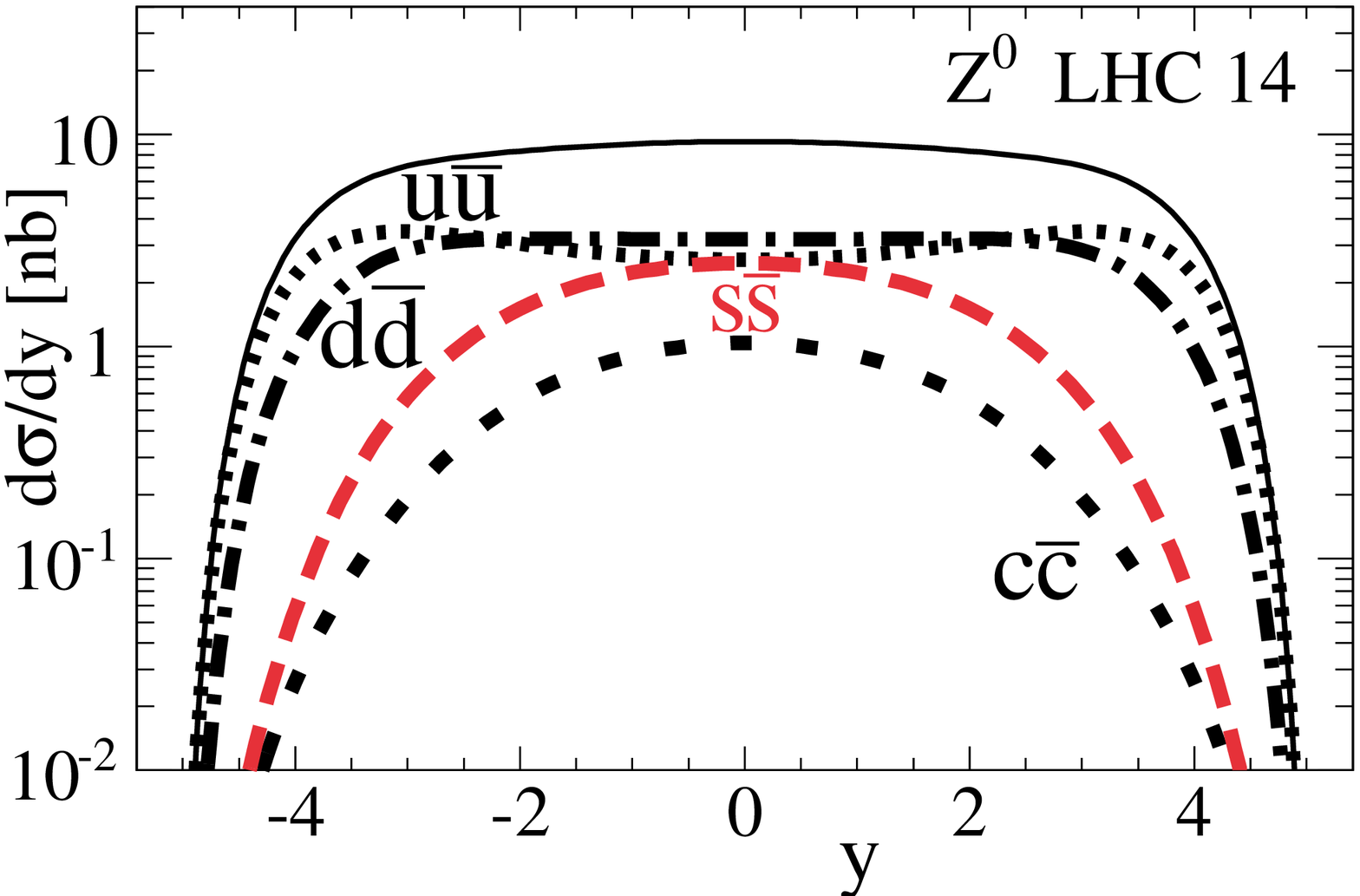}}\caption{Partonic contributions to the differential cross section of on-shell
$W^{\pm}/Z$ boson production at LO as a function of the vector boson
rapidity. Partonic contributions containing a strange or anti-strange
quark are denoted by (red) dashed and (blue) dot-dashed lines. The
solid lines show the total contribution.\label{fig:LOlum}}
\end{figure*}

The Drell-Yan production of $W^{\pm}$ and $Z$ bosons at hadron colliders
can provide precise measurements for electroweak observables such
as the $W$ boson mass~\cite{Besson:2008zs,Krasny:2010vd} and width,
the weak mixing angle in $\gamma^{*}/Z$ production~\cite{Chatrchyan:2011ya},
and the lepton asymmetry in $W$ production. These results can measure
fundamental parameters of the Standard Model (SM) and constrain the
Higgs boson mass.\textcolor{black}{{} If a Higgs boson is found at the
LHC, Drell-Yan $W/Z$ boson production will help in the search for
deviations of the SM and to reveal new physics signals \cite{Tarrade:1320912,Chatrchyan:2011nv,NNPDF:2011aa}.
For instance, new heavy gauge bosons could be discovered in the invariant
lepton distribution or new particles and interactions might leave
a footprint in the Peskin-Takeuchi $S$ and $T$ parameters \cite{Peskin:1991sw}.}

Furthermore, the $W$ and $Z$ boson cross section {}``benchmark''
processes are intended to be used for detector calibration and luminosity
monitoring~\cite{Dittmar:1997md}; to perform these tasks it is essential
that we know the impact of the PDF uncertainties on these measurements.
The impact on these benchmark processes, and the Higgs boson production,
were studied in Refs.~\cite{Alekhin:2010dd,Blumlein:2011zu,Alekhin:2012ig}.
In the following, we will investigate the influence of the PDFs on
the rapidity distributions of the Drell-Yan production process. Conversely,
it may be possible to use the $W/Z$ production process to further
constrain the parton distribution functions in general, and the strange
quark PDF in particular.\textcolor{black}{{} As noted in Ref.~\cite{Chatrchyan:2011ya},
when looking for new physics signals it is important not to mix the
information used to constrain the PDFs and the new physics as this
would lead to circular reasoning.}

As we move from the Tevatron to the LHC scattering processes, the
kinematics of the incoming partons changes considerably; in Fig.~\ref{fig:kinematics}
we show the momentum fractions $x_{A}$ and $x_{B}$ of the incoming
parton $A$ and parton $B$ for the Tevatron Run-2 ($\sqrt{S}=1.96$~TeV)
and the LHC with $\sqrt{S}=7$~TeV and $\sqrt{S}=14$~TeV. The solid
(red) lines show the range of $x_{A}$ and $x_{B}$ probed by $W^{\pm}$
and $Z$ boson production. At the Tevatron, values of $x_{A,B}$ down
to $2\times10^{-3}$ are probed for large rapidities of $y_{W/Z}=3$.
However, at the LHC much smaller values of $x_{A}$ and $x_{B}$ become
important due to the larger CMS energy and broader rapidity span.
For $\sqrt{S}=7$~TeV, the PDFs are probed for $x$-values as small
as $2\times10^{-4}$ for rapidities up to $\sim4.5$. With $\sqrt{S}=14$~TeV,
even larger rapidities of $y\sim5$ and smaller values of $x_{A/B}$
of $4\times10^{-5}$ might be reached.

\subsection{LHC Measurements}

The importance of the PDF uncertainties to the LHC measurements was
already evident in the 2010 and preliminary 2011 data. 

\textcolor{black}{ATLAS presented measurements of the Drell-Yan $W/Z$
production at the $\sqrt{S}=7$~TeV with }$35\, pb^{-1}$\textcolor{black}{{}
\cite{Aad:2011dm}. These results include not only the measurement
of total cross section and transverse distributions, but also a first
measurement of the rapidity distributions for $Z\to l^{+}l^{-}$ as
well as $W^{+}\to l^{+}\nu_{l}$ and $W^{-}\to l^{-}\bar{\nu}_{l}$.
Additionally, ATLAS has used $W/Z$ production to infer constraints
on the strange quark distribution, and they measure $r_{s}=0.5(s+\bar{s})/\bar{d}=1.00_{-0.28}^{+0.25}$
at $Q^{2}=1.9$~GeV$^{2}$ and $x=0.023$ \cite{Collaboration:2012sb}.}

\textcolor{black}{CMS has measured the rapidity and transverse momentum
distributions for $Z\to l^{+}l^{-}$ production \cite{Chatrchyan:2011wt}
}and inclusive $W/Z$ production \cite{CMS:2011aa} using $36\, pb^{-1}$
of data. Additionally CMS has measured the weak mixing angle~\cite{Chatrchyan:2011ya},
the forward-backward asymmetry in $\gamma^{*}/Z$ production~\cite{CMS-PAS-EWK-10-011},
and the lepton charge asymmetry in $W$ production~\cite{CMS-PAS-EWK-11-005,Chatrchyan:2011jz}. 

LHCb has measured the $W$ charge asymmetry in Refs.~\cite{Amhis:2012gj}
and~\cite{Shears:1394600}. These measurements show, already with
these data samples, the PDF uncertainties are important and can be
the leading source of measurement uncertainty.

Additionally, CMS has analyzed $W+c$ production which is directly
sensitive to the $s$ and $\bar{s}$ contribution of the proton; the
results for the $36\, pb^{-1}$ data sample are given in Ref.~\cite{CMS11013}.

\subsection{Strange Contribution to $W/Z$ Production}

\begin{figure*}
\subfloat[$d^{2}\sigma/dM/dy$ in pb/GeV for $pp\to W^{-}+X$~(left), $pp\to W^{+}+X$~(middle)
and $pp\to Z,\gamma^{*}+X$~(right) production at the LHC for 7 TeV
with CTEQ6.6 using the VRAP program~\cite{Anastasiou:2003ds} at
NNLO (for $M=m_{W}$ and $M=m_{Z}$ respectively). \label{fig:dyStrange66_7tev}]{\includegraphics[width=0.3\textwidth]{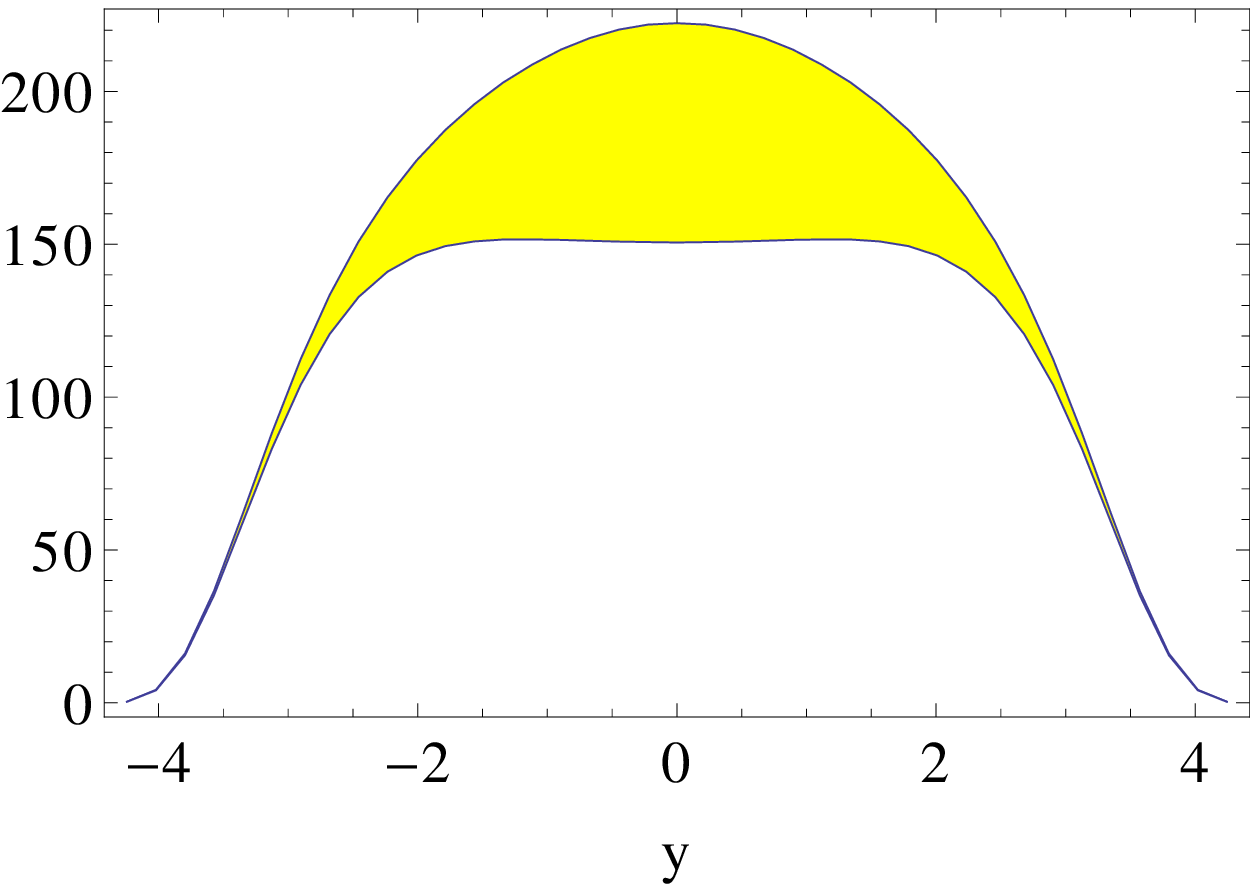}\quad{}\includegraphics[width=0.3\textwidth]{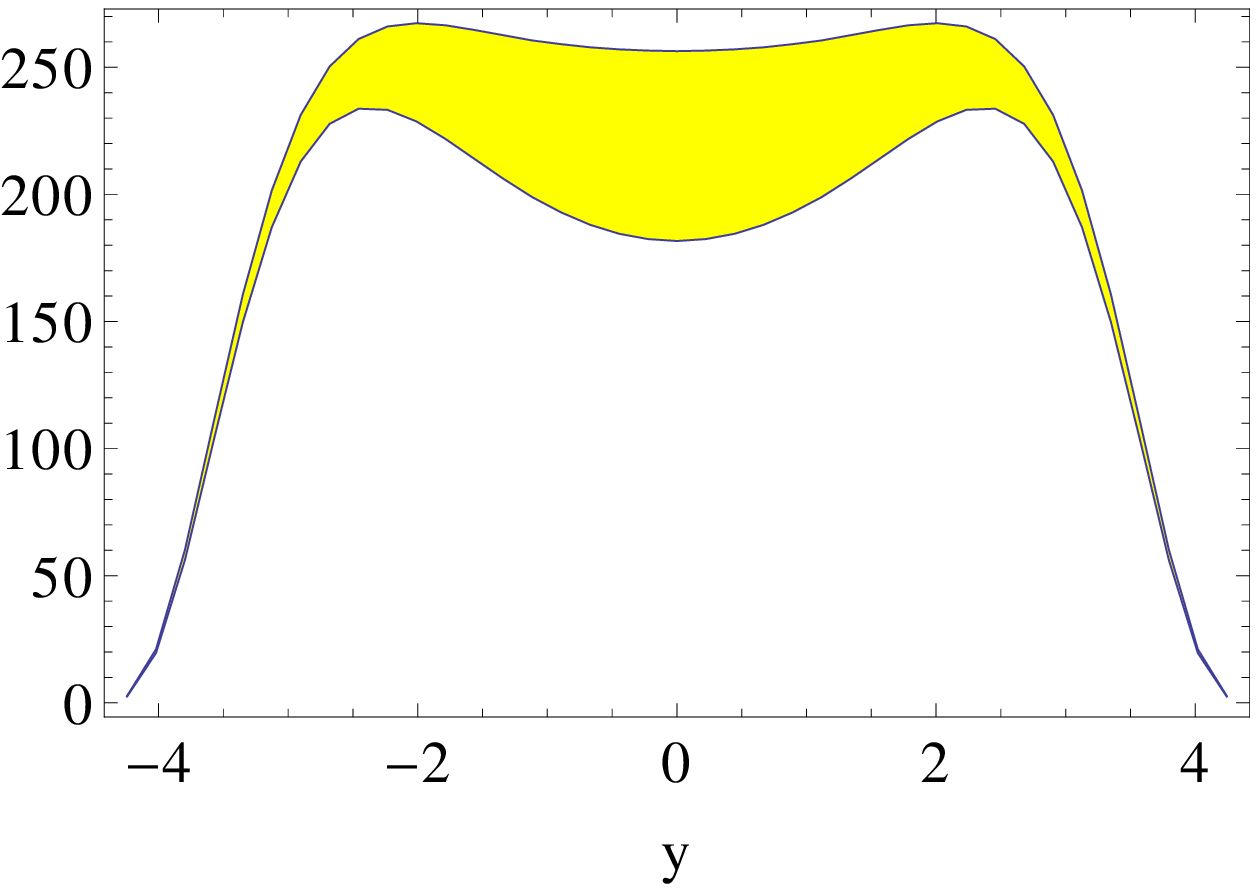}\quad{}\includegraphics[width=0.293\textwidth]{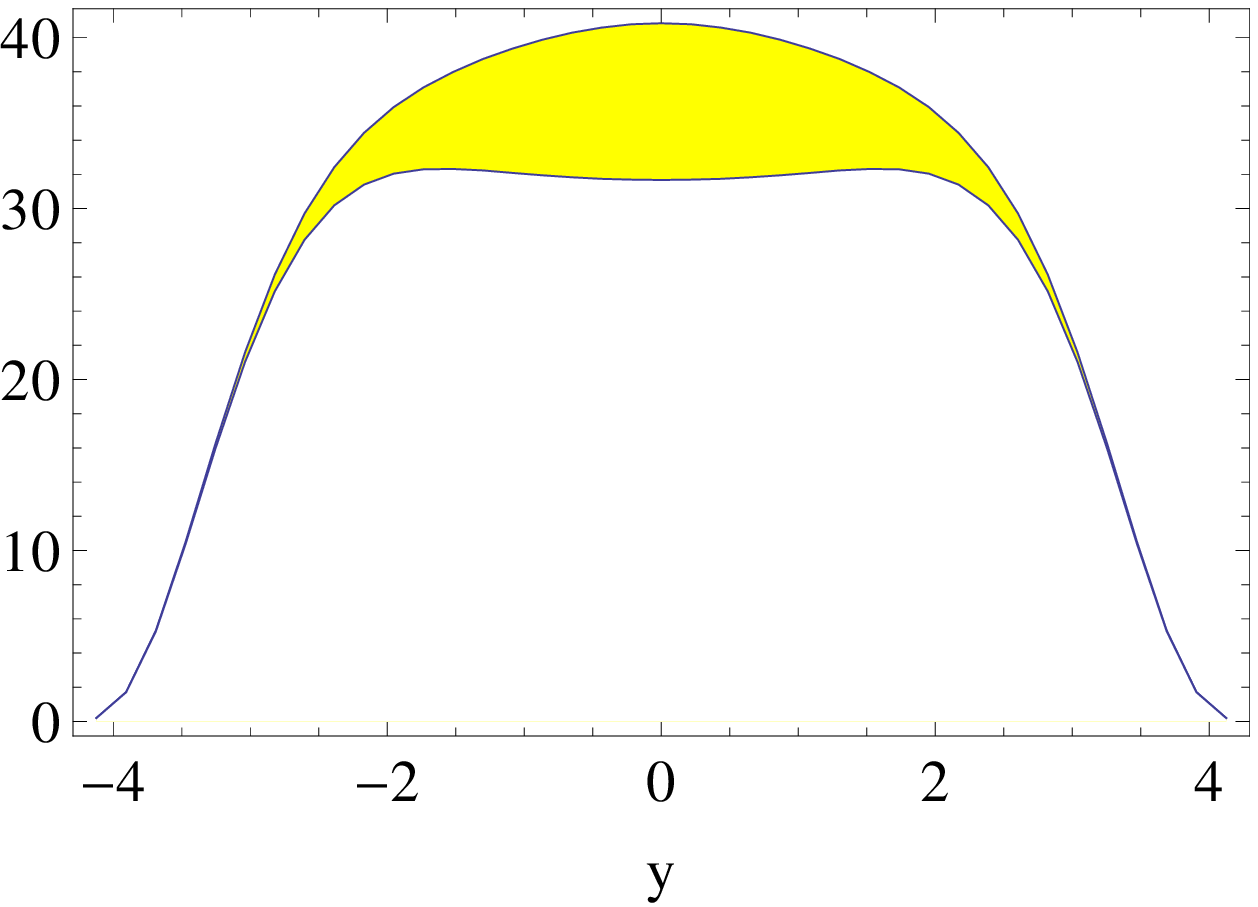}

}

\subfloat[$d^{2}\sigma/dM/dy$ in pb/GeV for $pp\to W^{-}+X$~(left), $pp\to W^{+}+X$~(middle)
and $pp\to Z,\gamma^{*}+X$~(right) production at the LHC for 14
TeV with CTEQ6.6 using the VRAP program~\cite{Anastasiou:2003ds}
at NNLO (for $M=m_{W}$ and $M=m_{Z}$ respectively). \label{fig:dyStrange66}]{\includegraphics[width=0.3\textwidth]{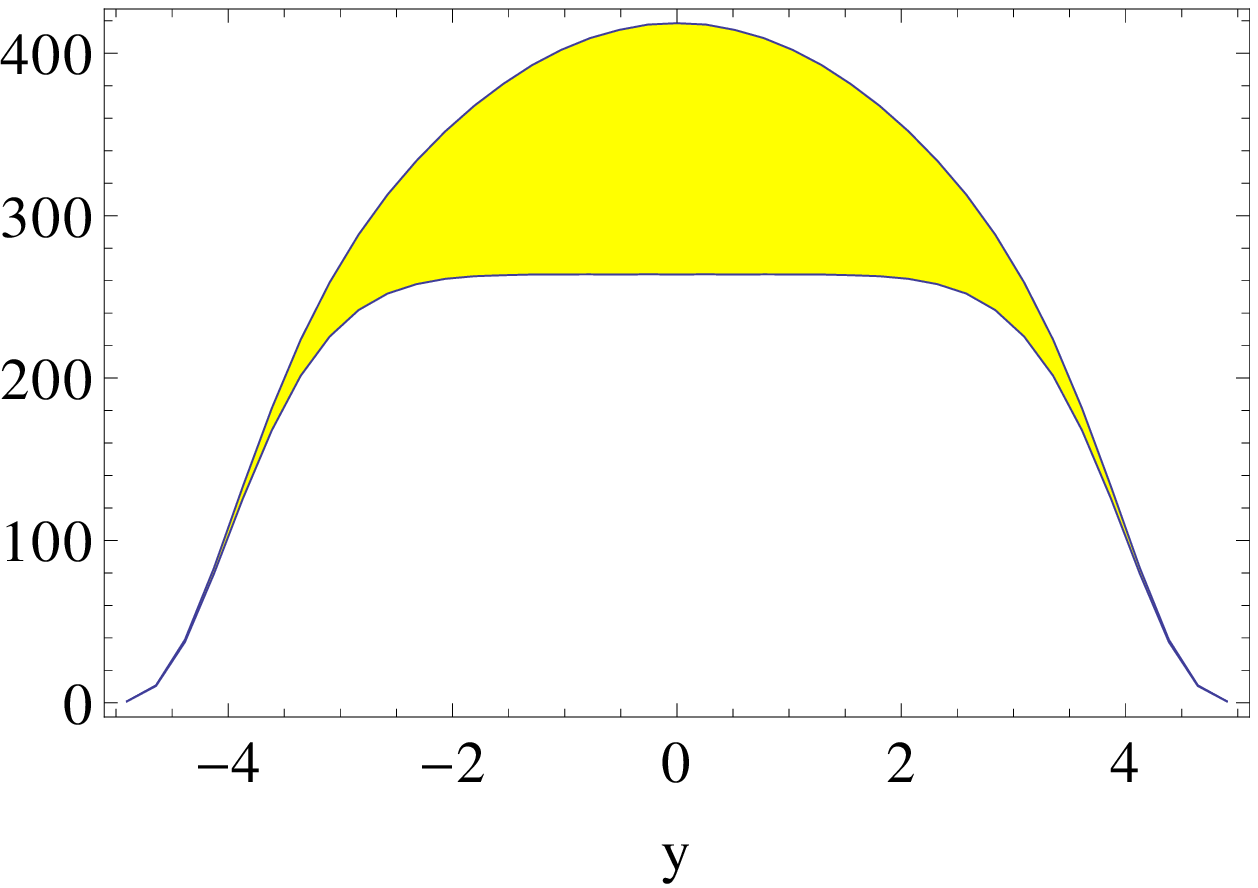}\quad{}\includegraphics[width=0.3\textwidth]{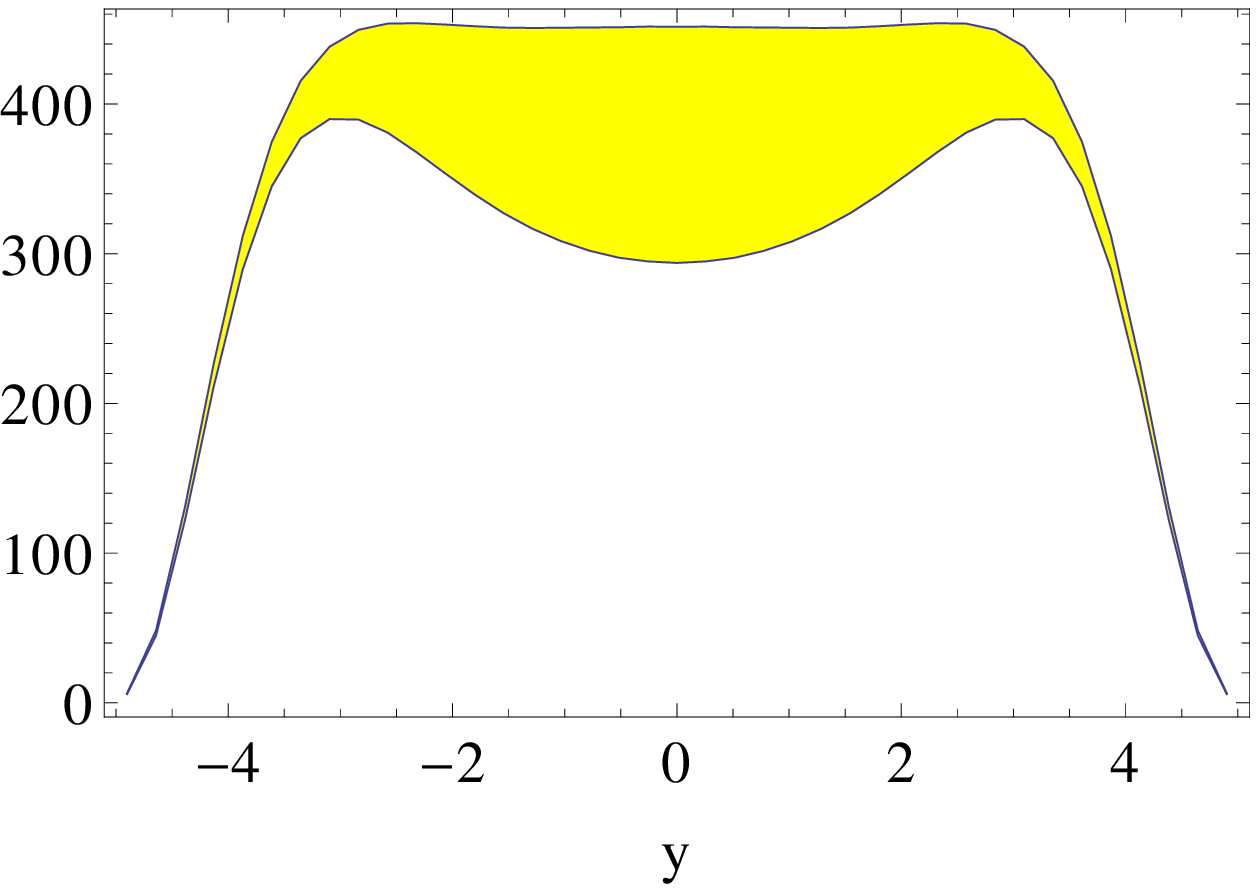}\quad{}\includegraphics[width=0.295\textwidth]{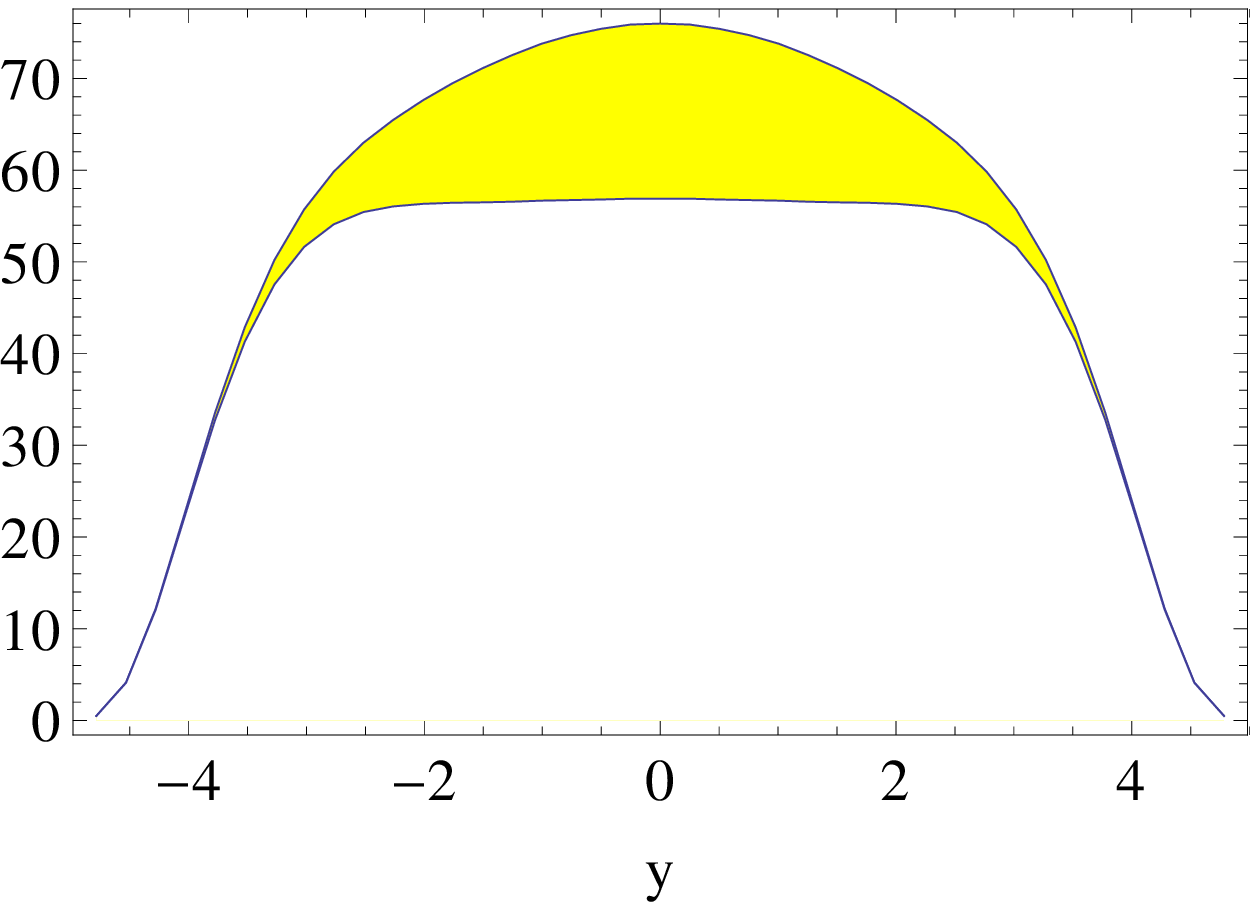}}

\caption{Contribution of the strange quark to $W^{\pm}/Z$ production at the
LHC.\label{fig:nnlo}}
\end{figure*}

Because the proton-proton LHC has a different initial state and a
higher CMS energy than the Tevatron, the relative contributions of
the partonic subprocesses of the $W^{\pm}/Z$ production change significantly.
At the LHC, the contributions of the second generation quarks $\{s,c\}$
are greatly enhanced. Additionally, the $W^{+}$ and $W^{-}$ rapidity
distributions are no longer related by a simple $y\to-y$ reflection
symmetry due to the $pp$ initial state. In Figure~\ref{fig:LOlum}
we display the contributions from the different partonic cross sections
which contribute to $W^{\pm}$ and $Z$ production at LO. 

Figure~\ref{fig:rapTEV} shows the rapidity distribution at the Tevatron.
For $W^{+}$~($W^{-}$) production, the $u\bar{d}$~($d\bar{u}$)
channel (dotted black lines) contributes $90\%$ of the cross section,
while in $Z$ production the $u\bar{u}$~(dotted black line) and
$d\bar{d}$~(dot-dashed black line) subprocesses contribute $93\%$
of the cross section. The first generation quarks $\{u,d\}$ therefore
dominate the production process while contributions from strange quarks
(red dashed and blue dot-dashed lines) are comparably small with $9\%$
for $W^{\pm}$ and $5\%$ ($s\bar{s}$) for $Z$ boson production.

At the LHC, subprocesses containing strange quarks are considerably
more important as shown in Fig.~\ref{fig:rapLHC7} for a CMS energy
of $7$~TeV and in Fig.~\ref{fig:rapLHC14} for $14$~TeV. For
$W^{-}$ production (left plots), the (blue) dot-dashed lines show
the $s\bar{u}$ channel while the (red) dashed lines show the $s\bar{c}$
contribution. At $14$~TeV the $s\bar{c}\to W^{-}$ subprocess contributes
$28\%$ to the cross section, while the $s\bar{u}\to W^{-}$ subprocess
contributes only $2\%$ as this is suppressed by the off-diagonal
CKM matrix entry. For $W^{+}$ production channels, the $\bar{s}u$
channel (blue dot-dashed lines) contributes only $2\%$, while the
$\bar{s}c$ channel (red dashed lines) yields $21\%$. Notice, the
absolute value of the $s\bar{c}\to W^{-}$ and $\bar{s}c\to W^{+}$
contributions are the same, however the relative contribution is smaller
for $W^{+}$ production due to the larger up-quark valence contribution
in the $u\bar{d}\to W^{+}$ subprocess as compared to $\bar{u}d\to W^{-}$. 

The rapidity distributions of the total $W^{-}$ and $W^{+}$ boson
production differ markedly at the LHC because of the different valence
quark contributions from $u$ and $d$. This effect is also present
in the $s\bar{u}\to W^{-}$ and $\bar{s}u\to W^{+}$ (blue dot-dashed
lines) subprocess. We will comment more on this feature in the following
subsection. 

Comparing Figs.~\ref{fig:rapTEV} with Figs.~\ref{fig:rapLHC14},
we note the LHC explores a much larger rapidity range. For channels
containing strange quarks, $|y_{W/Z}|$ can be measured up to $y\approx4.5$
at the LHC, compared to $y\approx2.5$ at the Tevatron; therefore
smaller values of $x$ of the strange quark distribution can be probed. 

Additionally, as at the Tevatron, we can use $W^{-}$ production to
probe the strange quark PDF while using $W^{+}$ production to probe
the anti-strange PDF.

While the LO illustration of Fig.~\ref{fig:LOlum} provides a useful
guide, in Fig.~\ref{fig:nnlo} we display the strange quark contribution
to the differential cross section $d^{2}\sigma/dM/dy$ of on-shell
$W^{-},\, W^{+},\, Z$ boson production computed at NNLO using the
VRAP program~\cite{Anastasiou:2003ds}. We display the LHC results
for $W^{\pm}$ and $Z$ with $\sqrt{S}$ of both 7~TeV and 14~TeV,
where the (yellow) band represents the strange-quark initiated contributions
to the total differential cross section. 

The figures impressively highlight the large contribution of the strange
and anti-strange quark subprocesses at the LHC. Consequently it is
essential to constrain the strange PDF if we are to make accurate
predictions and to perform precision measurements. Figure~\ref{fig:nnlo}
also demonstrates clearly the very different rapidity profiles of
the strange quark (arising from the sea-distribution) compared to
the $u$ and $d$ quark terms which are dominated by the valence distributions.
This property is most evident for the case of $W^{+}$ production.
Here, the dominant $u\bar{d}$ contribution has a twin-peak structure
due to the harder valence distribution, while the $c\bar{s}$ distribution
has a single-peak centered at $y=0$. The total distribution is then
a linear combination of the twin-peak and single-peak distributions,
and these are weighted by the corresponding PDF. 

Therefore, a detailed measurement of the rapidity distribution of
the $W^{\pm}/Z$ bosons can yield information about the contributions
of the $s$ quark relative to the $u,d$ quarks. As this is a relative
measurement, rather than an absolute cross section measurement, it
is reasonable to expect that this could be achieved with high precision
once sufficient statistics are collected. Consequently, this is an
ideal measurement where the LHC data could lead to stronger constraints
on the PDFs.

\subsection{PDF Uncertainty of the $W/Z$ rapidity distributions \label{sub:PDFuncertainty}}

\begin{figure*}
\subfloat[$d^{2}\sigma/dM/dy$ in pb/GeV for $pp\to W^{-}+X$~(left), $pp\to W^{+}+X$~(middle)
and $pp\to Z,\gamma^{*}+X$~(right) production at the LHC for 7 TeV
with CTEQ6.6 using the VRAP program~\cite{Anastasiou:2003ds} at
NNLO.\label{fig:dy}]{\includegraphics[width=0.3\textwidth]{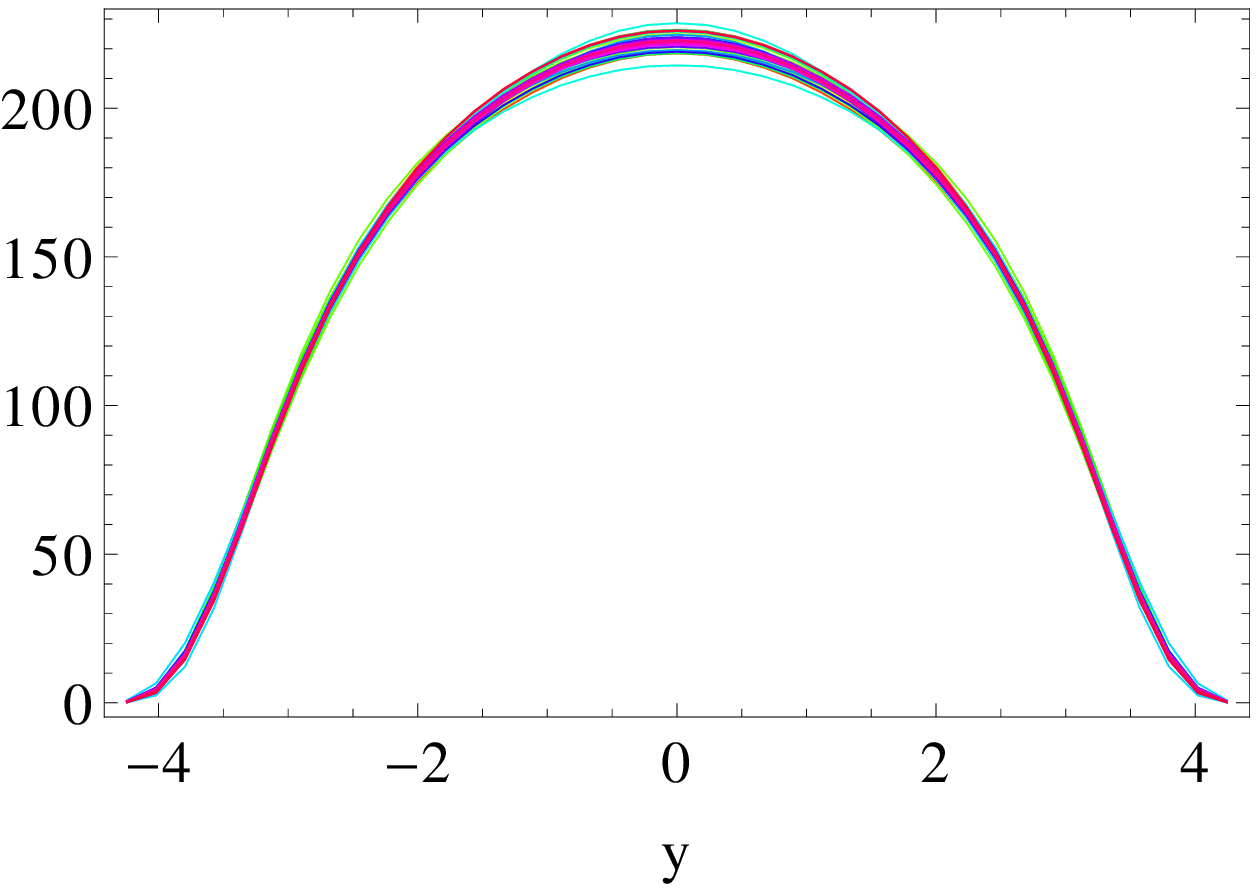}\quad{}\includegraphics[width=0.3\textwidth]{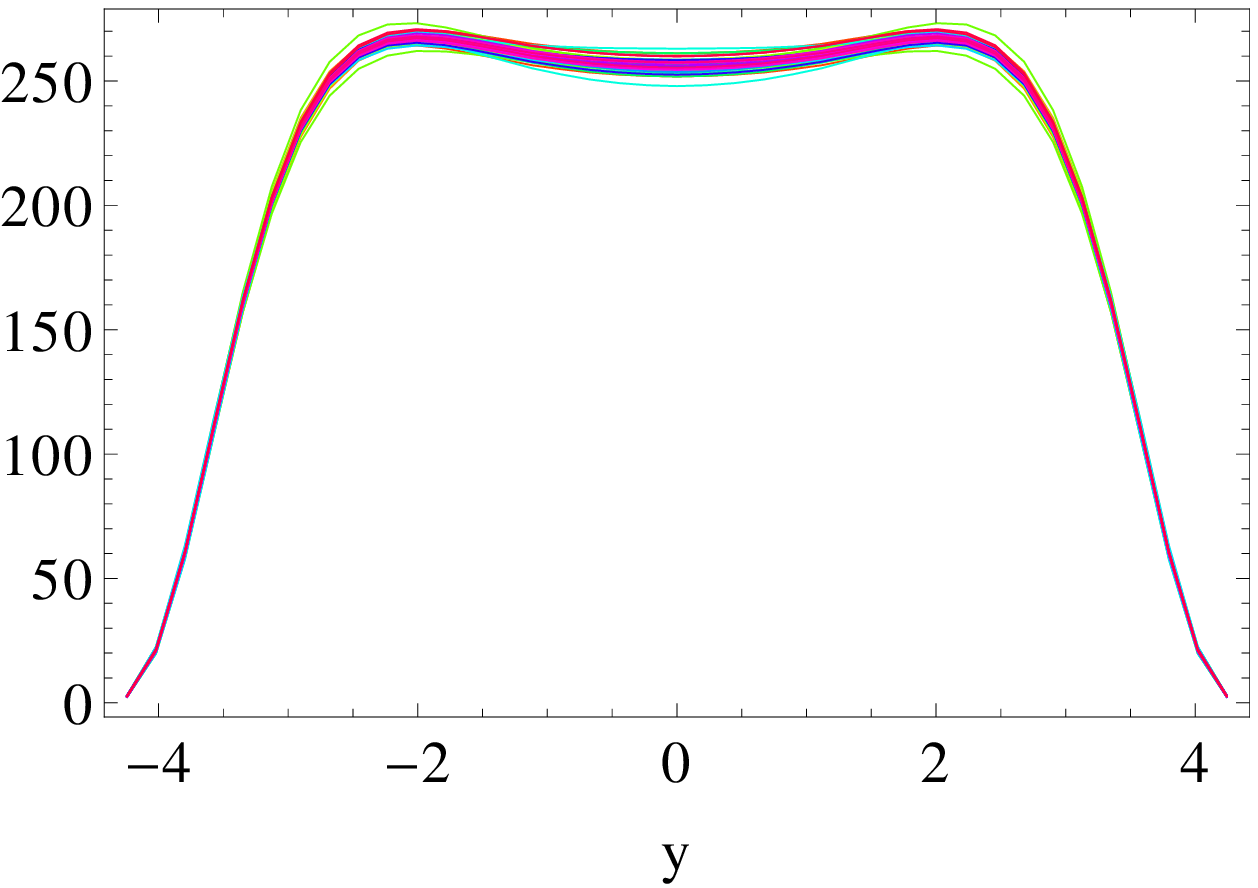}\quad{}\includegraphics[width=0.293\textwidth]{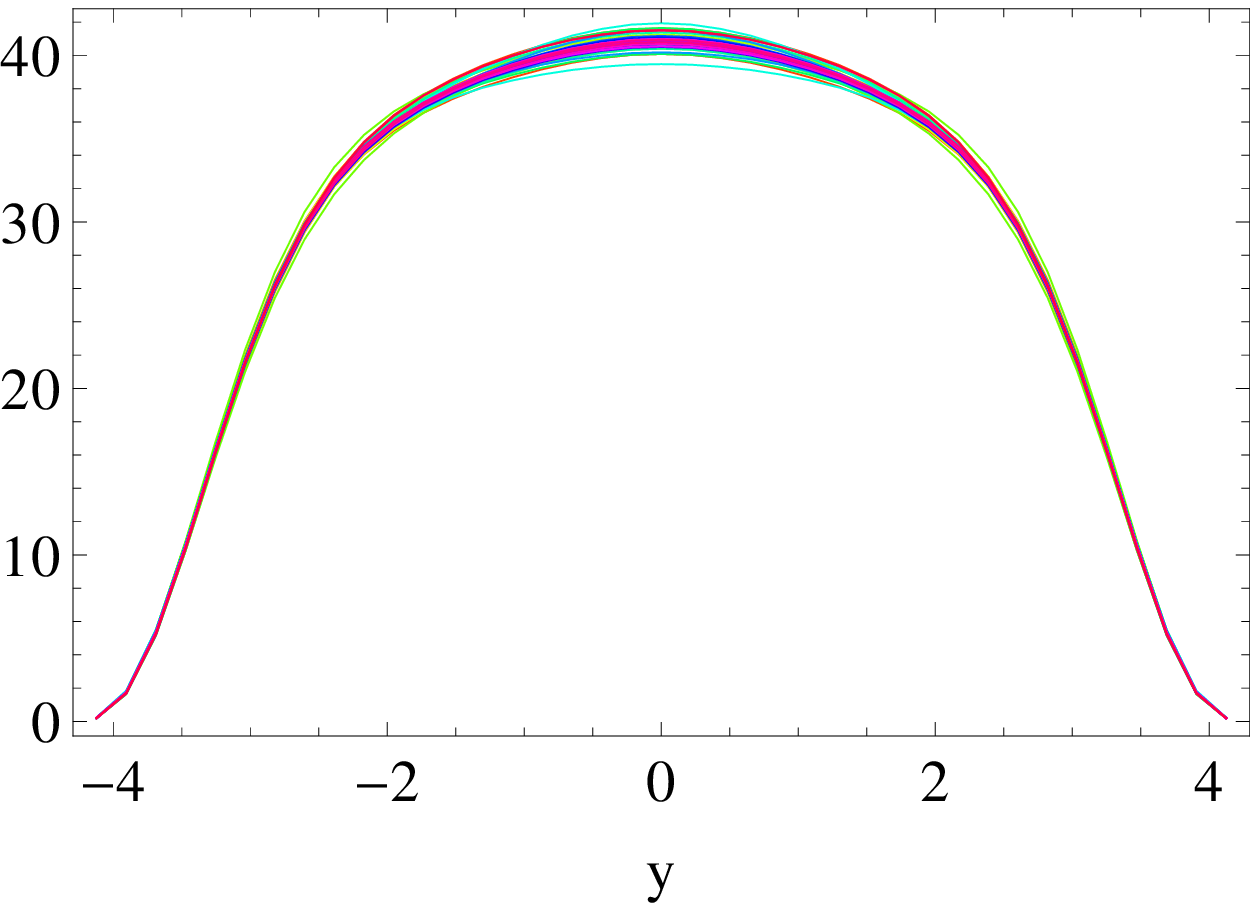}}

\subfloat[$d^{2}\sigma/dM/dy$ for on-shell $\{W^{-},W^{+},Z\}$ production
at the LHC for $\sqrt{S}=7$~TeV with CTEQ6.6 using the VRAP program
at NNLO, scaled by the central value. \label{fig:dySingle66}]{\includegraphics[width=0.3\textwidth]{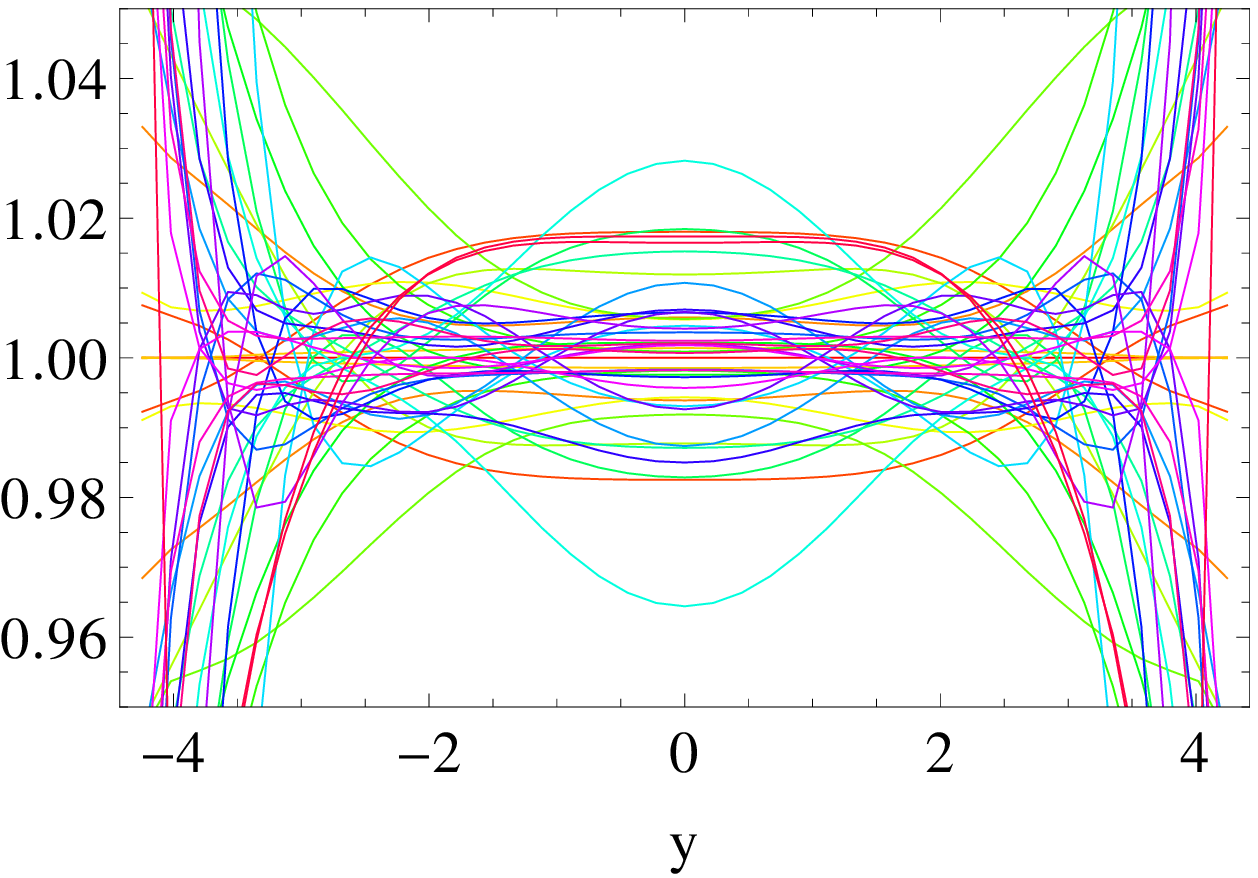}\quad{}\includegraphics[width=0.3\textwidth]{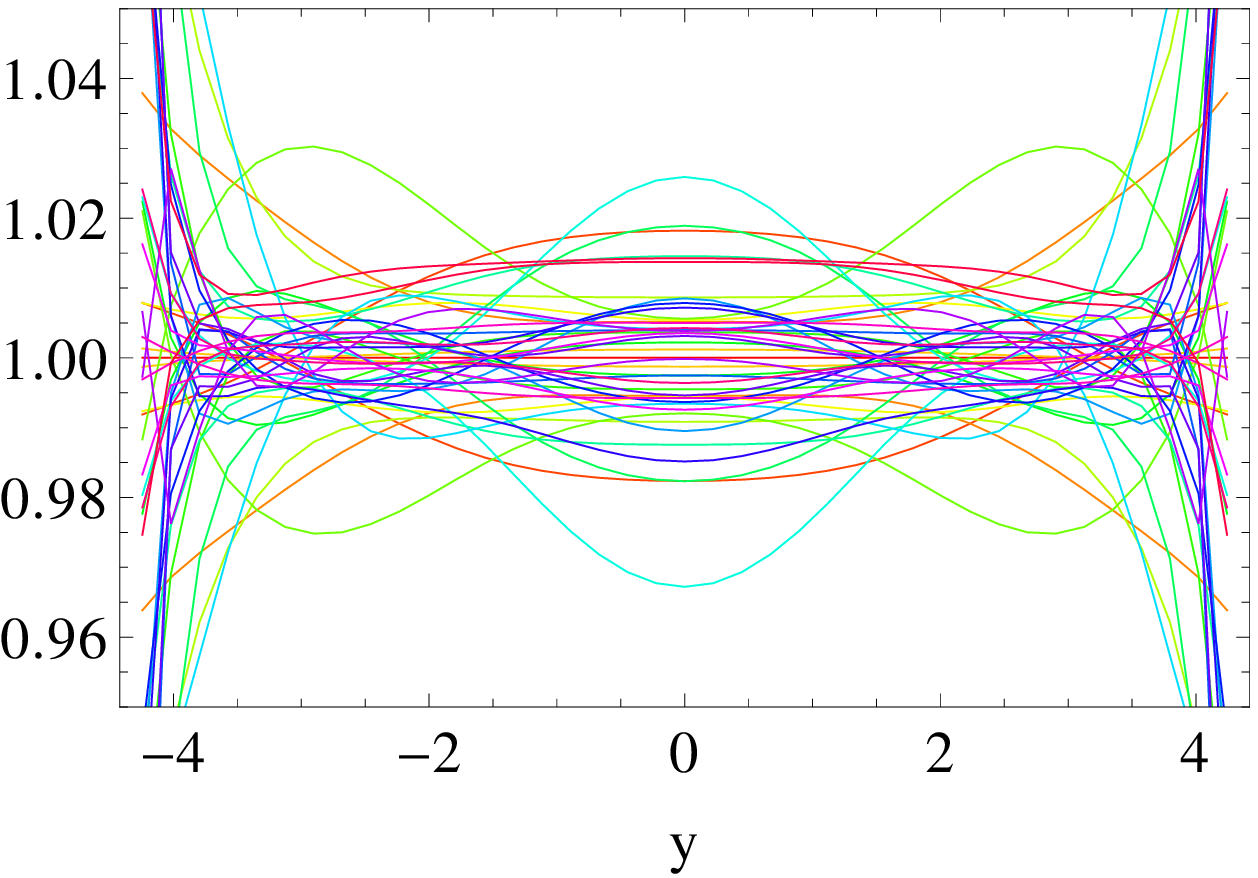}\quad{}\includegraphics[width=0.3\textwidth]{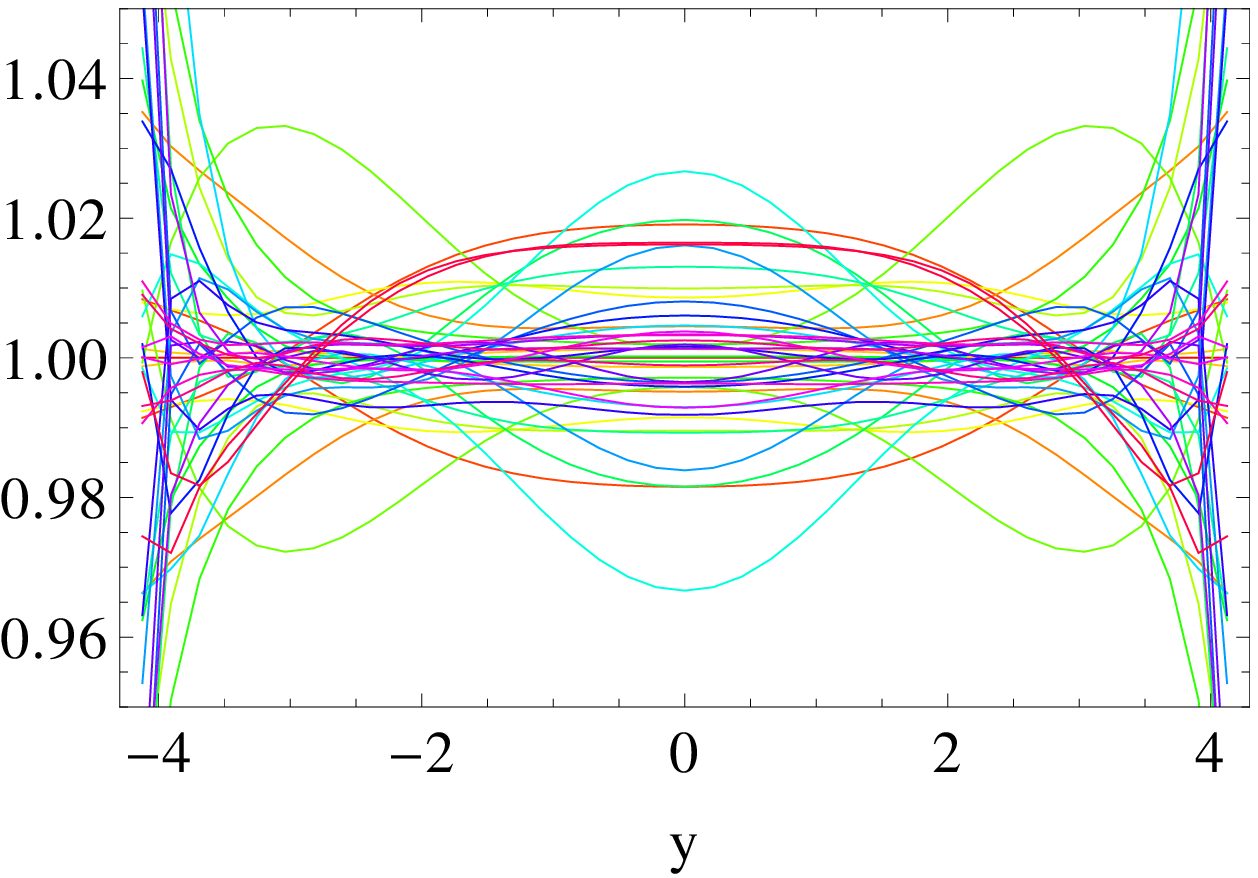}}

\subfloat[$d^{2}\sigma/dM/dy$ for $\{W^{-},W^{+},Z\}$ production at the LHC
for $\sqrt{S}=7$~TeV with a selection of PDFs using the VRAP program
at NNLO. The (yellow) band is for the CTEQ6.6 set~\cite{Nadolsky:2008zw},
and the other curves are for the central values of different PDF sets
(\emph{see text}). All plots are scaled by the central value for the
CETQ6.6 set. Note the scale of this figure is larger than for Fig.~\ref{fig:dySingle66}.
\label{fig:other7}]{\includegraphics[width=0.3\textwidth]{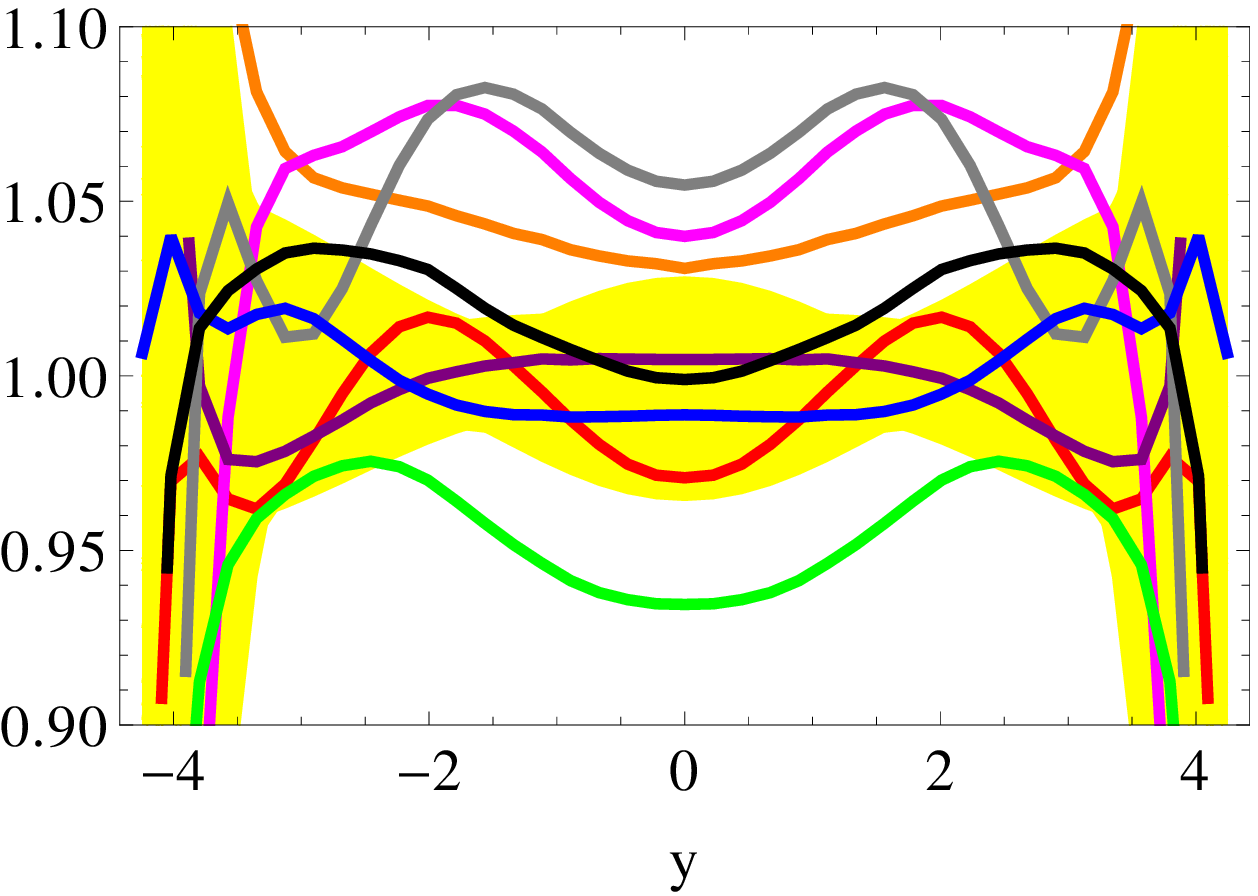}\quad{}\includegraphics[width=0.3\textwidth]{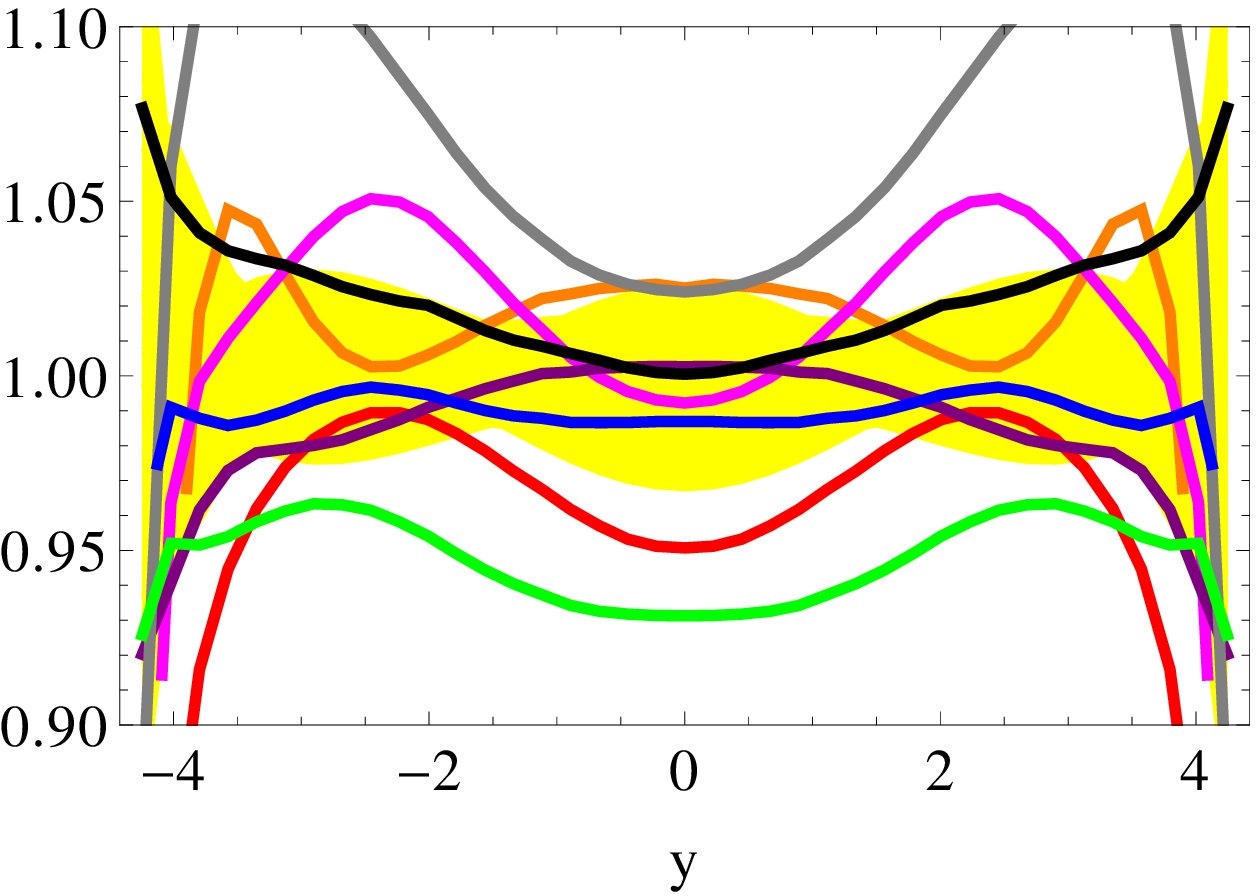}\quad{}\includegraphics[width=0.3\textwidth]{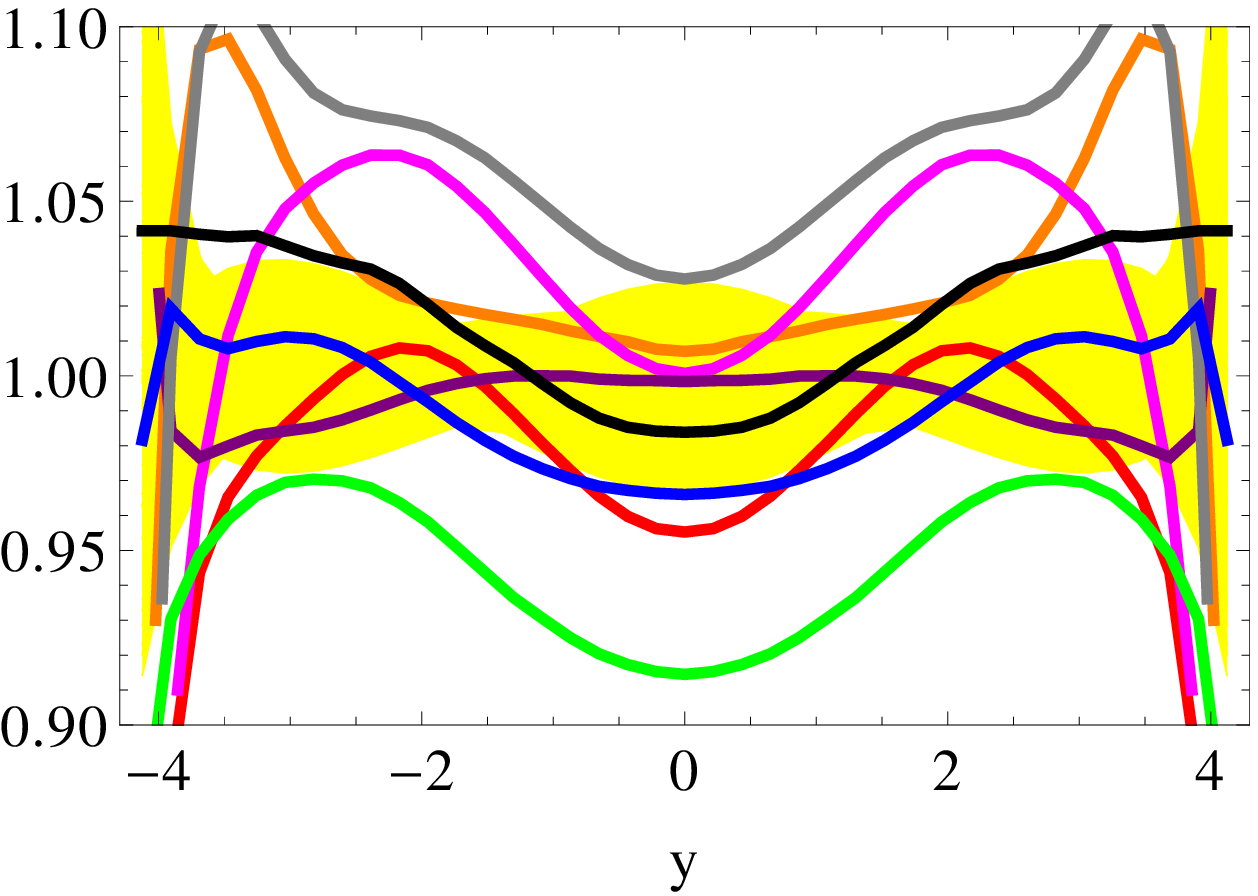}}

\caption{PDF uncertainty bands for on-shell $W^{-}$(left plots) , $W^{+}$(middle
plots) and $Z$(right plots) production at the LHC for $\sqrt{S}=7$~TeV.}
\end{figure*}

To estimate the influence of the PDF uncertainties (and in particular
the strange quark PDF) on the $W/Z$ production process and its differential
distributions at the LHC, we will use the different PDF sets within
CTEQ6.6 as well as compare the sets of different PDF groups. 

In Figure~\ref{fig:dy}, we display the differential cross section
$d^{2}\sigma/dM/dy$ for $W^{\pm}/Z$ boson production at the LHC
at $\sqrt{S}=7$~TeV using the 44 error PDF sets of CTEQ6.6. To better
resolve these PDF uncertainties, we plot the ratio of the differential
cross section $d^{2}\sigma/dM/dy$  compared to the central value
in Fig.~\ref{fig:dySingle66}. We observe that the uncertainty due
to the PDFs as measured by this band is between $\pm3\%$ and $\pm4\%$
for central boson rapidities of $-3\le y_{W/Z}\le+3$. For larger
rapidities, the PDF uncertainties increase dramatically, but the cross
section vanishes. 

For comparison, in Fig.~\ref{fig:other7} we display the (yellow)
band of CTEQ6.6 error PDFs together with the results using other contemporary
PDF sets. The (yellow) band shows the span of the 44 CTEQ6.6 error
PDFs of Fig.~\ref{fig:dySingle66}, and the solid lines show the
rapidity distribution from the selection of PDFs; all have been scaled
to the central value for the CETQ6.6 set.%
\footnote{Here, we are more interested in the general span of these different
PDFs rather than the specific sets and values. For reference the specific
curves are: MSTW2008~\cite{Martin:2009iq} (magenta), NNPDF~\cite{Ball:2011mu}
(blue), ABKM09~\cite{Alekhin:2009ni} (gray), CT10~\cite{Lai:2010vv}
(purple), CTEQ6.5~\cite{Tung:2006tb} (black), CTEQ6.1~\cite{Stump:2003yu}
(green), HERAPDF10~\cite{:2009wt} (orange), MRST2004~\cite{Martin:2004ir}
(red). %
} We observe that the choice of PDF sets can result in differences
ranging up to $\pm\hspace{0.3mm}8\hspace{0.3mm}\%$ for $-2\le y_{W/Z}\le+2$
and even up to $\pm\hspace{0.3mm}10\hspace{0.3mm}\%$ for $-3\le|y_{W/Z}|\le3$,
which is well beyond the $\pm\hspace{0.3mm}3\hspace{0.3mm}\%$ and
$\pm\hspace{0.3mm}4\hspace{0.3mm}\%$ range displayed in Fig.~\ref{fig:dySingle66};
note the different scales used in Fig.~\ref{fig:dySingle66} and
Fig.~\ref{fig:other7}. However, if we compute the PDF uncertainty
band using Eq.~\eqref{eq:error} as specified by Ref.~\cite{Pumplin:2002vw}
we find an estimated uncertainty of $\sim15\%$ (depending on the
rapidity) which generally does encompass the range of PDFs displayed
in Fig.~\ref{fig:other7}. 

While the band of error PDFs provides an efficient method to quantify
the uncertainty, the range spanned by the different PDF sets illustrates
there are other important factors which must be considered to encompass
the full range of possibilities.

\subsection{Correlations of the $W/Z$ rapidity distributions }

\begin{figure*}[t]
\subfloat[Double ratio $R^{-}$ as defined in Eq.~(\ref{eq:ratio_RWZpm}) for
the LHC for $\sqrt{S}=7$~TeV with CTEQ6.5 (left) and CTEQ6.6 (right)
at NNLO.\label{fig:dyDoubleMinus7} ]{\includegraphics[width=0.3\textwidth]{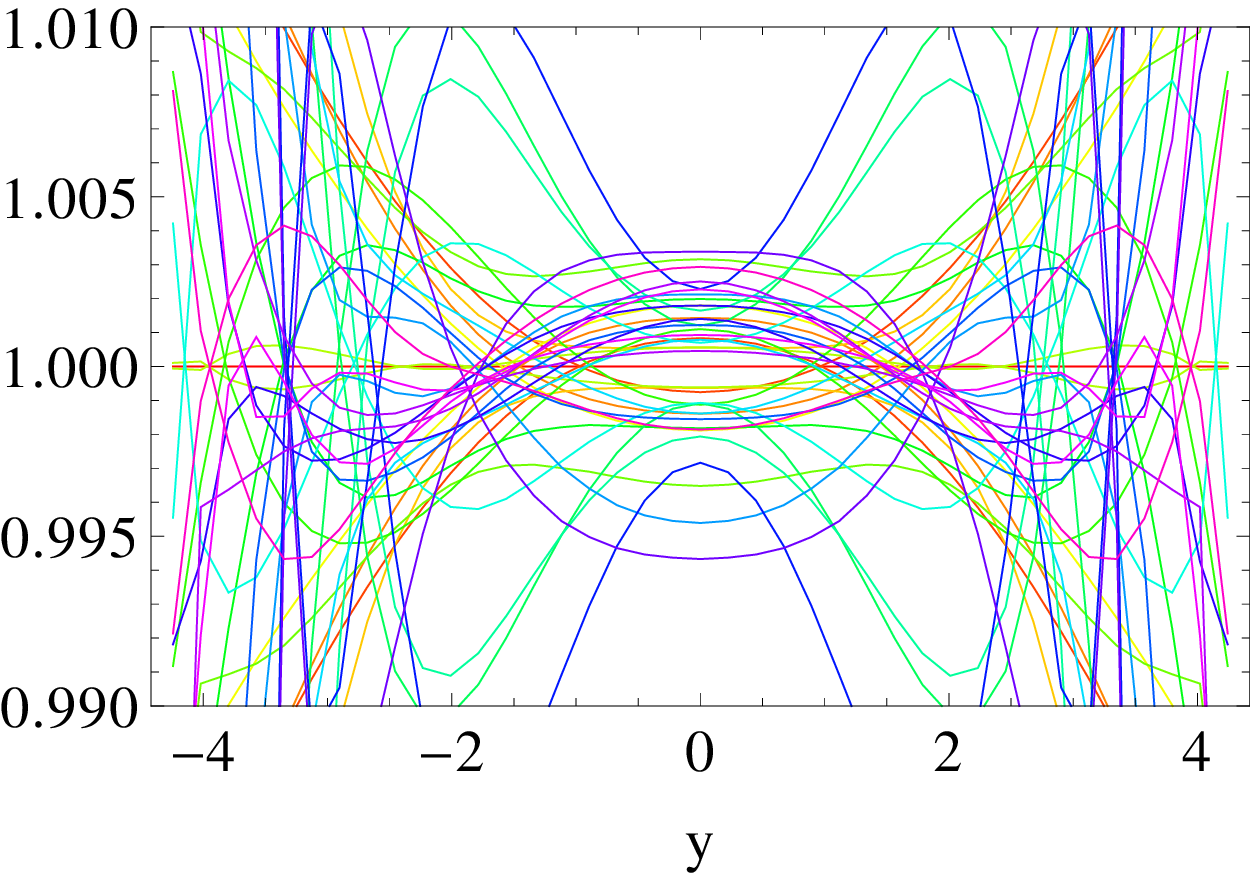}\quad{}\includegraphics[width=0.3\textwidth]{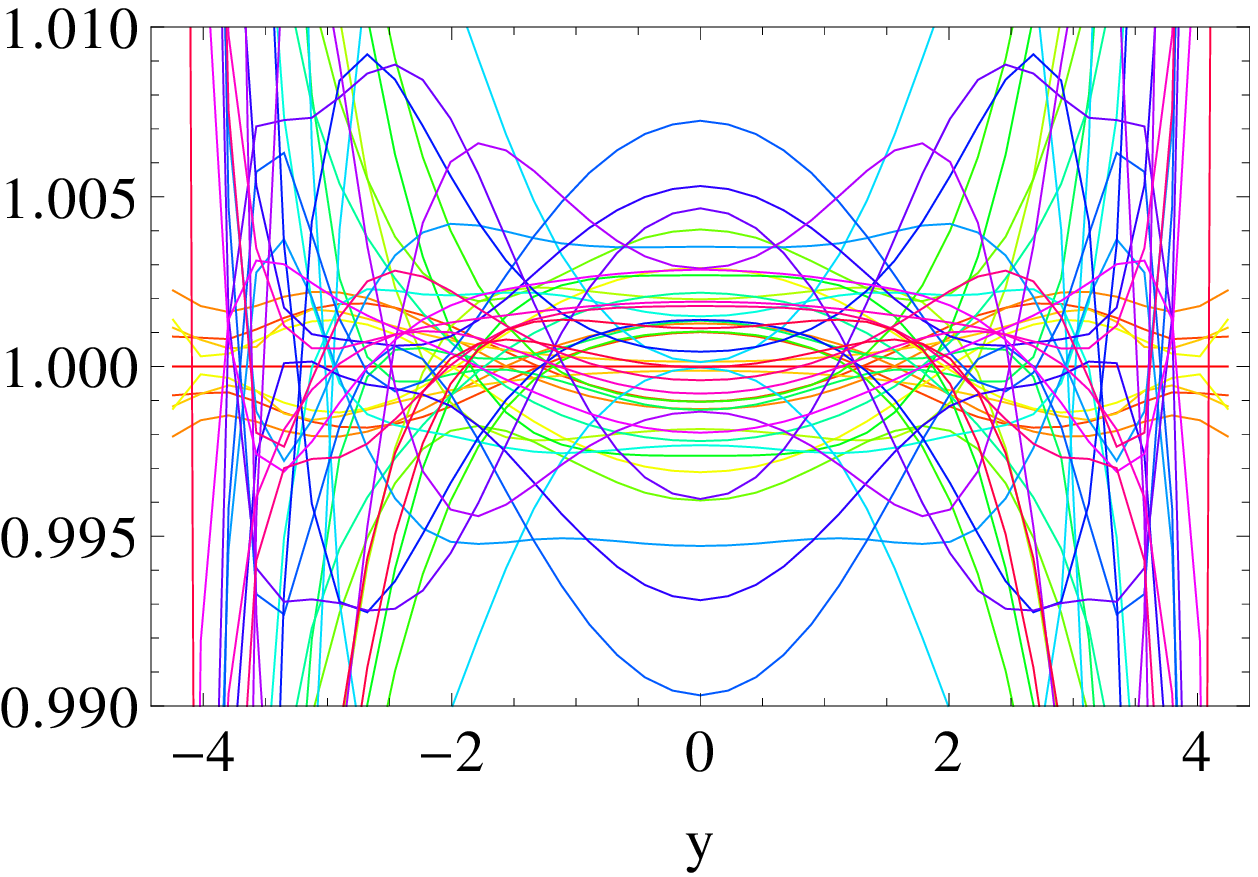}}

\subfloat[Double ratio $R^{+}$ as defined in Eq.~(\ref{eq:ratio_RWZpm}) for
the LHC for $\sqrt{S}=7$~TeV with CTEQ6.5 (left) and CTEQ6.6 (right)
at NNLO. \label{fig:dyDoublePlus7} ]{\includegraphics[width=0.3\textwidth]{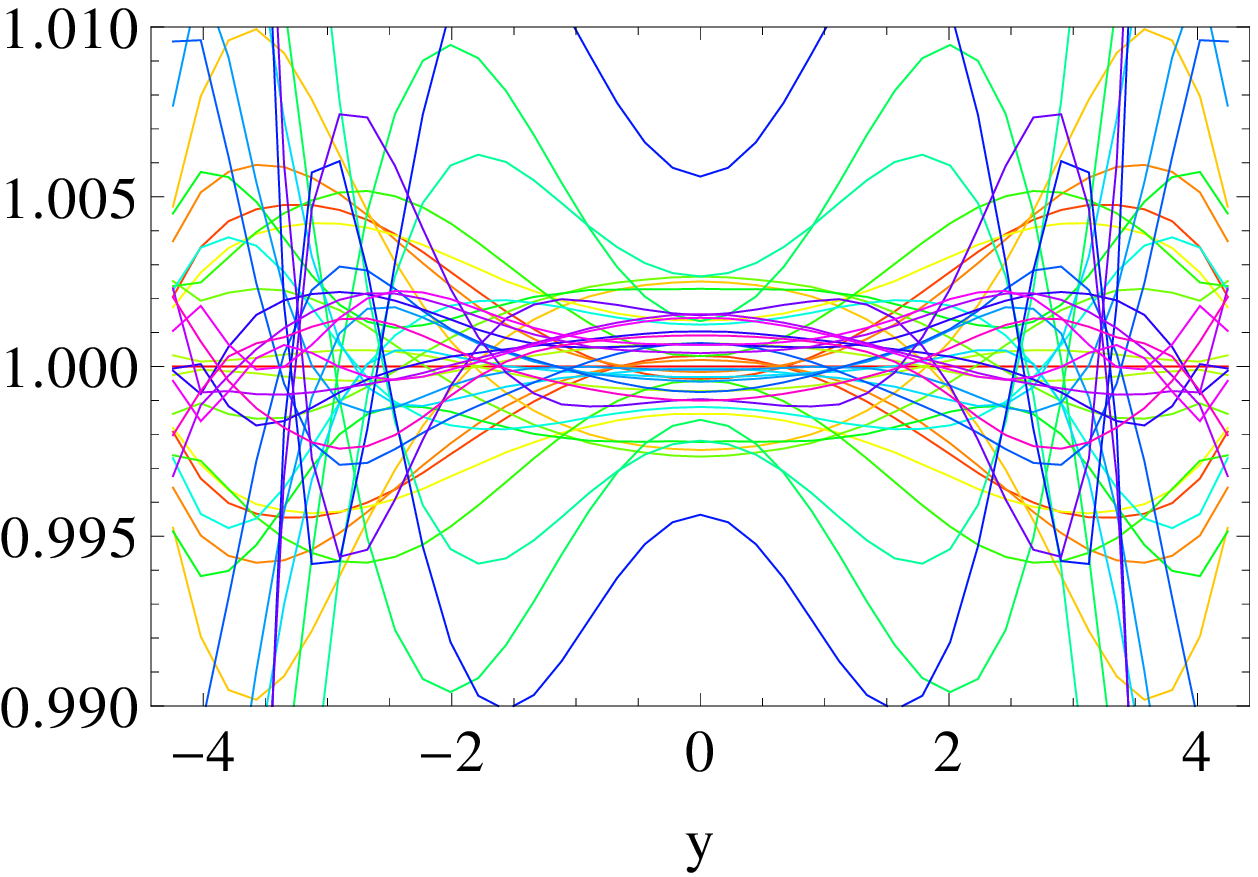}\quad{}\includegraphics[width=0.3\textwidth]{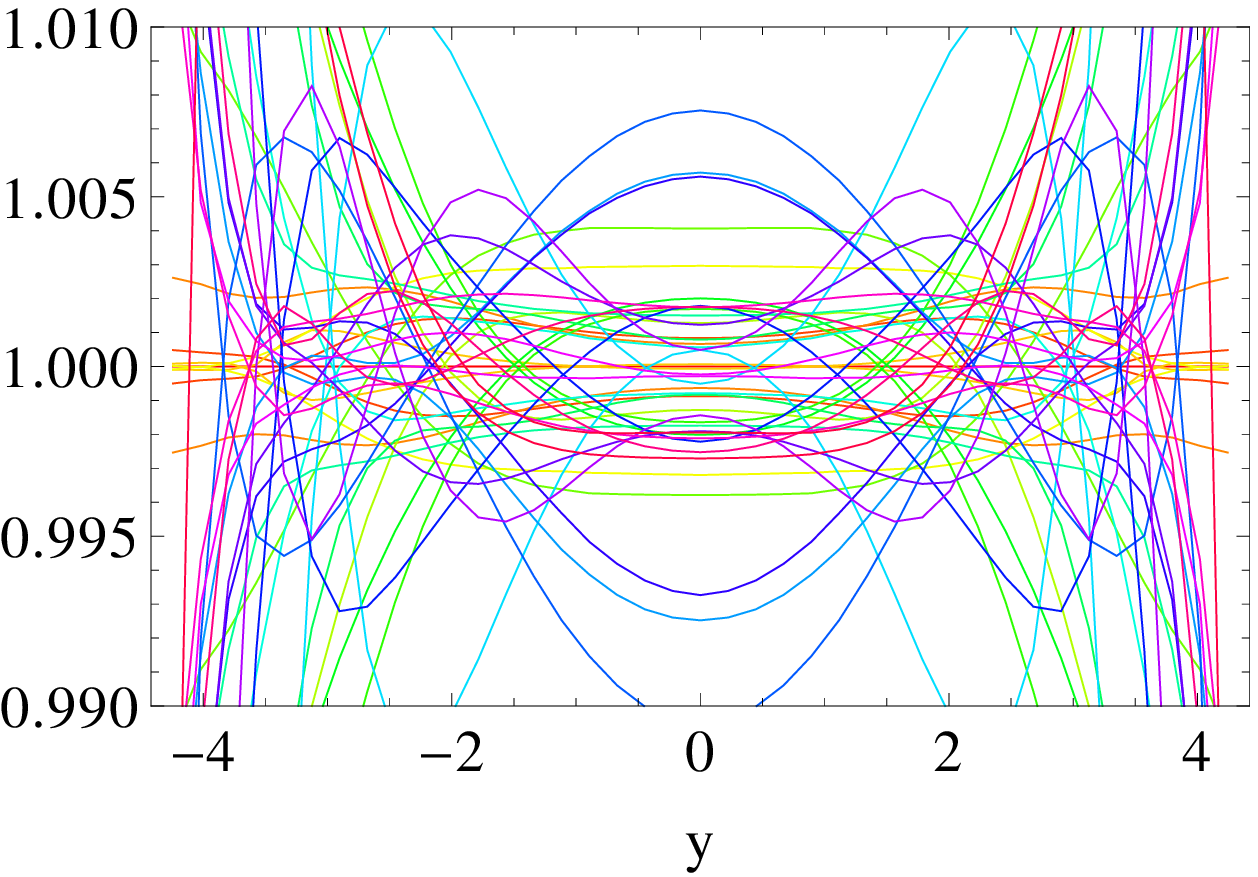}}

\subfloat[Double ratio $R$ as defined in Eq.~(\ref{eq:Ratio_RWZ}) for the
LHC for $\sqrt{S}=7$~TeV with CTEQ6.5 (left) and CTEQ6.6 (right)
at NNLO. \label{fig:DoubleLHC7}]{\includegraphics[width=0.3\textwidth]{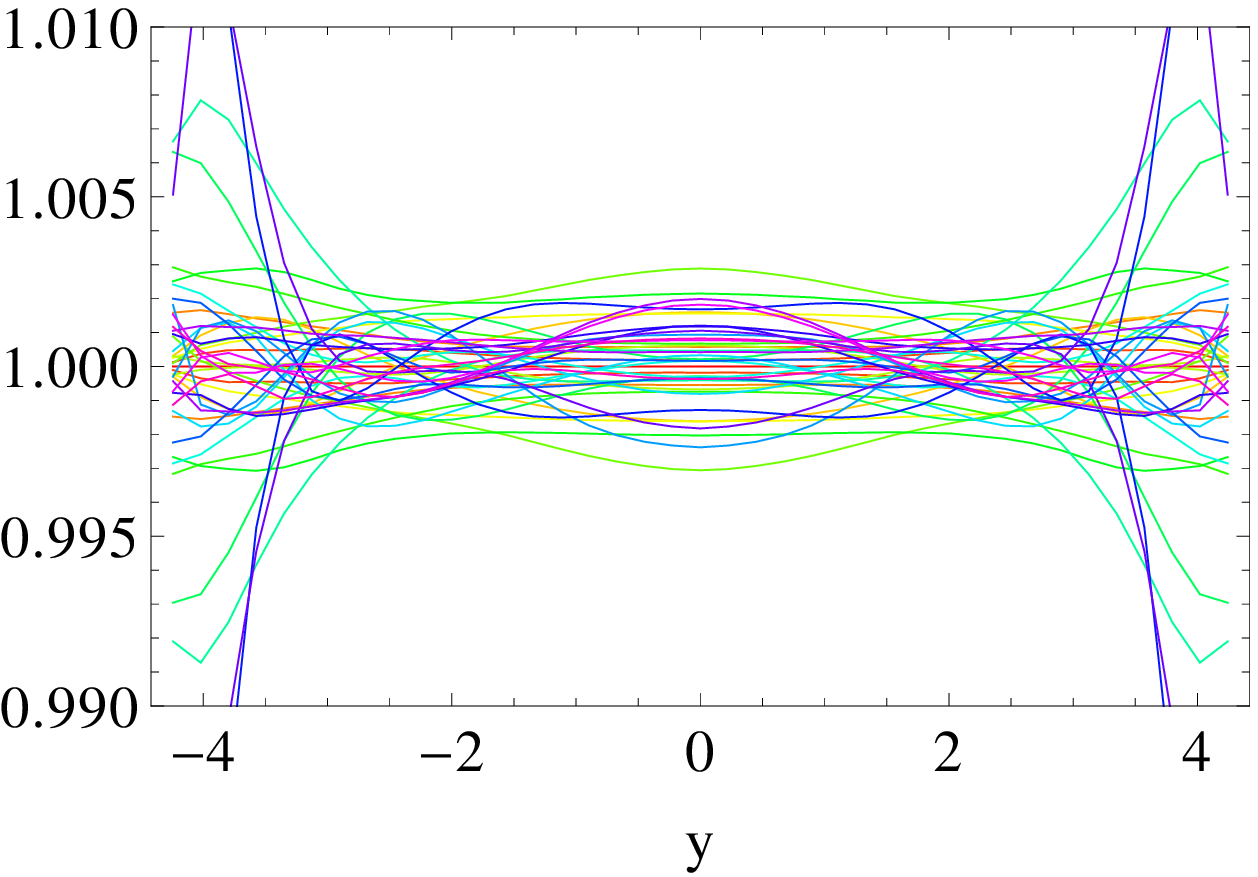}\quad{}\includegraphics[width=0.3\textwidth]{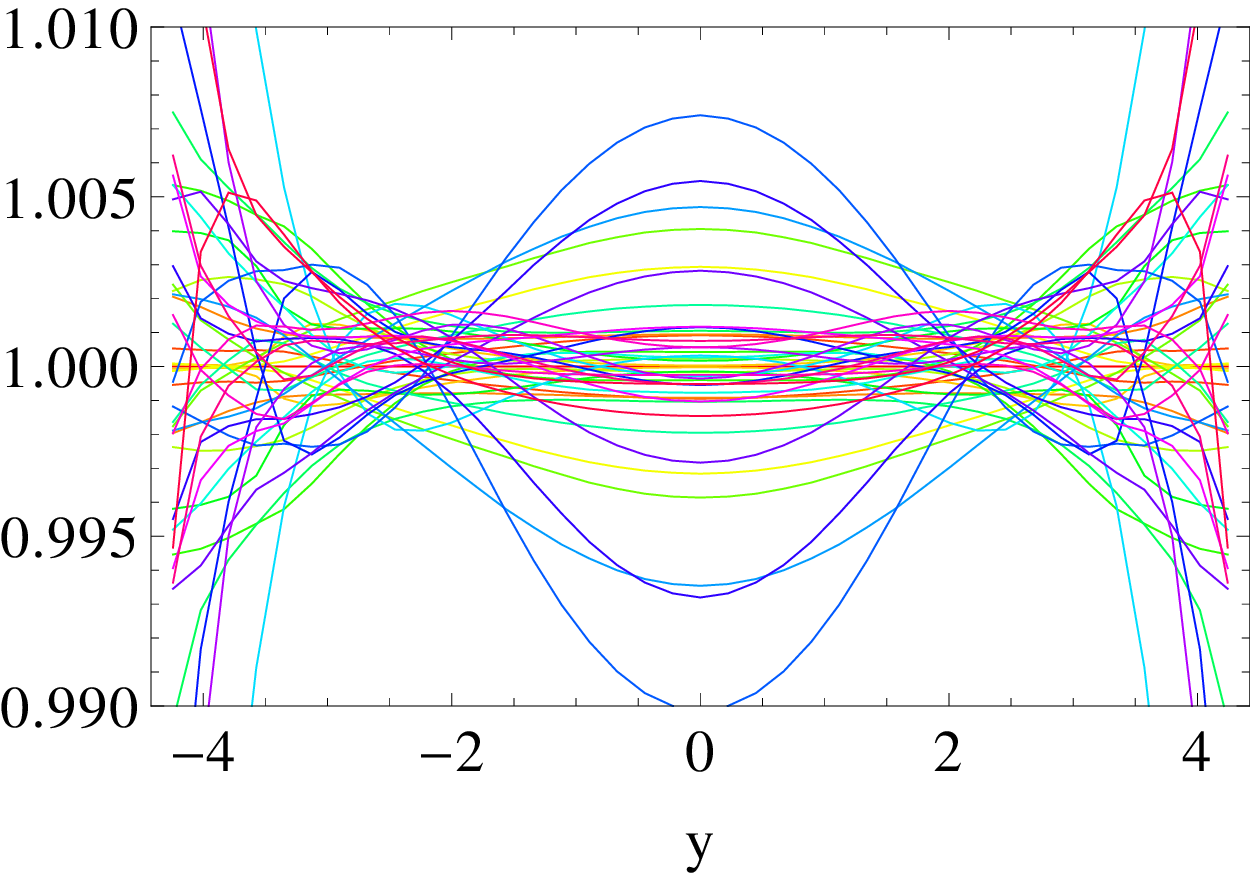}}\caption{Ratios of the differential $W^{\pm}$ and $Z$ production cross section
as defined in Eq.~(\ref{eq:ratio_RWZpm}) and Eq.~(\ref{eq:Ratio_RWZ})
at the LHC for $\sqrt{S}=7$~TeV.}
\end{figure*}

\begin{figure}
\includegraphics[width=0.45\textwidth]{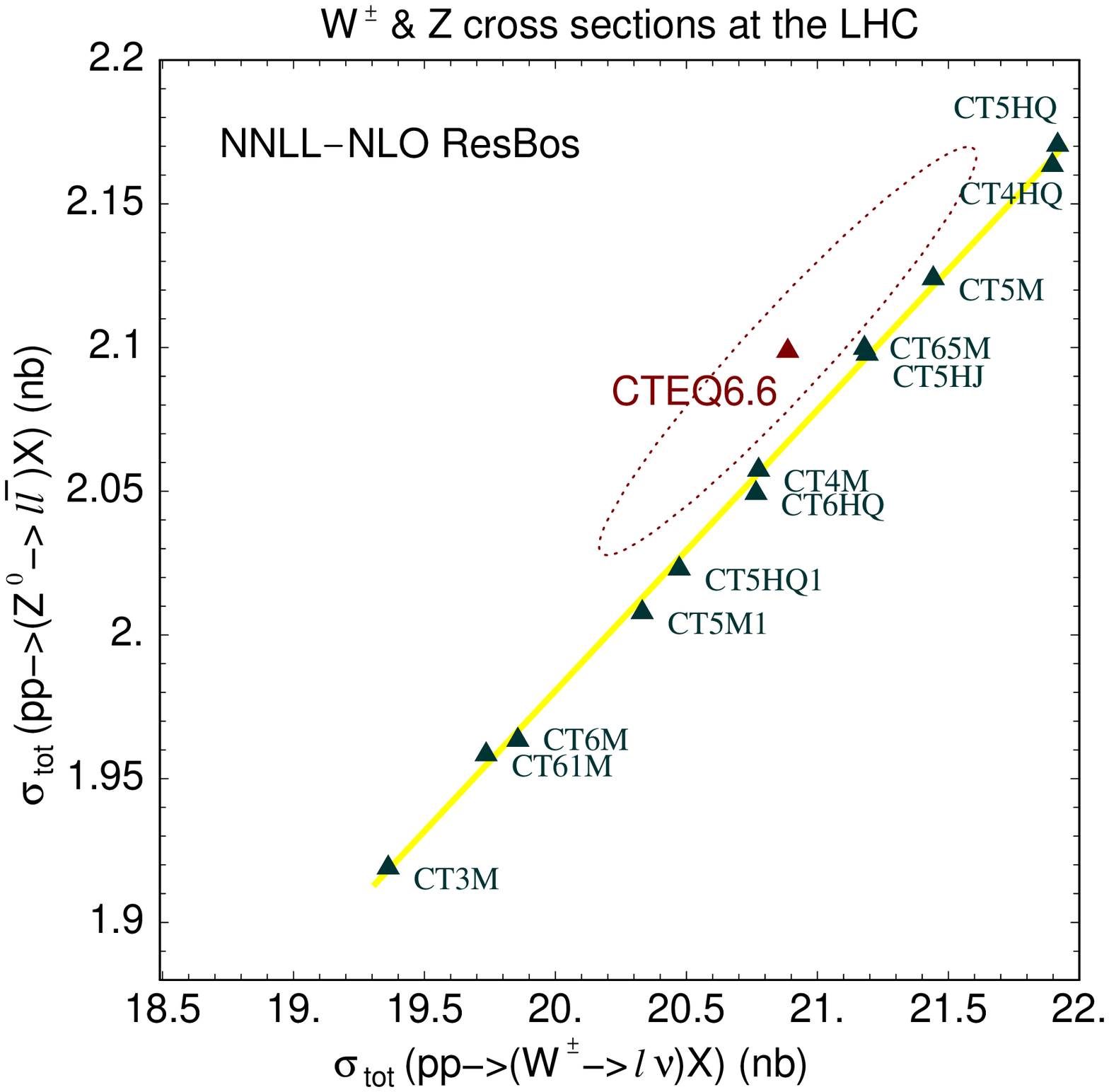}

\caption{Correlation of the $W^{\pm}$ and $Z$ cross sections for a selection
of PDF sets. Figure taken from Ref.~\cite{Nadolsky:2008zw}.\label{fig:SigWZ}}
\end{figure}

The leptonic decay modes of the $W/Z$ bosons provide a powerful tool
for precision measurements of electroweak parameters such as the $W$
boson mass. As the leptonic decay of the $W$ boson contains a neutrino
($W\to\ell\nu$), this process must be modeled to account for the
missing neutrino. The $W$ mass can then be measured by studying the
transverse momentum distribution of the decay lepton $\ell$ or the
transverse mass of the $\ell\nu$ pair. Performing this measurement,
the Drell-Yan $Z$ boson production process is used to calibrate the
leptonic $W$ process because the $Z$ can decay into two visible
leptons $Z\to\ell^{+}\ell^{-}$. This method works to the extent that
the production processes of the $W$ and $Z$ bosons are correlated. 

One possible measure to gauge the correlation of the PDF uncertainty
is the ratio of the $W$ and $Z$ boson differential cross section.
We compute $d^{2}\sigma/dM/dy$ for $W^{\pm}$ compared to $Z$, and
divide by the central PDF results to see the uncertainty band on a
relative scale. Schematically we define:\begin{eqnarray}
R^{\pm} & = & \left[\frac{d\sigma(W^{\pm})}{d\sigma(Z)}\right]/\left[\frac{d\sigma(W^{\pm})}{d\sigma(Z)}\right]_{0}\label{eq:ratio_RWZpm}\end{eqnarray}
where the {}``0'' subscript denotes the {}``central'' PDF set.
The resulting distributions are displayed in Fig.~\ref{fig:dyDoubleMinus7}
for $W^{-}$ production and in Fig.~\ref{fig:dyDoublePlus7} for
$W^{+}$ production. The left plot in each figure shows the distributions
for the CTEQ6.5 PDF set, and the right plots the distributions for
CTEQ6.6. We observe that the uncertainty band is generally $\pm\hspace{0.3mm}1\hspace{0.3mm}\%$
for central rapidities of $-2\le y_{W/Z}<+2$; this is smaller than
in the previous case, where the absolute uncertainty was investigated.
For larger rapidity ($\left|y_{W/Z}\right|>2$) the uncertainty band
exceeds the $\pm\hspace{0.3mm}1\hspace{0.3mm}\%$ range of the plot.

In Fig.~\ref{fig:DoubleLHC7}, we plot the sum of the differential
$W^{+}$ and $W^{-}$ cross sections with respect to the differential
$Z$ boson cross section, again normalized to the distribution of
the central PDF set. We define: 

\begin{eqnarray}
R & = & \left[\frac{d\sigma(W^{+}+W^{-})}{d\sigma(Z)}\right]/\left[\frac{d\sigma(W^{+}+W^{-})}{d\sigma(Z)}\right]_{0}\label{eq:Ratio_RWZ}\end{eqnarray}
for both the CTEQ6.5 and CTEQ6.6 PDFs. 

The contrast in Fig.~\ref{fig:DoubleLHC7} is striking. For the CTEQ6.5
PDFs, we observe that $W^{\pm}$ and $Z$ processes are strongly correlated,
while for the CTEQ6.6 the spread of the PDF band is substantially
larger.\textcolor{red}{\emph{ }}For example, the double ratio for
CTEQ6.5 has a spread of approximately $\pm0.2\%$ within the central
rapidity range of $-3\le y_{W/Z}\le+3$, while the uncertainty for
CTEQ6.6 is much wider in this rapidity region. 

The primary difference that is driving this result is the different
strange PDF. For CTEQ6.5 the strange quark was defined by Eq.~\eqref{eq:kappa}
while CTEQ6.6 introduced two extra fitting parameters which allowed
the strange PDF to vary independently from the up and down sea. Thus,
the uncertainty of the CTEQ6.6 distributions more accurately reflects
the true uncertainty. 

Another means to see how the additional freedom of the strange quark
introduces a decorrelation of the $W^{\pm}$ and $Z$ processes is
evident in Fig.~\ref{fig:SigWZ} which displays the correlation of
the $W^{\pm}$ and $Z$ boson cross sections for a selection of CTEQ
PDFs. Except for CTEQ6.6, all the PDFs make use of Eq.~\eqref{eq:kappa}
and yield results that lie along a straight line in the $\{\sigma_{W},\sigma_{Z}\}$
plane. Because CTEQ6.6 does not use Eq.~\eqref{eq:kappa}, the freedom
of the strange quark PDF is reflected in the freedom of the $W^{\pm}$
and $Z$ cross sections values. 

The above examples demonstrate the subtle features inherent in evaluating
the PDF uncertainties. For precision measurements it is important
to better constrain the parton distributions at the LHC, in particular
the strange and anti-strange quark PDFs.

\section{Conclusion}

We have investigated the constraints of the strange and anti-strange
PDFs and their impact on the Drell-Yan $W/Z$ boson production at
the LHC. 

Specifically, we observe that the strange quark is rather poorly constrained,
particularly in the low $x$ region which is sensitive to $W/Z$ production
at the LHC. Improved analyses from neutrino DIS measurements could
help reduce this uncertainty. Conversely, precision measurements of
$W/Z$ production at the LHC may provide input to the global PDF analyses
which could further constrain these distributions. 

In particular, the rapidity distribution of the $W/Z$ bosons provides
an incisive measure of the mix of valence and sea quarks, and the
prospect of measuring this at the LHC in the near future is excellent.

\section*{Appendix}

\subsection{PDF Fits with the Dimuon Data}

\begin{table*}[t]
\begin{tabular}{c|c|c|c|c|c|c}
\hline 
LO Dimuon & \# pts  & $B^{+}$  & $A$  & $B$  & $C$  & $B^{-}$\tabularnewline
\hline
\hline 
$[S^{-}]\times10^{4}$  & -  & 54.85  & 31.18  & 15.98  & 10.32  & -17.72\tabularnewline
\hline 
Entire Data  & 2465  & 1.015 (1.100)  & 1.001 (1.085)  & 1.000 (\noun{1.084})  & 1.002 (1.086)  & 1.018 (1.103)\tabularnewline
\hline 
Dimuon  & 174  & 1.289 (0.941)  & 1.018 (0.743)  & 1.000 (0.730)  & 0.996 (0.727)  & 1.241 (0.906)\tabularnewline
\hline 
Inclusive I  & 194  & 0.978 (0.711)  & 0.961 (0.699)  & 1.000 (0.727)  & 1.029 (0.748)  & 1.085 (0.789)\tabularnewline
\hline 
Dimuon+I  & 368  & 1.126 (0.820)  & 0.989 (0.720)  & 1.000 (0.728)  & 1.014 (0.738)  & 1.160 (0.844)\tabularnewline
\hline 
Inclusive II  & 2097  & 1.012 (1.120)  & 1.014 (1.123)  & 1.000 (1.107)  & 1.011 (1.119)  & 1.011 (1.119)\tabularnewline
\hline
\end{tabular}

\caption{Fit results using the LO dimuon calculation. We present the integrated
strange-quark asymmetry $[S^{-}]$, and the cells display $\chi^{2}/\chi_{0}^{2}$,
and the values in parentheses are the $\chi^{2}/DoF$ for each data
subset. Note $\chi_{0}^{2}$ is calibrated from the LO $B$-fit. \label{tab:lo}}
\end{table*}

\begin{table*}[t]
\begin{tabular}{c|c|c|c|c|c|c}
\hline 
NLO Dimuon & \# pts & $B^{+}$ & $A$ & $B$ & $C$ & $B^{-}$\tabularnewline
\hline
\hline 
$[S^{-}]\times10^{4}$ & - & 63.75 & 23.92 & 13.72 & 12.81 & -18.59\tabularnewline
\hline 
Entire Data & 2465 & 1.033 (1.120) & 0.995 (1.079) & 0.994 (1.077) & 0.994 (1.077) & 1.025 (1.111)\tabularnewline
\hline 
Dimuon & 174 & 1.660 (1.212) & 0.960 (0.701) & 0.936 (0.683) & 0.937 (0.684) & 1.416 (1.034)\tabularnewline
\hline 
Inclusive I & 194 & 0.977 (0.710) & 0.974 (0.708) & 1.008 (0.733) & 1.017 (0.739) & 1.088 (0.791)\tabularnewline
\hline 
Dimuon+I & 368 & 1.301 (0.947) & 0.968 (0.705) & 0.974 (0.709) & 0.979 (0.713) & 1.245 (0.906)\tabularnewline
\hline 
Inclusive II & 2097 & 1.012 (1.120) & 1.009 (1.117) & 1.007 (1.115) & 1.005 (1.113) & 1.010 (1.118)\tabularnewline
\hline
\end{tabular}

\caption{Fit results using the NLO dimuon calculation. We present the integrated
strange-quark asymmetry $[S^{-}]$, and the cells display $\chi^{2}/\chi_{0}^{2}$,
and the values in parentheses are the $\chi^{2}/DoF$ for each data
subset. \label{tab:nlo}}
\end{table*}

We have repeated the LO analysis of Ref.~\cite{Olness:2003wz} and
extended this using the NLO calculation for dimuon production~\cite{Kretzer:1998ju,Kretzer:2001tc}.
We have performed a series of fits to the data which includes the
dimuon data. The results of the $\{A,B^{+},B,B^{-},C\}$ fits%
\footnote{We follow the methodology and notation of Ref.~\cite{Olness:2003wz}.
See this reference for details. %
} using the LO dimuon analysis are shown in Table~\ref{tab:lo}, and
those with the NLO analysis are in Table~\ref{tab:nlo}. The fits
are sorted left-to-right by the integrated strange-quark asymmetry
$[S^{-}]$ (scaled by $10^{4}$). The cells display the $\chi^{2}$
relative to $\chi_{0}^{2}$ for the indicated data set where we choose
$\chi_{0}^{2}$ to be the $\chi^{2}$ value from the LO-$B$ fit;
this allows us to compare the incremental changes as we shift $[S^{-}]$
and alter the constraints. The values in parentheses are the $\chi^{2}/DoF$
for each data subset. The $B$-fit is the overall best fit to the
data. The $B^{+}$ and $B^{-}$ sets modify the $B$ fit using the
Lagrange multiplier method to determine the ranges of the $[S^{-}]$
parameter defined in Eq.~\eqref{eq:StrMom}. 

For example, the LO  $B^{+}$ fit demonstrates that we can increase
$[S^{-}]$ from 15.98 to 54.85, but the dimuon $\chi^{2}$ increases
by $33/174\sim21\%$ while the overall $\chi^{2}$ increases by only
$39/2465\sim2\%$; thus, the shift of $[S^{-}]$ is strongly constrained
by the dimuon data, and the remaining data are relatively insensitive
to this quantity. 

Comparing the NLO-$B$ fit to the LO-$B$ fit we note that $\chi^{2}/DoF$
has decreased both for the dimuon set and the entire data set; while
this decrease is not dramatic, it is encouraging to see that the proper
NLO treatment of the data results in an improved fit. As before, we
observe that for the NLO $B^{+}$ fit, we can increase $[S^{-}]$
from 13.72 to 63.75, but the dimuon $\chi^{2}$ increases by $92/174\sim53\%$
while the overall $\chi^{2}$ increases by only $106/2465\sim4\%$;
again, the shift of $[S^{-}]$ is primarily constrained by the dimuon
data.

In Figure~\ref{fig:chiDelS} we have plotted the ratio $\chi^{2}/\chi_{0}^{2}$
for the individual dimuon and ``Inclusive-I'' data sets~\cite{Olness:2003wz}
evaluated for a series of NLO $B$-fits as a function of the strange
asymmetry $[S^{-}]\times10^{4}$. This plot allows us to see the contribution
of each data set as we shift the strange asymmetry. Again the ``Inclusive-I''
data sets are essentially unchanged as the treatment of the dimuons
only affects these data indirectly. As before, this data set is mildly
sensitive to the dimuons, and weakly prefers larger values of $[S^{-}]$.

\section*{Acknowledgments}

We thank Sergey Alekhin, Tim Bolton, Maarten Boonekamp, Andrei Kataev,
Cynthia Keppel, Mieczyslaw Witold Krasny, Sergey Kulagin, Shunzo Kumano,
Dave Mason, Jorge Morfin, Pavel Nadolsky, Donna Naples, Jeff Owens,
Roberto Petti, Voica~A.~Radescu, Matthias Schott, Un-Ki Yang for
valuable discussions.

F.I.O., I.S., and J.Y.Y. acknowledge the hospitality of CERN, DESY,
Fermilab, and Les Houches where a portion of this work was performed.
This work was partially supported by the U.S. Department of Energy
under grant DE-FG02-04ER41299, and the Lightner-Sams Foundation. F.I.O
thanks the Galileo Galilei Institute for Theoretical Physics for their
hospitality and the INFN for partial support during the completion
of this work. The research of T.S. is supported by a fellowship from
the Th\'eorie LHC France initiative funded by the CNRS/IN2P3. This
work has been supported by \textit{Projet international de cooperation
scientifique} PICS05854 between France and the USA. The work of J.Y.Yu
was supported by the Deutsche Forschungsgemeinschaft (DFG) through
grant No. YU 118/1-1. The work of S.~B. is supported by the Initiative
and Networking Fund of the Helmholtz Association, contract HA-101
(`Physics at the Terascale') and by the Research Center `Elementary
Forces and Mathematical Foundations' of the Johannes-Gutenberg-Universit\"at
Mainz. 

\bibliographystyle{apsrev}
\bibliography{bibStrange}

\begin{thebibliography}{75}
\expandafter\ifx\csname natexlab\endcsname\relax\def\natexlab#1{#1}\fi
\expandafter\ifx\csname bibnamefont\endcsname\relax
  \def\bibnamefont#1{#1}\fi
\expandafter\ifx\csname bibfnamefont\endcsname\relax
  \def\bibfnamefont#1{#1}\fi
\expandafter\ifx\csname citenamefont\endcsname\relax
  \def\citenamefont#1{#1}\fi
\expandafter\ifx\csname url\endcsname\relax
  \def\url#1{\texttt{#1}}\fi
\expandafter\ifx\csname urlprefix\endcsname\relax\def\urlprefix{URL }\fi
\providecommand{\bibinfo}[2]{#2}
\providecommand{\eprint}[2][]{\url{#2}}

\bibitem[{\citenamefont{Abramowicz et~al.}(1982)}]{Abramowicz:1982zr}
\bibinfo{author}{\bibfnamefont{H.}~\bibnamefont{Abramowicz}}
  \bibnamefont{et~al.}, \bibinfo{journal}{Z. Phys.}
  \textbf{\bibinfo{volume}{C15}}, \bibinfo{pages}{19} (\bibinfo{year}{1982}).

\bibitem[{\citenamefont{Abramowicz et~al.}(1983)}]{Abramowicz:1982re}
\bibinfo{author}{\bibfnamefont{H.}~\bibnamefont{Abramowicz}}
  \bibnamefont{et~al.}, \bibinfo{journal}{Z. Phys.}
  \textbf{\bibinfo{volume}{C17}}, \bibinfo{pages}{283} (\bibinfo{year}{1983}).

\bibitem[{\citenamefont{Abramowicz et~al.}(1984)}]{Abramowicz:1984yk}
\bibinfo{author}{\bibfnamefont{H.}~\bibnamefont{Abramowicz}}
  \bibnamefont{et~al.}, \bibinfo{journal}{Z. Phys.}
  \textbf{\bibinfo{volume}{C25}}, \bibinfo{pages}{29} (\bibinfo{year}{1984}).

\bibitem[{\citenamefont{Berge et~al.}(1987)}]{Berge:1987zw}
\bibinfo{author}{\bibfnamefont{J.~P.} \bibnamefont{Berge}}
  \bibnamefont{et~al.}, \bibinfo{journal}{Z. Phys.}
  \textbf{\bibinfo{volume}{C35}}, \bibinfo{pages}{443} (\bibinfo{year}{1987}).

\bibitem[{\citenamefont{Aivazis et~al.}(1994)\citenamefont{Aivazis, Collins,
  Olness, and Tung}}]{Aivazis:1993pi}
\bibinfo{author}{\bibfnamefont{M.~A.~G.} \bibnamefont{Aivazis}},
  \bibinfo{author}{\bibfnamefont{J.~C.} \bibnamefont{Collins}},
  \bibinfo{author}{\bibfnamefont{F.~I.} \bibnamefont{Olness}},
  \bibnamefont{and} \bibinfo{author}{\bibfnamefont{W.-K.} \bibnamefont{Tung}},
  \bibinfo{journal}{Phys. Rev.} \textbf{\bibinfo{volume}{D50}},
  \bibinfo{pages}{3102} (\bibinfo{year}{1994}), \eprint{hep-ph/9312319}.

\bibitem[{\citenamefont{Bazarko et~al.}(1995)}]{Bazarko:1994tt}
\bibinfo{author}{\bibfnamefont{A.}~\bibnamefont{Bazarko}} \bibnamefont{et~al.}
  (\bibinfo{collaboration}{CCFR Collaboration}), \bibinfo{journal}{Z.Phys.}
  \textbf{\bibinfo{volume}{C65}}, \bibinfo{pages}{189} (\bibinfo{year}{1995}),
  \eprint{hep-ex/9406007}.

\bibitem[{\citenamefont{Gluck et~al.}(1996)\citenamefont{Gluck, Kretzer, and
  Reya}}]{Gluck:1996ve}
\bibinfo{author}{\bibfnamefont{M.}~\bibnamefont{Gluck}},
  \bibinfo{author}{\bibfnamefont{S.}~\bibnamefont{Kretzer}}, \bibnamefont{and}
  \bibinfo{author}{\bibfnamefont{E.}~\bibnamefont{Reya}},
  \bibinfo{journal}{Phys.Lett.} \textbf{\bibinfo{volume}{B380}},
  \bibinfo{pages}{171} (\bibinfo{year}{1996}), \bibinfo{note}{erratum-ibid.
  B405 (1997) 391}, \eprint{hep-ph/9603304}.

\bibitem[{\citenamefont{Alton et~al.}(2001)}]{Alton:2000ze}
\bibinfo{author}{\bibfnamefont{A.}~\bibnamefont{Alton}} \bibnamefont{et~al.}
  (\bibinfo{collaboration}{NuTeV Collaboration}),
  \bibinfo{journal}{Int.J.Mod.Phys.} \textbf{\bibinfo{volume}{A16S1B}},
  \bibinfo{pages}{764} (\bibinfo{year}{2001}), \eprint{hep-ex/0008068}.

\bibitem[{\citenamefont{Goncharov et~al.}(2001)}]{Goncharov:2001qe}
\bibinfo{author}{\bibfnamefont{M.}~\bibnamefont{Goncharov}}
  \bibnamefont{et~al.} (\bibinfo{collaboration}{NuTeV Collaboration}),
  \bibinfo{journal}{Phys.Rev.} \textbf{\bibinfo{volume}{D64}},
  \bibinfo{pages}{112006} (\bibinfo{year}{2001}), \eprint{hep-ex/0102049}.

\bibitem[{\citenamefont{Kretzer et~al.}(2004)\citenamefont{Kretzer, Lai,
  Olness, and Tung}}]{Kretzer:2003it}
\bibinfo{author}{\bibfnamefont{S.}~\bibnamefont{Kretzer}},
  \bibinfo{author}{\bibfnamefont{H.~L.} \bibnamefont{Lai}},
  \bibinfo{author}{\bibfnamefont{F.~I.} \bibnamefont{Olness}},
  \bibnamefont{and} \bibinfo{author}{\bibfnamefont{W.~K.} \bibnamefont{Tung}},
  \bibinfo{journal}{Phys. Rev.} \textbf{\bibinfo{volume}{D69}},
  \bibinfo{pages}{114005} (\bibinfo{year}{2004}), \eprint{hep-ph/0307022}.

\bibitem[{\citenamefont{Nadolsky et~al.}(2003)\citenamefont{Nadolsky,
  Kidonakis, Olness, and Yuan}}]{Nadolsky:2002jr}
\bibinfo{author}{\bibfnamefont{P.~M.} \bibnamefont{Nadolsky}},
  \bibinfo{author}{\bibfnamefont{N.}~\bibnamefont{Kidonakis}},
  \bibinfo{author}{\bibfnamefont{F.~I.} \bibnamefont{Olness}},
  \bibnamefont{and} \bibinfo{author}{\bibfnamefont{C.~P.} \bibnamefont{Yuan}},
  \bibinfo{journal}{Phys. Rev.} \textbf{\bibinfo{volume}{D67}},
  \bibinfo{pages}{074015} (\bibinfo{year}{2003}), \eprint{hep-ph/0210082}.

\bibitem[{\citenamefont{Pumplin et~al.}(2002)}]{Pumplin:2002vw}
\bibinfo{author}{\bibfnamefont{J.}~\bibnamefont{Pumplin}} \bibnamefont{et~al.},
  \bibinfo{journal}{JHEP} \textbf{\bibinfo{volume}{07}}, \bibinfo{pages}{012}
  (\bibinfo{year}{2002}), \eprint{hep-ph/0201195}.

\bibitem[{\citenamefont{Spentzouris}(2002)}]{Spentzouris:2002va}
\bibinfo{author}{\bibfnamefont{P.}~\bibnamefont{Spentzouris}}
  (\bibinfo{collaboration}{NuTeV}), \bibinfo{journal}{Acta Phys. Polon.}
  \textbf{\bibinfo{volume}{B33}}, \bibinfo{pages}{3843} (\bibinfo{year}{2002}).

\bibitem[{\citenamefont{Tung et~al.}(2002)\citenamefont{Tung, Kretzer, and
  Schmidt}}]{Tung:2001mv}
\bibinfo{author}{\bibfnamefont{W.-K.} \bibnamefont{Tung}},
  \bibinfo{author}{\bibfnamefont{S.}~\bibnamefont{Kretzer}}, \bibnamefont{and}
  \bibinfo{author}{\bibfnamefont{C.}~\bibnamefont{Schmidt}},
  \bibinfo{journal}{J. Phys.} \textbf{\bibinfo{volume}{G28}},
  \bibinfo{pages}{983} (\bibinfo{year}{2002}), \eprint{hep-ph/0110247}.

\bibitem[{\citenamefont{Tzanov et~al.}(2005)}]{Tzanov:2005dp}
\bibinfo{author}{\bibfnamefont{M.}~\bibnamefont{Tzanov}} \bibnamefont{et~al.}
  (\bibinfo{collaboration}{NuTeV}), \bibinfo{journal}{Int. J. Mod. Phys.}
  \textbf{\bibinfo{volume}{A20}}, \bibinfo{pages}{3759} (\bibinfo{year}{2005}).

\bibitem[{\citenamefont{Tzanov et~al.}(2006)}]{Tzanov:2005kr}
\bibinfo{author}{\bibfnamefont{M.}~\bibnamefont{Tzanov}} \bibnamefont{et~al.}
  (\bibinfo{collaboration}{NuTeV}), \bibinfo{journal}{Phys. Rev.}
  \textbf{\bibinfo{volume}{D74}}, \bibinfo{pages}{012008}
  (\bibinfo{year}{2006}), \eprint{hep-ex/0509010}.

\bibitem[{\citenamefont{Adams et~al.}(2010)}]{Adams:2009kp}
\bibinfo{author}{\bibfnamefont{T.}~\bibnamefont{Adams}} \bibnamefont{et~al.}
  (\bibinfo{collaboration}{NuSOnG}), \bibinfo{journal}{Int. J. Mod. Phys.}
  \textbf{\bibinfo{volume}{A25}}, \bibinfo{pages}{909} (\bibinfo{year}{2010}),
  \eprint{0906.3563}.

\bibitem[{\citenamefont{Stump et~al.}(2003)\citenamefont{Stump, Huston,
  Pumplin, Tung, Lai et~al.}}]{Stump:2003yu}
\bibinfo{author}{\bibfnamefont{D.}~\bibnamefont{Stump}},
  \bibinfo{author}{\bibfnamefont{J.}~\bibnamefont{Huston}},
  \bibinfo{author}{\bibfnamefont{J.}~\bibnamefont{Pumplin}},
  \bibinfo{author}{\bibfnamefont{W.-K.} \bibnamefont{Tung}},
  \bibinfo{author}{\bibfnamefont{H.}~\bibnamefont{Lai}}, \bibnamefont{et~al.},
  \bibinfo{journal}{JHEP} \textbf{\bibinfo{volume}{0310}}, \bibinfo{pages}{046}
  (\bibinfo{year}{2003}), \eprint{hep-ph/0303013}.

\bibitem[{\citenamefont{Nadolsky et~al.}(2008)\citenamefont{Nadolsky, Lai, Cao,
  Huston, Pumplin et~al.}}]{Nadolsky:2008zw}
\bibinfo{author}{\bibfnamefont{P.~M.} \bibnamefont{Nadolsky}},
  \bibinfo{author}{\bibfnamefont{H.-L.} \bibnamefont{Lai}},
  \bibinfo{author}{\bibfnamefont{Q.-H.} \bibnamefont{Cao}},
  \bibinfo{author}{\bibfnamefont{J.}~\bibnamefont{Huston}},
  \bibinfo{author}{\bibfnamefont{J.}~\bibnamefont{Pumplin}},
  \bibnamefont{et~al.}, \bibinfo{journal}{Phys.Rev.}
  \textbf{\bibinfo{volume}{D78}}, \bibinfo{pages}{013004}
  (\bibinfo{year}{2008}), \eprint{0802.0007}.

\bibitem[{\citenamefont{Martin et~al.}(2009)\citenamefont{Martin, Stirling,
  Thorne, and Watt}}]{Martin:2009iq}
\bibinfo{author}{\bibfnamefont{A.}~\bibnamefont{Martin}},
  \bibinfo{author}{\bibfnamefont{W.}~\bibnamefont{Stirling}},
  \bibinfo{author}{\bibfnamefont{R.}~\bibnamefont{Thorne}}, \bibnamefont{and}
  \bibinfo{author}{\bibfnamefont{G.}~\bibnamefont{Watt}},
  \bibinfo{journal}{Eur.Phys.J.} \textbf{\bibinfo{volume}{C63}},
  \bibinfo{pages}{189} (\bibinfo{year}{2009}), \eprint{0901.0002}.

\bibitem[{\citenamefont{Ball et~al.}(2010)}]{Ball:2010de}
\bibinfo{author}{\bibfnamefont{R.~D.} \bibnamefont{Ball}} \bibnamefont{et~al.},
  \bibinfo{journal}{Nucl. Phys.} \textbf{\bibinfo{volume}{B838}},
  \bibinfo{pages}{136} (\bibinfo{year}{2010}), \eprint{1002.4407}.

\bibitem[{\citenamefont{Alekhin et~al.}(2010)\citenamefont{Alekhin, Blumlein,
  Klein, and Moch}}]{Alekhin:2009ni}
\bibinfo{author}{\bibfnamefont{S.}~\bibnamefont{Alekhin}},
  \bibinfo{author}{\bibfnamefont{J.}~\bibnamefont{Blumlein}},
  \bibinfo{author}{\bibfnamefont{S.}~\bibnamefont{Klein}}, \bibnamefont{and}
  \bibinfo{author}{\bibfnamefont{S.}~\bibnamefont{Moch}},
  \bibinfo{journal}{Phys.Rev.} \textbf{\bibinfo{volume}{D81}},
  \bibinfo{pages}{014032} (\bibinfo{year}{2010}), \eprint{0908.2766}.

\bibitem[{\citenamefont{Jimenez-Delgado and
  Reya}(2009)}]{JimenezDelgado:2008hf}
\bibinfo{author}{\bibfnamefont{P.}~\bibnamefont{Jimenez-Delgado}}
  \bibnamefont{and} \bibinfo{author}{\bibfnamefont{E.}~\bibnamefont{Reya}},
  \bibinfo{journal}{Phys.Rev.} \textbf{\bibinfo{volume}{D79}},
  \bibinfo{pages}{074023} (\bibinfo{year}{2009}), \eprint{0810.4274}.

\bibitem[{\citenamefont{Mason et~al.}(2007)}]{Mason:2007zz}
\bibinfo{author}{\bibfnamefont{D.}~\bibnamefont{Mason}} \bibnamefont{et~al.}
  (\bibinfo{collaboration}{NuTeV Collaboration}),
  \bibinfo{journal}{Phys.Rev.Lett.} \textbf{\bibinfo{volume}{99}},
  \bibinfo{pages}{192001} (\bibinfo{year}{2007}).

\bibitem[{\citenamefont{Zeller et~al.}(2002)}]{Zeller:2002du}
\bibinfo{author}{\bibfnamefont{G.}~\bibnamefont{Zeller}} \bibnamefont{et~al.}
  (\bibinfo{collaboration}{NuTeV Collaboration}), \bibinfo{journal}{Phys.Rev.}
  \textbf{\bibinfo{volume}{D65}}, \bibinfo{pages}{111103}
  (\bibinfo{year}{2002}), \eprint{hep-ex/0203004}.

\bibitem[{\citenamefont{Yang et~al.}(2001)}]{Yang:2000ju}
\bibinfo{author}{\bibfnamefont{U.-K.} \bibnamefont{Yang}} \bibnamefont{et~al.}
  (\bibinfo{collaboration}{CCFR/NuTeV Collaboration}),
  \bibinfo{journal}{Phys.Rev.Lett.} \textbf{\bibinfo{volume}{86}},
  \bibinfo{pages}{2742} (\bibinfo{year}{2001}), \eprint{hep-ex/0009041}.

\bibitem[{\citenamefont{Olness et~al.}(2005)\citenamefont{Olness, Pumplin,
  Stump, Huston, Nadolsky et~al.}}]{Olness:2003wz}
\bibinfo{author}{\bibfnamefont{F.}~\bibnamefont{Olness}},
  \bibinfo{author}{\bibfnamefont{J.}~\bibnamefont{Pumplin}},
  \bibinfo{author}{\bibfnamefont{D.}~\bibnamefont{Stump}},
  \bibinfo{author}{\bibfnamefont{J.}~\bibnamefont{Huston}},
  \bibinfo{author}{\bibfnamefont{P.~M.} \bibnamefont{Nadolsky}},
  \bibnamefont{et~al.}, \bibinfo{journal}{Eur.Phys.J.}
  \textbf{\bibinfo{volume}{C40}}, \bibinfo{pages}{145} (\bibinfo{year}{2005}),
  \eprint{hep-ph/0312323}.

\bibitem[{\citenamefont{Airapetian et~al.}(2008)}]{Airapetian:2008qf}
\bibinfo{author}{\bibfnamefont{A.}~\bibnamefont{Airapetian}}
  \bibnamefont{et~al.} (\bibinfo{collaboration}{HERMES}),
  \bibinfo{journal}{Phys. Lett.} \textbf{\bibinfo{volume}{B666}},
  \bibinfo{pages}{446} (\bibinfo{year}{2008}), \eprint{0803.2993}.

\bibitem[{\citenamefont{Onengut et~al.}(2004)}]{Onengut:2004dy}
\bibinfo{author}{\bibfnamefont{G.}~\bibnamefont{Onengut}} \bibnamefont{et~al.}
  (\bibinfo{collaboration}{CHORUS}), \bibinfo{journal}{Phys. Lett.}
  \textbf{\bibinfo{volume}{B604}}, \bibinfo{pages}{11} (\bibinfo{year}{2004}).

\bibitem[{\citenamefont{Onengut et~al.}(2006)}]{Onengut:2005kv}
\bibinfo{author}{\bibfnamefont{G.}~\bibnamefont{Onengut}} \bibnamefont{et~al.}
  (\bibinfo{collaboration}{CHORUS}), \bibinfo{journal}{Phys. Lett.}
  \textbf{\bibinfo{volume}{B632}}, \bibinfo{pages}{65} (\bibinfo{year}{2006}).

\bibitem[{\citenamefont{Kayis-Topaksu et~al.}(2008)}]{KayisTopaksu:2008xp}
\bibinfo{author}{\bibfnamefont{A.}~\bibnamefont{Kayis-Topaksu}}
  \bibnamefont{et~al.} (\bibinfo{collaboration}{CHORUS}),
  \bibinfo{journal}{Nucl. Phys.} \textbf{\bibinfo{volume}{B798}},
  \bibinfo{pages}{1} (\bibinfo{year}{2008}), \eprint{0804.1869}.

\bibitem[{\citenamefont{Owens et~al.}(2007)\citenamefont{Owens, Huston, Keppel,
  Kuhlmann, Morfin et~al.}}]{Owens:2007kp}
\bibinfo{author}{\bibfnamefont{J.}~\bibnamefont{Owens}},
  \bibinfo{author}{\bibfnamefont{J.}~\bibnamefont{Huston}},
  \bibinfo{author}{\bibfnamefont{C.}~\bibnamefont{Keppel}},
  \bibinfo{author}{\bibfnamefont{S.}~\bibnamefont{Kuhlmann}},
  \bibinfo{author}{\bibfnamefont{J.}~\bibnamefont{Morfin}},
  \bibnamefont{et~al.}, \bibinfo{journal}{Phys.Rev.}
  \textbf{\bibinfo{volume}{D75}}, \bibinfo{pages}{054030}
  (\bibinfo{year}{2007}), \eprint{hep-ph/0702159}.

\bibitem[{\citenamefont{Astier et~al.}(2000)}]{Astier:2000us}
\bibinfo{author}{\bibfnamefont{P.}~\bibnamefont{Astier}} \bibnamefont{et~al.}
  (\bibinfo{collaboration}{NOMAD}), \bibinfo{journal}{Phys. Lett.}
  \textbf{\bibinfo{volume}{B486}}, \bibinfo{pages}{35} (\bibinfo{year}{2000}).

\bibitem[{\citenamefont{Petti}(2006)}]{Petti:2006tu}
\bibinfo{author}{\bibfnamefont{R.}~\bibnamefont{Petti}}
  (\bibinfo{collaboration}{NOMAD}), \bibinfo{journal}{Nucl. Phys. Proc. Suppl.}
  \textbf{\bibinfo{volume}{159}}, \bibinfo{pages}{56} (\bibinfo{year}{2006}),
  \eprint{hep-ex/0602022}.

\bibitem[{\citenamefont{Wu et~al.}(2008)}]{Wu:2007rv}
\bibinfo{author}{\bibfnamefont{Q.}~\bibnamefont{Wu}} \bibnamefont{et~al.}
  (\bibinfo{collaboration}{NOMAD}), \bibinfo{journal}{Phys. Lett.}
  \textbf{\bibinfo{volume}{B660}}, \bibinfo{pages}{19} (\bibinfo{year}{2008}),
  \eprint{0711.1183}.

\bibitem[{\citenamefont{de~Florian and Sassot}(2004)}]{deFlorian:2003qf}
\bibinfo{author}{\bibfnamefont{D.}~\bibnamefont{de~Florian}} \bibnamefont{and}
  \bibinfo{author}{\bibfnamefont{R.}~\bibnamefont{Sassot}},
  \bibinfo{journal}{Phys.Rev.} \textbf{\bibinfo{volume}{D69}},
  \bibinfo{pages}{074028} (\bibinfo{year}{2004}), \eprint{hep-ph/0311227}.

\bibitem[{\citenamefont{Hirai et~al.}(2007)\citenamefont{Hirai, Kumano, and
  Nagai}}]{Hirai:2007sx}
\bibinfo{author}{\bibfnamefont{M.}~\bibnamefont{Hirai}},
  \bibinfo{author}{\bibfnamefont{S.}~\bibnamefont{Kumano}}, \bibnamefont{and}
  \bibinfo{author}{\bibfnamefont{T.-H.} \bibnamefont{Nagai}},
  \bibinfo{journal}{Phys.Rev.} \textbf{\bibinfo{volume}{C76}},
  \bibinfo{pages}{065207} (\bibinfo{year}{2007}), \eprint{0709.3038}.

\bibitem[{\citenamefont{Eskola et~al.}(2009)\citenamefont{Eskola, Paukkunen,
  and Salgado}}]{Eskola:2009uj}
\bibinfo{author}{\bibfnamefont{K.}~\bibnamefont{Eskola}},
  \bibinfo{author}{\bibfnamefont{H.}~\bibnamefont{Paukkunen}},
  \bibnamefont{and} \bibinfo{author}{\bibfnamefont{C.}~\bibnamefont{Salgado}},
  \bibinfo{journal}{JHEP} \textbf{\bibinfo{volume}{0904}}, \bibinfo{pages}{065}
  (\bibinfo{year}{2009}), \eprint{0902.4154}.

\bibitem[{\citenamefont{Schienbein et~al.}(2008)\citenamefont{Schienbein, Yu,
  Keppel, Morfin, Olness et~al.}}]{Schienbein:2007fs}
\bibinfo{author}{\bibfnamefont{I.}~\bibnamefont{Schienbein}},
  \bibinfo{author}{\bibfnamefont{J.}~\bibnamefont{Yu}},
  \bibinfo{author}{\bibfnamefont{C.}~\bibnamefont{Keppel}},
  \bibinfo{author}{\bibfnamefont{J.}~\bibnamefont{Morfin}},
  \bibinfo{author}{\bibfnamefont{F.}~\bibnamefont{Olness}},
  \bibnamefont{et~al.}, \bibinfo{journal}{Phys.Rev.}
  \textbf{\bibinfo{volume}{D77}}, \bibinfo{pages}{054013}
  (\bibinfo{year}{2008}), \eprint{0710.4897}.

\bibitem[{\citenamefont{Schienbein et~al.}(2009)}]{Schienbein:2009kk}
\bibinfo{author}{\bibfnamefont{I.}~\bibnamefont{Schienbein}}
  \bibnamefont{et~al.}, \bibinfo{journal}{Phys. Rev.}
  \textbf{\bibinfo{volume}{D80}}, \bibinfo{pages}{094004}
  (\bibinfo{year}{2009}), \eprint{0907.2357}.

\bibitem[{\citenamefont{Kovarik et~al.}(2011{\natexlab{a}})}]{Kovarik:2010uv}
\bibinfo{author}{\bibfnamefont{K.}~\bibnamefont{Kovarik}} \bibnamefont{et~al.},
  \bibinfo{journal}{Phys. Rev. Lett.} \textbf{\bibinfo{volume}{106}},
  \bibinfo{pages}{122301} (\bibinfo{year}{2011}{\natexlab{a}}),
  \eprint{1012.0286}.

\bibitem[{\citenamefont{Kovarik et~al.}(2011{\natexlab{b}})}]{Kovarik:2011dr}
\bibinfo{author}{\bibfnamefont{K.}~\bibnamefont{Kovarik}} \bibnamefont{et~al.}
  (\bibinfo{year}{2011}{\natexlab{b}}), \eprint{1111.1145}.

\bibitem[{\citenamefont{Kovarik et~al.}(2011{\natexlab{c}})}]{Kovarik:2011zz}
\bibinfo{author}{\bibfnamefont{K.}~\bibnamefont{Kovarik}} \bibnamefont{et~al.},
  \bibinfo{journal}{AIP Conf. Proc.} \textbf{\bibinfo{volume}{1369}},
  \bibinfo{pages}{80} (\bibinfo{year}{2011}{\natexlab{c}}).

\bibitem[{\citenamefont{Aaltonen et~al.}(2008)}]{Aaltonen:2007dm}
\bibinfo{author}{\bibfnamefont{T.}~\bibnamefont{Aaltonen}} \bibnamefont{et~al.}
  (\bibinfo{collaboration}{CDF Collaboration}),
  \bibinfo{journal}{Phys.Rev.Lett.} \textbf{\bibinfo{volume}{100}},
  \bibinfo{pages}{091803} (\bibinfo{year}{2008}), \eprint{0711.2901}.

\bibitem[{\citenamefont{Abazov et~al.}(2008)}]{Abazov:2008qz}
\bibinfo{author}{\bibfnamefont{V.}~\bibnamefont{Abazov}} \bibnamefont{et~al.}
  (\bibinfo{collaboration}{D0 Collaboration}), \bibinfo{journal}{Phys.Lett.}
  \textbf{\bibinfo{volume}{B666}}, \bibinfo{pages}{23} (\bibinfo{year}{2008}),
  \eprint{0803.2259}.

\bibitem[{\citenamefont{Kawamura et~al.}(2011)\citenamefont{Kawamura, Kumano,
  and Kurihara}}]{Kawamura:2011dt}
\bibinfo{author}{\bibfnamefont{H.}~\bibnamefont{Kawamura}},
  \bibinfo{author}{\bibfnamefont{S.}~\bibnamefont{Kumano}}, \bibnamefont{and}
  \bibinfo{author}{\bibfnamefont{Y.}~\bibnamefont{Kurihara}},
  \bibinfo{journal}{Phys.Rev.} \textbf{\bibinfo{volume}{D84}},
  \bibinfo{pages}{114003} (\bibinfo{year}{2011}), \eprint{1110.6243}.

\bibitem[{\citenamefont{Besson et~al.}(2008)\citenamefont{Besson, Boonekamp,
  Klinkby, Mehlhase, and Petersen}}]{Besson:2008zs}
\bibinfo{author}{\bibfnamefont{N.}~\bibnamefont{Besson}},
  \bibinfo{author}{\bibfnamefont{M.}~\bibnamefont{Boonekamp}},
  \bibinfo{author}{\bibfnamefont{E.}~\bibnamefont{Klinkby}},
  \bibinfo{author}{\bibfnamefont{S.}~\bibnamefont{Mehlhase}}, \bibnamefont{and}
  \bibinfo{author}{\bibfnamefont{T.}~\bibnamefont{Petersen}}
  (\bibinfo{collaboration}{ATLAS Collaboration}),
  \bibinfo{journal}{Eur.Phys.J.} \textbf{\bibinfo{volume}{C57}},
  \bibinfo{pages}{627} (\bibinfo{year}{2008}), \eprint{0805.2093}.

\bibitem[{\citenamefont{Krasny et~al.}(2010)\citenamefont{Krasny, Dydak,
  Fayette, Placzek, and Siodmok}}]{Krasny:2010vd}
\bibinfo{author}{\bibfnamefont{M.~W.} \bibnamefont{Krasny}},
  \bibinfo{author}{\bibfnamefont{F.}~\bibnamefont{Dydak}},
  \bibinfo{author}{\bibfnamefont{F.}~\bibnamefont{Fayette}},
  \bibinfo{author}{\bibfnamefont{W.}~\bibnamefont{Placzek}}, \bibnamefont{and}
  \bibinfo{author}{\bibfnamefont{A.}~\bibnamefont{Siodmok}},
  \bibinfo{journal}{Eur. Phys. J.} \textbf{\bibinfo{volume}{C69}},
  \bibinfo{pages}{379} (\bibinfo{year}{2010}), \eprint{1004.2597}.

\bibitem[{\citenamefont{Chatrchyan
  et~al.}(2011{\natexlab{a}})}]{Chatrchyan:2011ya}
\bibinfo{author}{\bibfnamefont{S.}~\bibnamefont{Chatrchyan}}
  \bibnamefont{et~al.} (\bibinfo{collaboration}{CMS Collaboration}),
  \bibinfo{journal}{Phys.Rev.} \textbf{\bibinfo{volume}{D84}},
  \bibinfo{pages}{112002} (\bibinfo{year}{2011}{\natexlab{a}}),
  \eprint{1110.2682}.

\bibitem[{\citenamefont{Tarrade}(2011)}]{Tarrade:1320912}
\bibinfo{author}{\bibfnamefont{F.}~\bibnamefont{Tarrade}}, \bibinfo{type}{Tech.
  Rep.} \bibinfo{number}{ATL-PHYS-PROC-2011-003}, \bibinfo{institution}{CERN},
  \bibinfo{address}{Geneva} (\bibinfo{year}{2011}).

\bibitem[{\citenamefont{Chatrchyan
  et~al.}(2011{\natexlab{b}})}]{Chatrchyan:2011nv}
\bibinfo{author}{\bibfnamefont{S.}~\bibnamefont{Chatrchyan}}
  \bibnamefont{et~al.} (\bibinfo{collaboration}{CMS Collaboration}),
  \bibinfo{journal}{JHEP} \textbf{\bibinfo{volume}{1108}}, \bibinfo{pages}{117}
  (\bibinfo{year}{2011}{\natexlab{b}}), \eprint{1104.1617}.

\bibitem[{\citenamefont{Ball et~al.}(2011{\natexlab{a}})}]{NNPDF:2011aa}
\bibinfo{author}{\bibfnamefont{R.~D.} \bibnamefont{Ball}} \bibnamefont{et~al.}
  (\bibinfo{collaboration}{NNPDF Collaboration}), \bibinfo{journal}{Phys.Lett.}
  \textbf{\bibinfo{volume}{B704}}, \bibinfo{pages}{36}
  (\bibinfo{year}{2011}{\natexlab{a}}), \eprint{1102.3182}.

\bibitem[{\citenamefont{Peskin and Takeuchi}(1992)}]{Peskin:1991sw}
\bibinfo{author}{\bibfnamefont{M.~E.} \bibnamefont{Peskin}} \bibnamefont{and}
  \bibinfo{author}{\bibfnamefont{T.}~\bibnamefont{Takeuchi}},
  \bibinfo{journal}{Phys.Rev.} \textbf{\bibinfo{volume}{D46}},
  \bibinfo{pages}{381} (\bibinfo{year}{1992}).

\bibitem[{\citenamefont{Dittmar et~al.}(1997)\citenamefont{Dittmar, Pauss, and
  Zurcher}}]{Dittmar:1997md}
\bibinfo{author}{\bibfnamefont{M.}~\bibnamefont{Dittmar}},
  \bibinfo{author}{\bibfnamefont{F.}~\bibnamefont{Pauss}}, \bibnamefont{and}
  \bibinfo{author}{\bibfnamefont{D.}~\bibnamefont{Zurcher}},
  \bibinfo{journal}{Phys.Rev.} \textbf{\bibinfo{volume}{D56}},
  \bibinfo{pages}{7284} (\bibinfo{year}{1997}), \eprint{hep-ex/9705004}.

\bibitem[{\citenamefont{Alekhin et~al.}(2011)\citenamefont{Alekhin, Blumlein,
  Jimenez-Delgado, Moch, and Reya}}]{Alekhin:2010dd}
\bibinfo{author}{\bibfnamefont{S.}~\bibnamefont{Alekhin}},
  \bibinfo{author}{\bibfnamefont{J.}~\bibnamefont{Blumlein}},
  \bibinfo{author}{\bibfnamefont{P.}~\bibnamefont{Jimenez-Delgado}},
  \bibinfo{author}{\bibfnamefont{S.}~\bibnamefont{Moch}}, \bibnamefont{and}
  \bibinfo{author}{\bibfnamefont{E.}~\bibnamefont{Reya}},
  \bibinfo{journal}{Phys.Lett.} \textbf{\bibinfo{volume}{B697}},
  \bibinfo{pages}{127} (\bibinfo{year}{2011}), \eprint{1011.6259}.

\bibitem[{\citenamefont{Blumlein et~al.}(2011)\citenamefont{Blumlein,
  Hasselhuhn, Kovacikova, and Moch}}]{Blumlein:2011zu}
\bibinfo{author}{\bibfnamefont{J.}~\bibnamefont{Blumlein}},
  \bibinfo{author}{\bibfnamefont{A.}~\bibnamefont{Hasselhuhn}},
  \bibinfo{author}{\bibfnamefont{P.}~\bibnamefont{Kovacikova}},
  \bibnamefont{and} \bibinfo{author}{\bibfnamefont{S.}~\bibnamefont{Moch}},
  \bibinfo{journal}{Phys.Lett.} \textbf{\bibinfo{volume}{B700}},
  \bibinfo{pages}{294} (\bibinfo{year}{2011}), \eprint{1104.3449}.

\bibitem[{\citenamefont{Alekhin et~al.}(2012)\citenamefont{Alekhin, Blumlein,
  and Moch}}]{Alekhin:2012ig}
\bibinfo{author}{\bibfnamefont{S.}~\bibnamefont{Alekhin}},
  \bibinfo{author}{\bibfnamefont{J.}~\bibnamefont{Blumlein}}, \bibnamefont{and}
  \bibinfo{author}{\bibfnamefont{S.}~\bibnamefont{Moch}}
  (\bibinfo{year}{2012}), \eprint{1202.2281}.

\bibitem[{\citenamefont{Aad et~al.}(2011)}]{Aad:2011dm}
\bibinfo{author}{\bibfnamefont{G.}~\bibnamefont{Aad}} \bibnamefont{et~al.}
  (\bibinfo{collaboration}{ATLAS Collaboration}) (\bibinfo{year}{2011}),
  \eprint{1109.5141}.

\bibitem[{\citenamefont{Aad et~al.}(2012)}]{Collaboration:2012sb}
\bibinfo{author}{\bibfnamefont{G.}~\bibnamefont{Aad}} \bibnamefont{et~al.}
  (\bibinfo{collaboration}{ATLAS Collaboration}) (\bibinfo{year}{2012}),
  \eprint{1203.4051}.

\bibitem[{\citenamefont{Chatrchyan
  et~al.}(2011{\natexlab{c}})}]{Chatrchyan:2011wt}
\bibinfo{author}{\bibfnamefont{S.}~\bibnamefont{Chatrchyan}}
  \bibnamefont{et~al.} (\bibinfo{collaboration}{CMS Collaboration})
  (\bibinfo{year}{2011}{\natexlab{c}}), \eprint{1110.4973}.

\bibitem[{\citenamefont{Chatrchyan et~al.}(2011{\natexlab{d}})}]{CMS:2011aa}
\bibinfo{author}{\bibfnamefont{S.}~\bibnamefont{Chatrchyan}}
  \bibnamefont{et~al.} (\bibinfo{collaboration}{CMS Collaboration}),
  \bibinfo{journal}{JHEP} \textbf{\bibinfo{volume}{1110}}, \bibinfo{pages}{132}
  (\bibinfo{year}{2011}{\natexlab{d}}), \eprint{1107.4789}.

\bibitem[{\citenamefont{{\, }}(2011{\natexlab{a}})}]{CMS-PAS-EWK-10-011}
\bibinfo{author}{\bibnamefont{{\, }}} (\bibinfo{collaboration}{CMS
  Collaboration}), \bibinfo{journal}{CMS-PAS-EWK-10-011}
  (\bibinfo{year}{2011}{\natexlab{a}}).

\bibitem[{\citenamefont{{\, }}(2011{\natexlab{b}})}]{CMS-PAS-EWK-11-005}
\bibinfo{author}{\bibnamefont{{\, }}} (\bibinfo{collaboration}{CMS
  Collaboration}), \bibinfo{journal}{CMS-PAS-EWK-11-005}
  (\bibinfo{year}{2011}{\natexlab{b}}).

\bibitem[{\citenamefont{Chatrchyan
  et~al.}(2011{\natexlab{e}})}]{Chatrchyan:2011jz}
\bibinfo{author}{\bibfnamefont{S.}~\bibnamefont{Chatrchyan}}
  \bibnamefont{et~al.} (\bibinfo{collaboration}{CMS Collaboration}),
  \bibinfo{journal}{JHEP} \textbf{\bibinfo{volume}{1104}}, \bibinfo{pages}{050}
  (\bibinfo{year}{2011}{\natexlab{e}}), \eprint{1103.3470}.

\bibitem[{\citenamefont{{Amhis, Yasmine et al.}}(2012)}]{Amhis:2012gj}
\bibinfo{author}{\bibnamefont{{Amhis, Yasmine et al.}}}
  (\bibinfo{collaboration}{LHCb Collaboration}) (\bibinfo{year}{2012}),
  \eprint{1202.0654}.

\bibitem[{\citenamefont{Shears}(2011)}]{Shears:1394600}
\bibinfo{author}{\bibfnamefont{T.}~\bibnamefont{Shears}}
  (\bibinfo{collaboration}{LHCb Collaboration}),
  \bibinfo{journal}{LHCb-Proc-2011-076}  (\bibinfo{year}{2011}).

\bibitem[{\citenamefont{{\, }}(2011{\natexlab{c}})}]{CMS11013}
\bibinfo{author}{\bibnamefont{{\, }}} (\bibinfo{collaboration}{CMS
  Collaboration}), \bibinfo{journal}{CMS-PAS-EWK-11-013}
  (\bibinfo{year}{2011}{\natexlab{c}}).

\bibitem[{\citenamefont{Anastasiou et~al.}(2004)\citenamefont{Anastasiou,
  Dixon, Melnikov, and Petriello}}]{Anastasiou:2003ds}
\bibinfo{author}{\bibfnamefont{C.}~\bibnamefont{Anastasiou}},
  \bibinfo{author}{\bibfnamefont{L.~J.} \bibnamefont{Dixon}},
  \bibinfo{author}{\bibfnamefont{K.}~\bibnamefont{Melnikov}}, \bibnamefont{and}
  \bibinfo{author}{\bibfnamefont{F.}~\bibnamefont{Petriello}},
  \bibinfo{journal}{Phys.Rev.} \textbf{\bibinfo{volume}{D69}},
  \bibinfo{pages}{094008} (\bibinfo{year}{2004}), \eprint{hep-ph/0312266}.

\bibitem[{\citenamefont{Ball et~al.}(2011{\natexlab{b}})\citenamefont{Ball,
  Bertone, Cerutti, Del~Debbio, Forte et~al.}}]{Ball:2011mu}
\bibinfo{author}{\bibfnamefont{R.~D.} \bibnamefont{Ball}},
  \bibinfo{author}{\bibfnamefont{V.}~\bibnamefont{Bertone}},
  \bibinfo{author}{\bibfnamefont{F.}~\bibnamefont{Cerutti}},
  \bibinfo{author}{\bibfnamefont{L.}~\bibnamefont{Del~Debbio}},
  \bibinfo{author}{\bibfnamefont{S.}~\bibnamefont{Forte}},
  \bibnamefont{et~al.}, \bibinfo{journal}{Nucl.Phys.}
  \textbf{\bibinfo{volume}{B849}}, \bibinfo{pages}{296}
  (\bibinfo{year}{2011}{\natexlab{b}}), \eprint{1101.1300}.

\bibitem[{\citenamefont{Lai et~al.}(2010)\citenamefont{Lai, Guzzi, Huston, Li,
  Nadolsky et~al.}}]{Lai:2010vv}
\bibinfo{author}{\bibfnamefont{H.-L.} \bibnamefont{Lai}},
  \bibinfo{author}{\bibfnamefont{M.}~\bibnamefont{Guzzi}},
  \bibinfo{author}{\bibfnamefont{J.}~\bibnamefont{Huston}},
  \bibinfo{author}{\bibfnamefont{Z.}~\bibnamefont{Li}},
  \bibinfo{author}{\bibfnamefont{P.~M.} \bibnamefont{Nadolsky}},
  \bibnamefont{et~al.}, \bibinfo{journal}{Phys.Rev.}
  \textbf{\bibinfo{volume}{D82}}, \bibinfo{pages}{074024}
  (\bibinfo{year}{2010}), \eprint{1007.2241}.

\bibitem[{\citenamefont{Tung et~al.}(2007)\citenamefont{Tung, Lai, Belyaev,
  Pumplin, Stump et~al.}}]{Tung:2006tb}
\bibinfo{author}{\bibfnamefont{W.}~\bibnamefont{Tung}},
  \bibinfo{author}{\bibfnamefont{H.}~\bibnamefont{Lai}},
  \bibinfo{author}{\bibfnamefont{A.}~\bibnamefont{Belyaev}},
  \bibinfo{author}{\bibfnamefont{J.}~\bibnamefont{Pumplin}},
  \bibinfo{author}{\bibfnamefont{D.}~\bibnamefont{Stump}},
  \bibnamefont{et~al.}, \bibinfo{journal}{JHEP}
  \textbf{\bibinfo{volume}{0702}}, \bibinfo{pages}{053} (\bibinfo{year}{2007}),
  \eprint{hep-ph/0611254}.

\bibitem[{\citenamefont{Aaron et~al.}(2010)}]{:2009wt}
\bibinfo{author}{\bibfnamefont{F.}~\bibnamefont{Aaron}} \bibnamefont{et~al.}
  (\bibinfo{collaboration}{H1 and ZEUS Collaboration}), \bibinfo{journal}{JHEP}
  \textbf{\bibinfo{volume}{1001}}, \bibinfo{pages}{109} (\bibinfo{year}{2010}),
  \eprint{0911.0884}.

\bibitem[{\citenamefont{Martin et~al.}(2004)\citenamefont{Martin, Roberts,
  Stirling, and Thorne}}]{Martin:2004ir}
\bibinfo{author}{\bibfnamefont{A.}~\bibnamefont{Martin}},
  \bibinfo{author}{\bibfnamefont{R.}~\bibnamefont{Roberts}},
  \bibinfo{author}{\bibfnamefont{W.}~\bibnamefont{Stirling}}, \bibnamefont{and}
  \bibinfo{author}{\bibfnamefont{R.}~\bibnamefont{Thorne}},
  \bibinfo{journal}{Phys.Lett.} \textbf{\bibinfo{volume}{B604}},
  \bibinfo{pages}{61} (\bibinfo{year}{2004}), \eprint{hep-ph/0410230}.

\bibitem[{\citenamefont{Kretzer and Schienbein}(1998)}]{Kretzer:1998ju}
\bibinfo{author}{\bibfnamefont{S.}~\bibnamefont{Kretzer}} \bibnamefont{and}
  \bibinfo{author}{\bibfnamefont{I.}~\bibnamefont{Schienbein}},
  \bibinfo{journal}{Phys. Rev.} \textbf{\bibinfo{volume}{D58}},
  \bibinfo{pages}{094035} (\bibinfo{year}{1998}), \eprint{hep-ph/9805233}.

\bibitem[{\citenamefont{Kretzer et~al.}(2002)\citenamefont{Kretzer, Mason, and
  Olness}}]{Kretzer:2001tc}
\bibinfo{author}{\bibfnamefont{S.}~\bibnamefont{Kretzer}},
  \bibinfo{author}{\bibfnamefont{D.}~\bibnamefont{Mason}}, \bibnamefont{and}
  \bibinfo{author}{\bibfnamefont{F.}~\bibnamefont{Olness}},
  \bibinfo{journal}{Phys.Rev.} \textbf{\bibinfo{volume}{D65}},
  \bibinfo{pages}{074010} (\bibinfo{year}{2002}), \eprint{hep-ph/0112191}.

\end{thebibliography}

\end{document}